\documentclass[twocolumn,aps,prb,
10pt,
amssymb,superscriptaddress]{revtex4-2}

\usepackage{hyperref}
\usepackage{amsmath}
\usepackage{amsfonts}
\usepackage{graphicx}
\usepackage{appendix}
\usepackage{dsfont}
\usepackage{amssymb}
\usepackage{bm}
\usepackage{color}
\usepackage{soul}
\usepackage{natbib}
\setstcolor{red}
\usepackage{nccmath}
\usepackage{array}
\usepackage{footmisc}
\usepackage{comment}

\newcommand{\beq}{\begin{equation*}}
\newcommand{\eneq}{\end{equation*}}
\mathchardef\mhyphen="2D

\begin{document}
\title{Tunneling current and current correlations for anyonic quasiparticles\\ of $\nu=1/2$ chiral Luttinger liquid in multi-edge geometries}

\author{Gu Zhang}
\affiliation{Beijing Academy of Quantum Information Sciences, Beijing 100193, China}
\email{zhanggu@baqis.ac.cn}

\author{Domenico Giuliano}
\affiliation{
Dipartimento di Fisica, Universit\`a della Calabria Arcavacata di 
Rende I-87036, Cosenza, Italy}
\affiliation{I.N.F.N., Gruppo collegato di Cosenza, 
Arcavacata di Rende I-87036, Cosenza, Italy}

\author{Igor V. Gornyi}
\affiliation{Institute for Quantum Materials and Technologies and Institut f\"ur Theorie der Kondensierten Materie, Karlsruhe Institute of Technology, 76131 Karlsruhe, Germany}

\author{Gabriele Campagnano}
\affiliation{CNR-SPIN, c/o Complesso di Monte S. Angelo, via Cinthia, 80126 Napoli, Italy}

\begin{abstract}
We consider anyonic quasiparticles with charge $e/2$ described by the $\nu=1/2$ chiral Luttinger liquid, which collide in a Hong–Ou–Mandel-like interferometer. 
These colliding anyonic channels can be formally viewed as hosting Laughlin-like fractional $\nu=1/2$ quasiparticles.
More specifically, two possible geometries are considered: (i) a two-edge-channel setup where anyons originate from equilibrium reservoirs; (ii) a four-edge-channel setup where nonequilibrium anyons arrive at the collider in the form of diluted beams. For both setups, we calculate the tunneling current and the current correlations. 
For setup (i), our results provide analytically exact expressions for the tunneling current, tunneling-current noise, and cross-correlation noise. An exact relation between conductance and noises is explicitly demonstrated.
For setup (ii), we show that the tunneling current and the generalized Fano factor [defined in B. Rosenow \textit{et al.} (2016)] are finite for diluted streams of $\nu = 1/2$ anyons.  
This is due to the processes where nonequilibrium anyons, supplied via either source edge, directly tunnel at the central QPC. Thus, to obtain meaningful results in this case, one should go beyond the so-called time-domain braiding processes, where nonequilibrium anyons do not tunnel at the collider, but rather indirectly influence the tunneling by braiding with the quasiparticle-quasihole pairs created at the collider. This suggests that the effect of direct tunneling and collisions of diluted anyons in the Hong-Ou-Mandel interferometer can be important for various observables in physical quantum-Hall edges at Laughlin filling fractions.  
\end{abstract}

\maketitle

\section{introduction}

Anyons are particles living in a two-dimensional space; in contrast to particles existing in three dimensions, anyonic quantum statistics can differ from the bosonic and the fermionic cases~[\onlinecite{Leinaas:1977}]. It has been predicted that excitations of quantum Hall (QH) systems 
at fractional filling (fractional QH effect) should provide a physical realization of such anyons~[\onlinecite{Arovas:1984}].
This prediction has stimulated a large amount of work, through which the fractional charge was observed~[\onlinecite{de-Picciotto:1997, Saminadayar:1997}] decades ago, and anyonic statistics~[\onlinecite{Nakamura:2020, Bartolomei:2020, NakamuraNC22,
Glidic:2023,
Ruelle:2023,
Nakamura:2023,
LeeNature23}] in recent years.
Subsequent work suggested that, for some specific filling fractions, these systems may support non-Abelian anyons~[\onlinecite{MooreRead91}] that could be considered as a possible platform for topological quantum computation~[\onlinecite{Najak:2008}].

Access to the bulk properties of QH liquids is not easily carried out. However, as it was shown by Wen~[\onlinecite{Wen:1991}], there exists a correspondence between the bulk and the edges excitation ({\em anyons}) of a QH liquid. Because of the chiral propagation of excitations along the edges of the quantum Hall liquid, such solid-state devices may mimic optical interferometers. A crucial question here concerns the effect of anyonic statistics on interference, which is generically demanding to be probed in conventional optical setups (see, however, Ref.~\cite{FrancesconiQIAM2021}). To date, there have been theoretical and experimental investigations of Fabry-Perot~\cite{Chamon:1997, Camino:2005, Bonderson:2006, Bonderson:2006b, Kim:2008, Bishara:2009, Willett:2009, Halperin:2011, Rosenow:2020, Nakamura:2020, Nakamura:2023}, Mach-Zehnder~\cite{Feldman:2006, Feldman:2007, Ponomarenko:2007, Ponomarenko:2009, Bonderson:2008, Law:2008, Levkivskyi:2009, Ponomarenko:2010, Wang:2010, Levkivskyi:2012, Ganeshan:2012, Yang:2015, Batra:2023}, Hanbury Brown and Twiss~\cite{Safi:2001, Vishveshwara:2003, Kim:2005, Kim:2006, Campagnano:2012, Campagnano:2013}, and Hong–Ou–Mandel (HOM) interferometers~\cite{Rosenow:2016, Han:2016, Bartolomei:2020, Lee:2022, Jonckheere:2023, Schiller:2023, Glidic:2023, Ruelle:2023, AndreevX23, IyerX2023, ThammBerndPRL24} based on fractional QH edges, which have been providing promising signatures of fractional statistics.
Yet, often the intrinsic difficulties in the calculations and the lack of reliable numerical methods pose severe limitations to the theoretical analysis of these interferometric setups.

In this work, we address anyonic states within the model of chiral Luttinger liquid \cite{Wen:1991} with parameter $\nu=1/2$. While edge states of a Laughlin QH system are described by a chiral Luttinger liquid with $\nu=1/(2m+1)$, there are several good reasons to study the (seemingly unphysical) edge states at $\nu=1/2$. First, working at $\nu=1/2$ allows for employing the refermionization approach to obtain exact solutions that one can possibly use as a benchmark for ``physical’’ QH edge states~\cite{Fendley:1995c, Fendley:1996, Kane:2003}. Second, $\nu = 1/2$ states were suggested to emerge in ladder structures~\cite{PetrescuLeHurPRB15,
CornfeldSelaPRB15,
CalvaneseMazzaPRX17,
CooperDalibardSpielmanRMP19}. Further, the $\nu=1/2$ states were recently anticipated (with support from experiments) to exist at the edge of a BCS-paired exciton system~\cite{WangNature23}, making our work potentially relevant for real experimental setups. Finally, an edge-state channel characterized by $\nu = 1/2$ is a crucial composite element for candidates for the filling factor $\nu = 5/2$ edge state~\cite{MooreRead91,
LevinHalperinRosenowPRL07,
LeeRyuNayakFisherPRL07,
FidkowskiChenVishwanathPRX13,
ZuckerFeldmanPRL16,
AntoniPRB18}.

Here, we consider two possible setups hosting chiral $\nu=1/2$ anyonic states. The first one is an HOM interferometer realized by two edge states originating from equilibrium reservoirs, which impinge on a quantum point contact (QPC).
By means of a nonperturbative calculation of the tunneling current at the QPC and of the current-current correlations at the output arms, we get results consistent with Ref.~\cite{Campagnano:2016}, where the current correlations were studied for Laughlin states,
as well as with earlier works where refermionization and Bethe-ansatz solution were employed. We present explicit exact expressions for the current cross-correlations and the tunneling-current noise in this setup, which are valid at arbitrary temperatures and voltages. 

The second geometry we address is again an HOM interferometer, but in this case, the incoming quasiparticles form diluted streams originating from two QPCs (``diluters''). 
To implement such a situation, we consider a four-edge-channels geometry as initially theoretically analyzed for Abelian~[\onlinecite{Rosenow:2016}] and non-Abelian~[\onlinecite{Lee:2022}] anyons, and later experimentally realized in Ref.~\cite{Bartolomei:2020}. In this work, we consider only Abelian braiding that generates simply a statistical phase. For $\nu=1/2$ anyons considered here, we show that, in order to obtain meaningful results for quantities such as the tunneling current, it is necessary to include corrections to the correlation functions, which stem from the processes beyond those considered in previous works. Indeed, in Refs.~\cite{Rosenow:2016,
Bartolomei:2020,
Lee:2022,
Schiller:2023}, tunneling of nonequilibrium anyons from diluters at the central QPC was discarded. The tunneling current, obtained in previous references~\cite{Rosenow:2016,
Bartolomei:2020,
Lee:2022,
Schiller:2023} for an arbitrary filling factor, vanishes when the filling factor is set to $\nu=1/2$. Thus, one should go beyond the so-called time-domain braiding processes, where nonequilibrium anyons only indirectly influence the tunneling by braiding with the quasiparticle-quasihole pairs created at the collider.
In our work, after allowing the nonequilibrium anyons to directly tunnel at the central QPC, we obtain a finite tunneling current that is proportional to the difference between nonequilibrium currents of channels bridged by the central QPC.

The article is organized as follows. In Sec.~\ref{two-edge-setup}, we introduce the $\nu=1/2$ chiral Luttinger model for a simple beam splitter realized with two edge mode impinging on a QPC. We perform a non-perturbative calculation of the tunneling current at the QPC and of the cross-current correlation at the the outgoing arms. We verify that for such a geometry, in the absence of a voltage bias, the cross-correlation is zero when the two edges are kept at the same temperature. It is worth noting that similar setups have been studied within the $\nu = 1/2$ chiral Luttinger-liquid model in Refs.~\cite{Fendley:1995c, Fendley:1995, Kane:2003}. While Ref.~\cite{Fendley:1995c} obtained an exact expression for the current and conductance, Ref.~\cite{Fendley:1995} derived a relation between the nonequilibrium noise and current for zero temperature. 
In the present work, we go beyond these references and provide exact analytical expressions, valid under all system parameters (including temperature, bias. and tunneling amplitudes), for both the tunneling-current noise and cross-correlation. Further, with these analytical expressions, we obtain a general relation between the noise and current that is exact for the two-edge $\nu =1/2$ model.

The model of Ref.~\cite{Kane:2003} consists of three chiral channels and two QPCs, different from the two-edge model. The first QPC (diluter) supplies a nonequilibrium (diluted) beam of anyons to be scattered at the second QPC (collider). In the $\nu=1/2$ case, the second QPC was treated exactly by refermionization, whereas the diluter was introduced perturbatively. Dilution of a $\nu=1/2$ anyonic beam in the setup of Ref.~\cite{Kane:2003} is the common feature with our four-edge model. Specifically, in Sec.~\ref{four-edge-setup}, we consider an HOM interferometer with two diluters supplying quasiparticles to the central QPC and compute the tunneling current and the corresponding noise for $\nu=1/2$ anyons.
In contrast to Ref.~\cite{Kane:2003}, the central QPC is addressed perturbatively, whereas the transmission through the diluters is taken into account in all orders upon the resummation of higher-order processes, as in Refs.~\cite{Rosenow:2016,
Bartolomei:2020,
Lee:2022,
Schiller:2023}. 

Here, we go beyond the approach of these works, after including the subleading corrections to the correlation functions. These corrections are crucial, since for $\nu=1/2$ anyons, a peculiar particle-hole symmetry manifests itself in the braiding phase, which renders the tunneling current at the central QPC zero, irrespective of the currents coming from the upper and lower diluters. In this case, a nonperturbative calculation of the type used in Ref.~\cite{Kane:2003} appears to be not easily feasible when the exact treatment of diluters is concerned, and we employ a perturbative Keldysh nonequilibrium approach supplemented by the resummation of the important terms at each order. Full details, concerning the mathematical derivation of results reported in Secs.~\ref{two-edge-setup} and \ref{four-edge-setup}, are presented in the Appendices.
Finally, in Sec.~\ref{sec:summary}, we summarize our findings and discuss possible further developments of our work. Throughout the paper, we set $\hbar=1$ and $k_B=1$.

\section{Two-edge setup}
\label{two-edge-setup}
\begin{figure} \begin{center} 
\includegraphics[width=8cm]{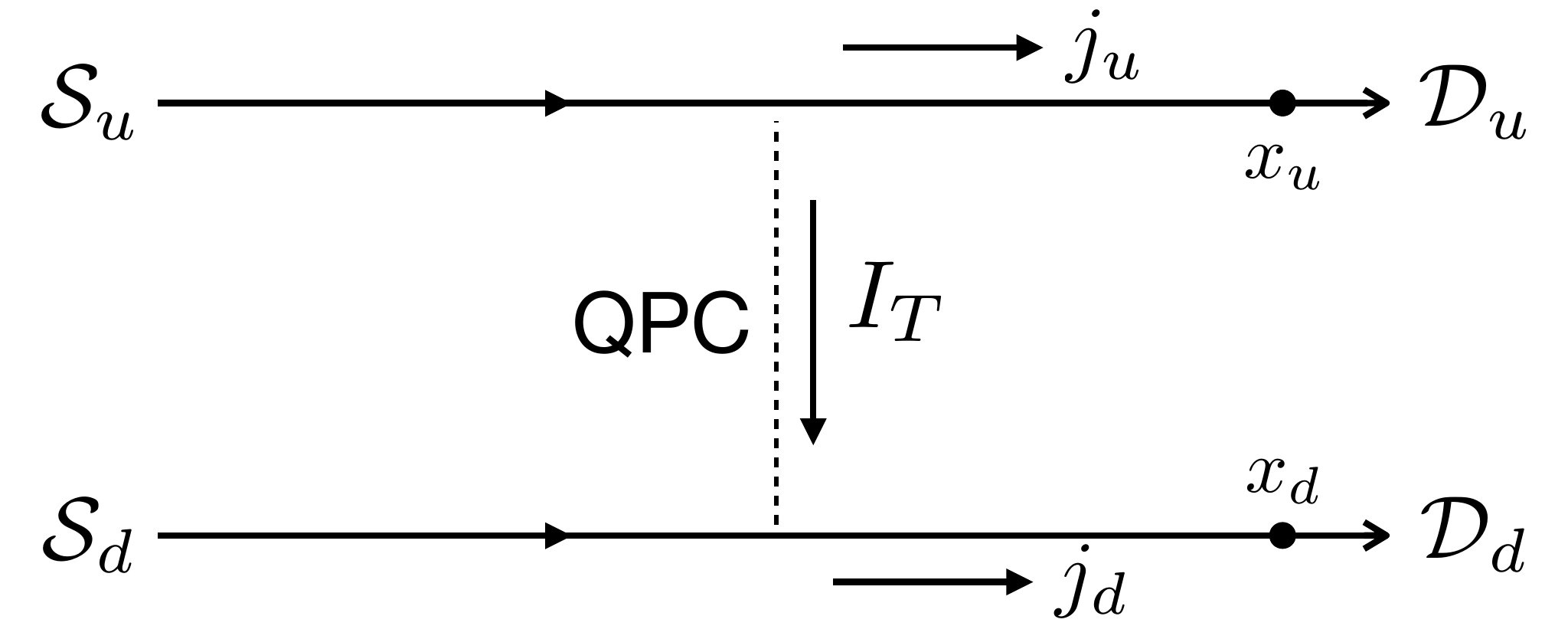}
\end{center}
\caption{
Schematic representation of the two-edge-channel setup. $\mathcal{S}_u$ and $\mathcal{S}_d$ are quasiparticle sources kept at thermodynamic equilibrium at the same temperature $T$. Bias $V$ is applied between $\mathcal{S}_u$ and $\mathcal{S}_d$.
The two sources communicate through a collider that is placed at the position $x = 0$.
Currents are measured in the drains $\mathcal{D}_u$ and $\mathcal{D}_d$ at points $x_u$ and  $x_d$, respectively. For a possible physical implementation, see Ref.~\cite{Campagnano:2016}.} \label{2-edge}
\end{figure}

In this Section, we consider a two-edge setup shown in  Fig.~\ref{2-edge}, such as the one discussed in Ref.~\cite{Campagnano:2016}, where the current-current correlations were studied perturbatively in the tunneling at the QPC for edge states of a Laughlin QH liquid.
Although the physical realization of the model in terms of Laughlin edge states corresponds to  $\nu=1/(2m+1)$ with integer $m$, we will term the chiral channels ``edges'' for generic values of $\nu$.

In the bosonic language, two chiral modes coupled by the QPC are described by the following Hamiltonian~\cite{Kane:2003}: 
\begin{multline}
H =  \frac{v}{4\pi} \sum_{j=u,d} \int_{-\mathcal{L}}^\mathcal{L} \:\! d x \: [ \partial_x \phi_j ( x )]^2 
 \\
+ \frac{2 w}{ ( 2 \pi l_c )^\nu }\: \cos \left\{ \sqrt{\nu}\left[ \phi_u (0) - \phi_d (0 ) \right] \right\}
\:,
\label{bos.1}
\end{multline}
\noindent
with $l_c$ being a short-distance cutoff, $v$ the edge plasmon velocity, $2\mathcal{L}$ the length of the channels ($\mathcal{L}\rightarrow \infty$, eventually), and $w$ describes the amplitude of tunneling of charge-$\nu$ quasiparticle between the two edges.
The chiral bosonic fields $\phi_u(x)$ and $\phi_d(x)$
obey the canonical commutation relation
\begin{equation}
\left[ \phi_j(x),\phi_k(x')\right]=i \pi \delta_{jk} \,\mbox{sgn} (x-x').
\label{commutation-rule}
\end{equation}
Charged anyonic excitations are described by the vertex operators, 
\begin{equation}
    \psi_j = F_j \exp(i\sqrt{\nu} \phi_j)/\sqrt{2\pi l_c},
    \label{psi-qp}
\end{equation}
where $F_j$ refers to the Klein factor that is responsible for the commutation relation between vertex operators from different channels.
Klein factors are important when considering a Mach–Zehnder interferometer~[\onlinecite{Feldman:2006}]; however, they can be discarded in our structures~[\onlinecite{Kane:2003}].
We thus omit the Klein factors in what follows.
In terms of the quasiparticle operators (\ref{psi-qp}),
the tunneling term in the Hamiltonian reads as
$(2\pi l_c)^{1-\nu} w\, \psi_u^\dagger\psi_d+\text{H.c.}$
Before moving on, we would like to point out that bosonic fields in Eqs.~\eqref{bos.1}-\eqref{psi-qp} differ from those of Ref.~\cite{Kane:2003} by a constant factor.
More specifically, $\sqrt{\nu} \phi_j$ equals the bosonic field used in Ref.~\cite{Kane:2003}. This choice, however, does not affect any results for observables.

With the above conventions, the current 
density operators in the two arms, $j_{u,d} (x)$,
are given by the spatial derivatives of the bosonic fields:
\begin{eqnarray}
j_{u} (x) &=& \frac{e v \sqrt{\nu}}{2 \pi }\, \partial_x \phi_{u} (x), \label{bos.2-1} \\
j_{d} (x)&=&  \frac{e v \sqrt{\nu}}{2 \pi }\, \partial_x \phi_{d} (x)
\:.
\label{bos.2}
\end{eqnarray}
We are interested in calculating the tunneling current $I_T$ at the QPC (which can be obtained as a commutator of the chiral density with the Hamiltonian),
\begin{equation}
    I_T = -i \nu e \frac{2w}{(2\pi l_c)^{\nu}} \sin \left\{ \sqrt{\nu}\left[ \phi_u (0) - \phi_d (0 ) \right] \right\},
    \label{eq:it_operator}
\end{equation}
and the zero-frequency current-current correlation function $S(0;x_u,x_d)$ (describing the current cross-correlations in the drains) given by 
\begin{equation}
S(0;x_u,x_d)=\frac{1}{2}\int_{-\infty}^\infty dt\, \big\langle \big\langle \{ j_u(t,x_u),j_d(0,x_d)\}\big\rangle \big\rangle,
\label{S0}
\end{equation}
where $\{ j_u(t,x_u),j_d(0,x_d)\}$ is the anticommutator and the double angular brackets denote an irreducible average 
\footnote{\label{note1} Note that, following the prescription of Ref.~\cite{Kane:2003}, an extra factor of $1/2$ is included in Eq.~\eqref{S0}, which differs from a more conventional definition of the current fluctuations. Consequently, the Fano factor, Eq.~\eqref{eq:Fg}, is defined as $\mathcal{F}_\text{CC} \equiv S(0)/ e I_T$, instead of $2S(0)/ e I_T$.
In addition, a factor of two appears in Eq.~\eqref{eq:diss_fluct_half}}.
With the tunneling current operator $I_T$ defined in Eq.~\eqref{eq:it_operator}, the tunneling-current noise reads as
\begin{equation}
    \begin{aligned}
        S_T(0) &\equiv \int dt \left[\big\langle I_T (t) I_T (0)\big\rangle -\big\langle I_T (t) \rangle \langle I_T (0)\big\rangle \right].
    \end{aligned}
    \label{eq:sT}
\end{equation}

In Ref.~[\onlinecite{Campagnano:2016}], the current-current correlations were calculated, by means of the Keldysh perturbative approach, for Laughlin filling fractions. In addition, for generic values of $\nu$, this model can be exactly solved via thermodynamic Bethe ansatz, see e.g., Refs.~\cite{Fendley:1995, Fendley:1995b, Fendley:1995c, Fendley:1996, SaleurWeissPRB01, TrauzettelPRB04, SchmitteckertPRL10}. However, the Bethe-ansatz results are usually represented in the form of series, i.e., they cannot be written in the form of compact analytical expressions. Here, we will solve the problem exactly at $\nu=1/2$ by means of the refermionization approach~\cite{Fendley:1995c}.

\subsection{Refermionization}
\label{sec:referm}

With refermionization at $\nu = 1/2$, we can obtain nonperturbative results that are exact in the QPC transmission.
Applications of this approach can be found in the literature on a broad class of interacting models, such as dissipative systems~\cite{GuineaPRB85, WeissSassettiNegeleWollensak91,ZamoumCrepieuxSafiPRB12,GuCrossoverPRB21} and various impurity models in non-chiral Luttinger-liquid models in Refs.~\cite{Kane:1992, EggerGrabertPRB98, GogolinBook, Vondelft:1998, KomnikGogolinPRL03}.
Refermionization was also widely applied to Kondo models, e.g., the two-channel Kondo model (for spin~\cite{EmeryKivelsonPRB92, SchillerHershfieldPRB98} and charge~\cite{MatveevPRB95, FurusakiMatveevPRB95, LandauCornfeldSelaPRL18} ones) and the two-impurity Kondo model~\cite{GanPRL95, GanPRB95, SelaAffleckPRL09}.
In addition, refermionization was also employed in Ref.~\cite{Kane:2003} to study the three-edge structure for $\nu=1/2$ chiral Luttinger liquids, where one of the QPC was treated exactly, while the tunneling at the other QPC (supplying diluted beams of anyons) was accounted perturbatively. We will address a related four-edge setup with diluted beams of anyons scattered at the collider QPC in Sec.~\ref{four-edge-setup} below. 

To start with, let us introduce the bosonic fields, $\phi_\rho (x)$ and $\phi_\sigma (x)$, 
defined as 
\begin{equation}
\begin{aligned}
\phi_\rho (x) =& \frac{\phi_u ( x ) + \phi_d (x)}{\sqrt{2}},  \\
\phi_\sigma (x) =& \frac{\phi_u (x) - \phi_d (x) }{\sqrt{2}}
\: . 
\end{aligned}
\label{bos.3}
\end{equation}
In terms of these fields, the Hamiltonian is rewritten as a sum of the two terms~\cite{Kane:2003}:
\begin{equation}
H=H_\rho+H_\sigma,
\end{equation}
with
\begin{equation}
 H_\sigma = \frac{v}{4 \pi }   \int_{-\mathcal{L}}^\mathcal{L} \: d x \: [ \partial_x \phi_\sigma (x )]^2 
+ \frac{2w}{(2 \pi l_c)^\frac{1}{2}} \: \cos [ \phi_\sigma (0) ],   
\label{bos.4}
\end{equation}
and
\begin{equation}
 H_\rho = \frac{v}{4 \pi }   \int_{-\mathcal{L}}^\mathcal{L} \: d x \: [ \partial_x \phi_\rho (x )]^2.
\end{equation}
Note that the Hamiltonian of the $\rho$ field describes a free field.
The currents $j_u$ and $j_d$ are given by
\begin{align}
 j_u ( x )  &=  \frac{ev}{4 \pi } \big[\partial_x \phi_\rho (x) + \partial_x \phi_\sigma ( x ) \big],   
\\
j_d ( x )  &=  \frac{ev}{4 \pi } \big[\partial_x \phi_\rho (x) - \partial_x \phi_\sigma ( x ) \big].  
\end{align}
We consider the case when a voltage bias $V$ is symmetrically applied between the two edges. 
This can be taken into account with the following transformation of the Hamiltonian
\begin{equation}
H \to H - \frac{e V}{4 \pi} \int_{-\mathcal{L}}^\mathcal{L} \: d x \: \partial_x \phi_\sigma ( x ) \,. 
\label{bos.5}
\end{equation}

Let us introduce the chiral fields $\psi_{\sigma} (x)$ and $\psi_{\rho} (x)$ as 
\begin{equation}
\psi_{\alpha} (x ) = F_{\alpha}  \frac{ e^{ i \phi_{\alpha} (x) }}{(2 \pi l_c )^\frac{1}{2}}, 
\label{bos.6}
\end{equation}
with $\alpha=\sigma,\rho$.
The fields $\psi_{\sigma} (x)$ and $\psi_{\rho} (x)$ obey conventional fermionic anticommutation relations, hence ``refermionization''.
The two terms  of the Hamiltonian, when written in terms of the chiral fermions (\ref{bos.6}), take the following forms: 
\begin{equation}
H_\rho = - i v \int_{-\mathcal{L}}^\mathcal{L} \: d x \: \psi_\rho^\dagger ( x ) \partial_x \psi_\rho ( x ), 
\end{equation}
and
\begin{align}
H_\sigma &= - i v \int_{-\mathcal{L}}^\mathcal{L} \: d x \: \psi_\sigma^\dagger ( x ) \partial_x \psi_\sigma ( x )
\notag 
\\
&- \frac{eV}{2} \int_{-\mathcal{L}}^\mathcal{L} 
\: d x \: \psi_\sigma^\dagger ( x ) \psi_\sigma ( x ) 
+ w\, \gamma\, \big[  \psi_\sigma (0 ) - \psi_\sigma^\dagger ( 0 ) \big]. 
\label{bos.7}
\end{align}
\noindent
The additional Majorana fermion mode $\gamma$ is introduced so as to have a bosonic operator 
describing the boundary interaction (cf. Ref.~\cite{Kane:2003} and references therein). 
The corresponding current operators are given by 
\begin{align}
j_\sigma ( x ) &= e v\, \psi_\sigma^\dagger ( x ) \psi_\sigma ( x ),
\label{bos.8}
\\
j_\rho ( x ) &= e v\, \psi_\rho^\dagger ( x ) \psi_\rho ( x ),
\end{align}
in terms of the refermionized fields.
The current operators 
in channels $u$ and $d$ are given by 
\begin{eqnarray}
j_u ( x ) &=& \frac{ev}{2}  \big[\psi_\rho^\dagger ( x ) \psi_\rho ( x ) + \psi_\sigma^\dagger ( x ) \psi_\sigma ( x ) \big],  \label{bos.10-1} \\
j_d ( x ) &=& \frac{ev}{2}  \big[  \psi_\rho^\dagger ( x ) \psi_\rho ( x ) - \psi_\sigma^\dagger ( x ) \psi_\sigma ( x ) \big]. 
\label{bos.10}
\end{eqnarray}

The Hamiltonian expressed in terms of the refermionized fields is readily diagonalized by a Bogoliubov-Valatin transformation. In terms of the eigenmodes of the problem (scattering-wave states with fermionic annihilation/creation operators $c_{\epsilon ,\lambda}$ and $c_{\epsilon, \lambda}^\dagger$ with energy $\epsilon$), we can rewrite the fermionic field operator $\psi_\sigma(x)$ as  
\begin{equation}
\psi_\sigma(x) = \sum_{\epsilon >0}  \sum_{\lambda = p,h} \: \{ \mathcal{P}^\lambda_\epsilon (x )\, c_{\epsilon ,\lambda}
+ [\mathcal{H}_\epsilon^\lambda (x) ]^*\, c_{\epsilon , \lambda}^\dagger \}. 
\label{eq:scattering_state_ops}
\end{equation}
Here 
\begin{eqnarray}
&& \mathcal{P}_\epsilon^p (x) = \frac{ e^{i \epsilon x/v}}{\sqrt{2 \mathcal{L}}} \: \big[   \Theta (-x) + t_\epsilon^p\: \Theta (x) \big], \label{ef.6-1} \\
&& \mathcal{H}_\epsilon^p (x) = \frac{ e^{i \epsilon x/v}}{\sqrt{2 \mathcal{L}}} \: r_\epsilon^p \: \Theta (x), 
\label{ef.6}
\end{eqnarray}
for an incoming particle state and 
\begin{eqnarray}
&& \mathcal{P}_\epsilon^h (x)  =  \frac{ e^{i \epsilon x/v}}{\sqrt{2 \mathcal{L}}} \: r_\epsilon^h\: \Theta (x),  
\label{ef.7-1} \\
&& \mathcal{H}_\epsilon^h  (x) = \frac{ e^{i \epsilon x/v}}{\sqrt{2 \mathcal{L}}} \: \big[   \Theta (-x) + t_\epsilon^h \Theta (x) \big], 
\label{ef.7}
\end{eqnarray}
for an incoming hole state,
with $\Theta(x)$ the Heaviside step function.

Equation \eqref{eq:scattering_state_ops} describes a superposition of particle- and hole-like states. In Eqs.~\eqref{ef.6-1}-\eqref{ef.7}, the amplitudes $\mathcal{P}_\epsilon^\lambda$ and $\mathcal{H}_\epsilon^\lambda$ denote the wave-function components in the particle and hole sectors of the scattering state, respectively, for the type of incoming states indicated by a superscript $\lambda=p,h$. The transmission amplitudes $t_\epsilon$ and $r_\epsilon$ refer to, respectively, the wave-function components that remain the same as the incoming particle or change from particle to hole ($r_\epsilon^p$ of $\mathcal{H}_\epsilon^p$), or the vice versa ($r_\epsilon^h$ of $\mathcal{P}_\epsilon^h$), like in Andreev-like scattering. Their explicit expressions read: 
\begin{equation}
\begin{aligned}
&t_\epsilon^p = t_\epsilon^h \equiv t_\epsilon= \frac{ v\: \epsilon}{ v\: \epsilon +2 i w^2},  \\
&r_\epsilon^p = r_\epsilon^h\equiv r_\epsilon  = \frac{2iw^2}{ v\: \epsilon + 2 i w^2},
\end{aligned}
\label{ef.8}
\end{equation}
following the convention of Ref.~\cite{Kane:2003}.
In performing the calculations, it is worth recalling that we have to employ different Fermi distribution functions for incoming particles and holes in the presence of the symmetric voltage bias introduced above. In particular, denoting them by $f_p ( \epsilon )$ and $f_h (\epsilon )$, respectively, we have 
\begin{equation}
\begin{aligned}
f_p ( \epsilon ) &=\frac{1}{1 + e^{\beta \left( \epsilon - eV/2 \right)}}, \\
f_h ( \epsilon ) &=\frac{1}{1 + e^{\beta \left( \epsilon + eV/2 \right)}}.
\end{aligned}
\label{tunc.4}
\end{equation}
For $w = 0$, Eq.~\eqref{tunc.4} also provides distributions of the refermionized $\sigma$ operators ($\psi_\sigma^\dagger $ and $ \psi_\sigma$), following the refermionized Hamiltonian~\eqref{bos.7}.
When $w\neq 0$, since the coupling between $\sigma$ operators and the boundary Majorana $\gamma$ is quadratic, the system energy spectrum is not modified.
As a consequence, scattering states given by fermionic annihilation/creation operators $c_{\epsilon ,\lambda}$ and $c_{\epsilon, \lambda}^\dagger$ are in the one-to-one correspondence to those of the $\sigma$ modes in the $w = 0$ limit. The distributions of $c_{\epsilon ,\lambda}$ and $c_{\epsilon, \lambda}^\dagger$ are thus also provided by Eq.~\eqref{tunc.4}.
In the following, we will use the above refermionized model to compute the tunneling current and the current-current correlations. 

\subsection{Tunneling current}
\label{tuncur}

The tunneling current $I_T$ can be represented
as a difference in the current flowing through channel $d$ before and after the quantum point contact, see Eq.~(\ref{bos.10}):
\begin{multline}
I_T^{(d)}  = \frac{ev}{2} \langle [ \psi_\rho^\dagger ( x ) \psi_\rho (x) - \psi_\rho^\dagger (-x) \psi_\rho (-x) ]
\rangle 
\\
- \frac{ev}{2}  \langle [  \psi_\sigma^\dagger ( x ) \psi_\sigma (x ) - \psi_\sigma^\dagger (-x ) \psi_\sigma (-x ) ] 
\rangle,
\label{tunc.1}
\end{multline}
where $x>0$. At the same time, it can also be computed as minus the analogous difference, but for channel $u$ [see Eq.~(\ref{bos.10-1})], which yields
\begin{multline}
I_T^{(u)}  = -   \frac{ev}{2} \langle [ \psi_\rho^\dagger ( x ) \psi_\rho (x) - \psi_\rho^\dagger (-x) \psi_\rho (-x) ]
\rangle
\\
- \frac{ev}{2}  \langle [  \psi_\sigma^\dagger ( x ) \psi_\sigma (x ) - \psi_\sigma^\dagger (-x ) \psi_\sigma (-x ) ] 
\rangle.
\label{tunc.2}
\end{multline}
To simplify our calculations, we compute  $I_T$ as the mean value of Eqs.(\ref{tunc.1}) and (\ref{tunc.2}):
\begin{equation}
I_T = - \frac{ev}{2}   \langle [  \psi_\sigma^\dagger ( x ) \psi_\sigma (x ) - \psi_\sigma^\dagger (-x ) \psi_\sigma (-x ) ] 
\rangle.
\label{tunc.3}
\end{equation}

Inserting the expansion of fermion operators given by Eq.~\eqref{eq:scattering_state_ops} into the right-hand side of Eq.~\ref{tunc.3}, we obtain the expectation value of the tunneling current:
\begin{align}
&\langle I_T \rangle \! =\! - \frac{e v}{2}  \sum_{\epsilon > 0 }
\sum_{\lambda = p , h } \!  
\big[|\mathcal{P}_\epsilon^\lambda (x)|^2+| \mathcal{P}_\epsilon^\lambda ( - x ) |^2\big] f_\lambda ( \epsilon ) 
\notag
\\ &+\frac{e v}{2}\, \sum_{\epsilon > 0 } \sum_{\lambda = p , h } 
\big[|\mathcal{H}_\epsilon^\lambda ( x )|^2 + | \mathcal{H}_\epsilon^\lambda ( - x ) |^2 \big]
[ 1 - f_\lambda ( \epsilon ) ], 
\label{tunc.5}
\end{align}
\noindent
for $x>0$. Taking into account Eqs.~(\ref{ef.6-1}-\ref{ef.7}), we arrive at 
\begin{multline}
\langle I_T \rangle = \frac{e  v }{2 \mathcal{L}} \sum_\epsilon \: | r_\epsilon^p |^2 \: \big[f_p (\epsilon) - f_h (\epsilon) \big] 
\\ =\! \frac{e}{2 \pi}\!\!\! \: \int_{0}^{\infty} \!\!\! \: d \epsilon \, \frac{4 w^4}{v^2 \epsilon^2 \!+\! 4 w^4} \,
 \frac{\sinh \left( \frac{\beta\: e V}{2}  \right)}{\cosh ( \beta\: \epsilon )\! +\! \cosh 
\left( \frac{ \beta\: e V}{2}  \right) }.
\end{multline}
The integral over energies yields a compact explicit
expression for the average tunneling current:
\begin{equation}
\langle I_T \rangle= \frac{e w^2}{\pi v} \text{Im}\, \psi \!\left( \frac{1}{2} + \frac{w^2}{\pi v}\beta + i\frac{ e V \beta}{4\pi} \right), 
\label{tunc.6}
\end{equation}
\noindent
where $\psi(z)$ is the digamma function.
It is worth noting that Eq.~\eqref{tunc.6} perfectly agrees with the tunneling current computed for this setup at $\nu=1/2$ in Ref.~\cite{Fendley:1995c} 
In later sections, we go beyond Ref.~\cite{Fendley:1995c} and provide compact analytical expressions of the tunneling-current noise, current-current correlation function, as well as the general noise-current relation. All these results are exact for arbitrary temperature, bias, and the QPC transmission amplitude of the $\nu=1/2$ two-edge model. Another relevant work, Ref.~\cite{Kane:2003}, also discussed the tunneling current and noise in the $\nu=1/2$ case but in a three-edge structure with a diluted anyonic beam. We will return to Ref.~\cite{Kane:2003} in Sec.~\ref{four-edge-setup}, where we will consider tunneling of non-equilibrium anyons (diluted beams) in the four-edge setup.

\begin{figure} \begin{center} 
\includegraphics[width=8.5cm]{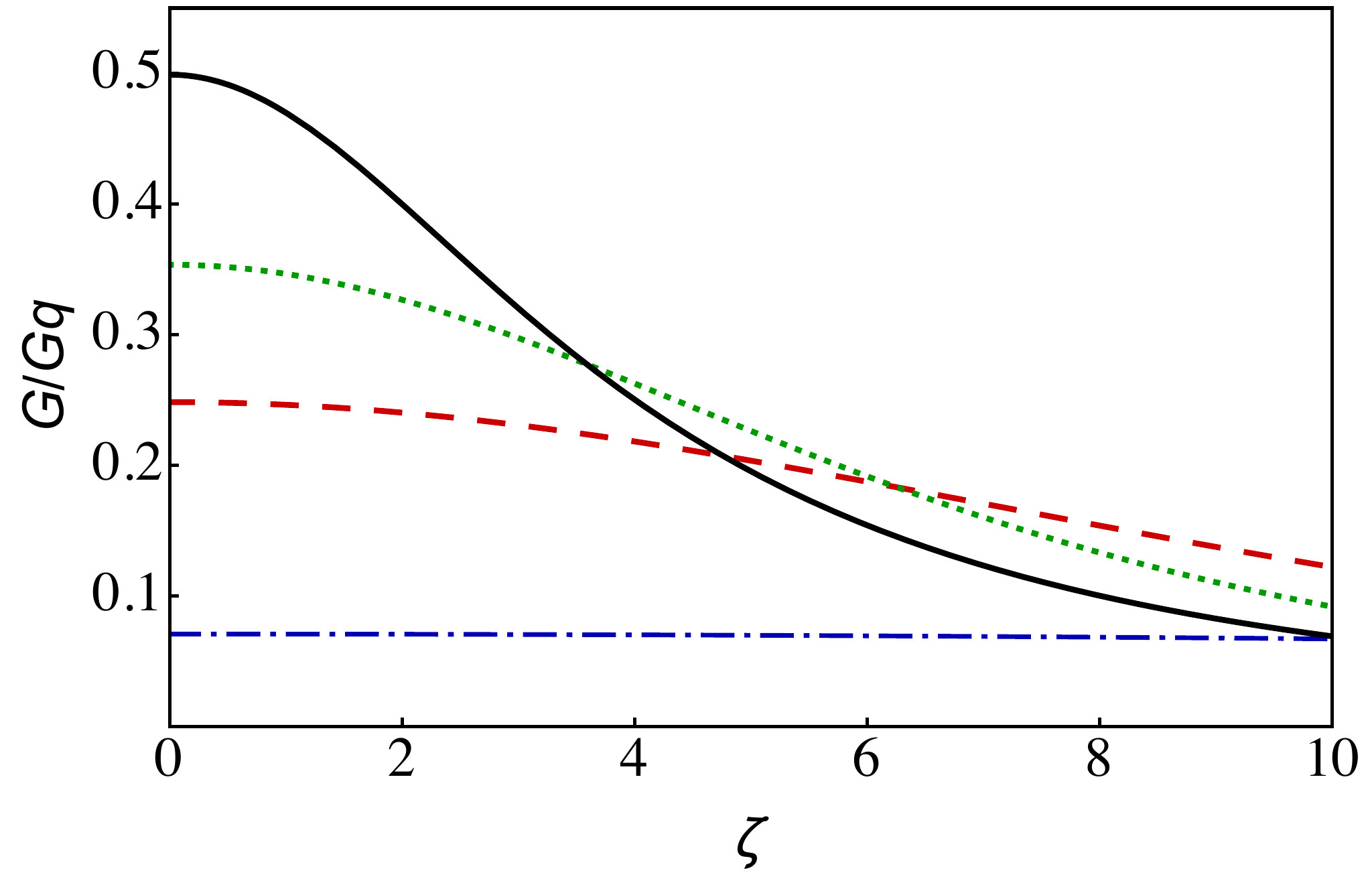}
\end{center}
\caption{Dimensionless differential conductance $G/G_q$ [$G(V)\equiv d \langle I_T(V) \rangle/dV$ and $G_q \equiv e^2/h$], as a function of $\zeta=veV/w^2$ for several values of dimensionless inverse temperature $\tilde{\beta}=\beta w^2/v$:
$\tilde{\beta}= 0.1$ for blue (dot-dashed), 0.5 for red (dashed), and 1 for green (dotted) curves. The black solid curve shows the zero-temperature result, $G/G_q=1/(2+\zeta^2/8)$.}  \label{differentialconductance}
\end{figure}

In the zero-temperature limit, assuming $V>0$, the 
tunneling current (\ref{tunc.6}) simplifies:
\begin{equation}
\langle I_T \rangle \!=\!\frac{e}{2 \pi}\int_{0}^{eV/2}\!\!\!\! d \epsilon \frac{4 w^4}{v^2 \epsilon^2 \!+\! 4 w^4}\!=\!\frac{e w^2}{\pi v}\!
\arctan \!\left( \!\frac{v e V }{4w^2}\!\right).
\label{eq:it_two_channel}
\end{equation}
In the weak- and strong-tunneling limits (or, equivalently, high- and low-bias limits, respectively), the tunneling current takes the approximate form:
    \begin{align}
     \text{weak tunneling, }\,\, \frac{w^2}{v} &\!\ll  eV:\notag \\
         \langle I_T \rangle&\simeq\frac{e w^2}{2v} \left( 1- \frac{8}{\pi} \frac{w^2}{v eV} \right),
         \label{eq:it_small_tunneling}\\
      \text{strong tunneling, }\, \frac{w^2}{v} &\!\gg  eV:\notag \\ 
      \langle I_T \rangle&\simeq\frac{e^2 V}{4\pi}\!\! \left[ 1 \!-\! \frac{1}{3} \!\left(\! \frac{veV}{4w^2} \!\right)^2 \right].
        \label{eq:it_strong_tunneling} 
    \end{align}
In the following, we will also refer to the above limits as strong- and weak-tunneling fixed points.

In the weak-tunneling limit [i.e., $w^2/v \ll  eV$, cf. Eq.~\eqref{eq:it_small_tunneling}], the tunneling current $I_T$ does not depend on the applied bias to the leading order of tunneling. The corresponding conductance then approaches zero in this limit.
In great contrast, in the opposite strong-tunneling limit [i.e., $w^2/v \gg  eV$, cf. Eq.~\eqref{eq:it_strong_tunneling}], $\langle I_T \rangle \propto V$ to the leading order, with the corresponding conductance approaching a constant value. To illustrate our results, we plot the differential conductance $$G(V)=d\langle I_T (V) \rangle/dV$$ in units of the quantum of conductance $G_q = e^2/h$, for several values of the temperature in Fig.~\ref{differentialconductance}.

\begin{figure} \begin{center} 
\includegraphics[width=7.5cm]{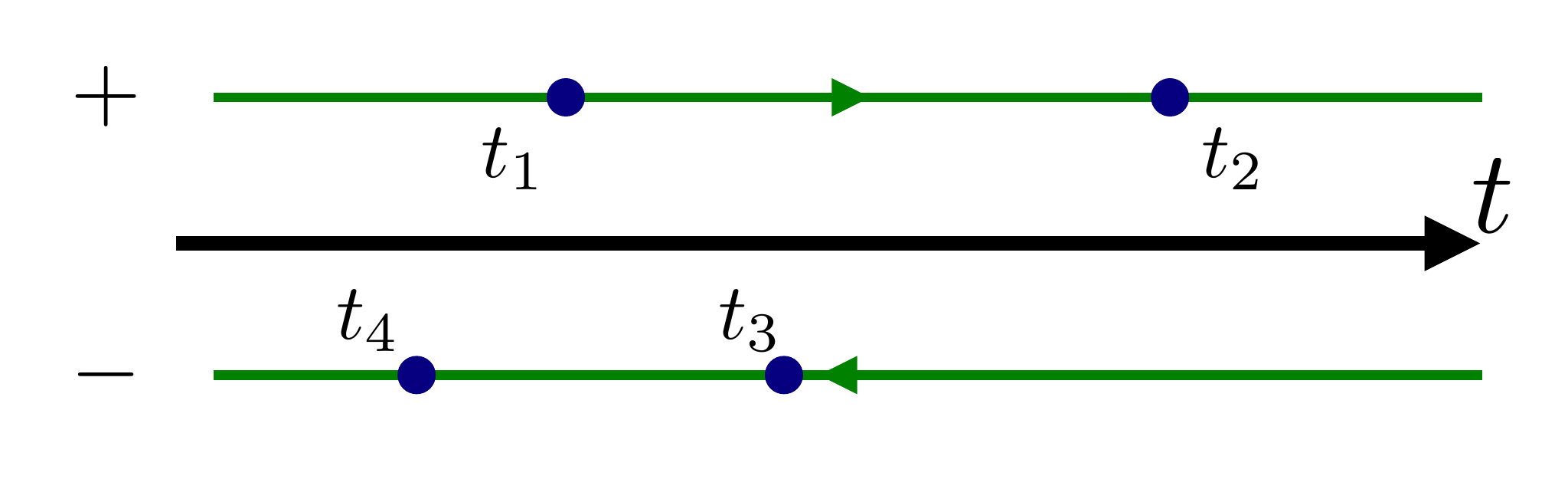}
\end{center}
\caption{The Keldysh time-ordering operator ${\bf T}_K$ is such that times on the upper branch are ordered according to the standard time-ordering operator, whereas times on the lower branch are instead arranged in the opposite order. By this convention, times on the lower branch appear later than those on the upper branch. The times shown in the figure obey $t_4<t_1<t_3<t_4$ but are ordered along the Keldysh contour as $t_1,\, t_2,\, t_3, \ t_4$. 
} \label{keldysh-contour}
\end{figure}

\begin{widetext}

\subsection{Current-current correlations}

To compute the current-current correlation, we start with the correlation function $s ( t_1-t_2 ;  x_u , x_d )$, defined as
\begin{align}
s ( t_1\! -\! t_2 ; x_u , x_d ) \!=\! \frac{1}{2} \sum_{\eta = \pm } \big\{ \langle {\bf T}_K  j_u ( x_u , t_1 , \eta ) j_d ( x_d , t_2 , - \eta ) \rangle
-\langle j_u ( x_u , t_1 , \eta ) \rangle \langle  j_d ( x_d , t_2 , - \eta ) \rangle \big\}, 
\label{tunc.7}
\end{align}
\noindent
with ${\bf T}_K$ referring to the Keldysh ordering (Fig.~\ref{keldysh-contour}), 
$j_i ( x , t , \eta )$ being the current density operator for branch-$i$ in Heisenberg representation, and 
$\eta$ being the Keldysh index. The integral of $s ( t_1-t_2 ;  x_u , x_d )$ over the time difference $t_1-t_2$ produces the zero-frequency cross-correlation $S(0;  x_u , x_d )$, see Eq.~(\ref{S0}).

In the following, we consider the case $x_u,\,x_d>0$ (i.e., at positions after the collider that is located at $x=0$) and assume $\mathcal{L} \to \infty$. 
Accounting for obvious cancellations, we obtain 
\begin{equation}
s( t_1 - t_2 ; x_u , x_d )=s_\rho( t_1 - t_2 ; x_u , x_d )-s_\sigma( t_1 - t_2 ; x_u , x_d )
\end{equation}
where each of the two terms is given by the irreducible correlators of the four fermionic operators that belong to the same sector $\alpha=\rho,\sigma$:
\begin{multline}
 s_\alpha( t_1 - t_2 ; x_u , x_d ) =   \frac{e^2v^2}{8}
  \sum_{\eta = \pm 1}
 \Big\{ 
 \big\langle {\bf T}_K 
\psi_\alpha^\dagger ( x_u , t_1 , \eta ) \psi_\alpha ( x_u , t_1 , \eta )
\psi_\alpha^\dagger ( x_d , t_2 , - \eta ) 
\psi_\alpha ( x_d , t_2 , -\eta ) \big\rangle  
\\
-\big\langle  \psi_\alpha^\dagger ( x_u , t_1, \eta  ) \psi_\alpha ( x_u , t_1, \eta  ) 
\big\rangle\, 
\big\langle 
\psi_\alpha^\dagger ( x_d , t_2, -\eta  ) 
\psi_\alpha ( x_d , t_2, -\eta ) \big\rangle \Big\},
\end{multline}
where the Keldysh structure is irrelevant to the product of averages (i.e., the second line) of the equation above.
The zero-frequency limit of the Fourier transform of the functions $s_{\rho,\sigma}(t;x_u,x_d)$ yields
\begin{align}
S_\rho  (\Omega \to 0; x_u , x_d ) = \frac{e^2}{4} \int \frac{d \epsilon}{2 \pi}
  \big[ f_{h,\rho} (\epsilon ) f_{p,\rho} ( - \epsilon ) 
   + f_{h,\rho} (  - \epsilon ) f_{p,\rho} ( \epsilon ) \big], 
\label{tunc.x11}
\end{align}
and
\begin{equation}
\begin{aligned}
    S_\sigma (\Omega \to 0; x_u , x_d )& = \frac{e^2}{4} \int_0^\infty \frac{d\epsilon}{2 \pi} \left\{ 4 |t_\epsilon|^2 |r_\epsilon|^2 \left[f_{h,\sigma} (\epsilon) f_{h,\sigma} (-\epsilon) + f_{p,\sigma} (\epsilon) f_{p,\sigma} (-\epsilon) \right] \right.\\
    &\quad\left. (|t_\epsilon|^2 - |r_\epsilon|^2)^2 \left[ f_{h,\sigma} (\epsilon) f_{p,\sigma} (-\epsilon) + f_{h,\sigma} (-\epsilon) f_{p,\sigma} (\epsilon) \right] \right\},
\end{aligned}
\label{tunc.13}
\end{equation}
where we have used Eq.~\eqref{ef.8} to simplify the formulas.
Note that Eq.~\eqref{tunc.13} reduces to Eq.~\eqref{tunc.x11} after taking $w = 0$ and replacing $\sigma$ by $\rho$.
Indeed, in this case, $S_\rho$ and $S_\sigma$ should have the same shape, as $\sigma$ and $\rho$ are both isolated channels when $w = 0$ (i.e., they do not couple to each other or to the Majorana operator $\gamma$ at the boundary).
In addition, since $\rho$ is always (i.e., irrespective of the value of $w$) an isolated channel, we can freely choose its bias, when evaluating measurable quantities, e.g., Eq.~\eqref{tunc.x11}.

Adding up these expressions (following discussions above, we choose $\rho$ as biased by $V/2$, for later convenience), we obtain:
\begin{align}
S (\Omega \to 0; x_u , x_d  )= \frac{e^2 v^2}{\pi}\int_{0}^{\infty} d\epsilon \frac{w^4 \epsilon^2}{(4 w^4+v^2\epsilon^2)^2}
\,
\frac{1-\cosh(\beta e V)}{\left[\cosh(\beta eV/2)+\cosh(\beta\epsilon)\right]^2}.
\label{eq:sud}
\end{align}
The integral in Eq.~\eqref{eq:sud} can be exactly performed, leading to 
\begin{equation}
    \begin{aligned}
       S (\Omega \to 0; x_u , x_d  ) & = - \frac{e^2 w^2}{4 \pi^2 v} \left[ 2\pi \coth\left( \frac{\beta eV}{2} \right) \text{Im}\, \psi \left( \frac{1}{2} + \frac{w^2 \beta}{\pi v} + i\frac{e V \beta}{4\pi} \right) -  \text{Re}\, \psi'\left( \frac{1}{2} + \frac{w^2 \beta}{\pi v} + i\frac{ e V \beta}{4\pi} \right)  \right. \\
       & \left. + \frac{2\beta w^2}{v} \coth\left(\frac{\beta eV}{2} \right) \text{Im}\, \psi' \left( \frac{1}{2} + \frac{w^2 \beta}{\pi v} + i\frac{e V \beta}{4\pi} \right) - \frac{\beta w^2}{\pi v} \text{Re}\, \psi'' \left( \frac{1}{2} + \frac{w^2 \beta}{\pi v} + i\frac{ e V \beta}{4\pi} \right)\right],
    \end{aligned}
    \label{eq:s0_expression}
\end{equation}
\end{widetext}
where $\psi'(z)$ and $\psi''(z)$ denote, respectively, the first and second derivatives of the digamma function $\psi(z)$. Details of the derivation of Eq.~(\ref{eq:s0_expression}) are presented in Appendix~\ref{app:tunneling current and noise}.

\begin{figure} \begin{center} 
\includegraphics[width=8cm]{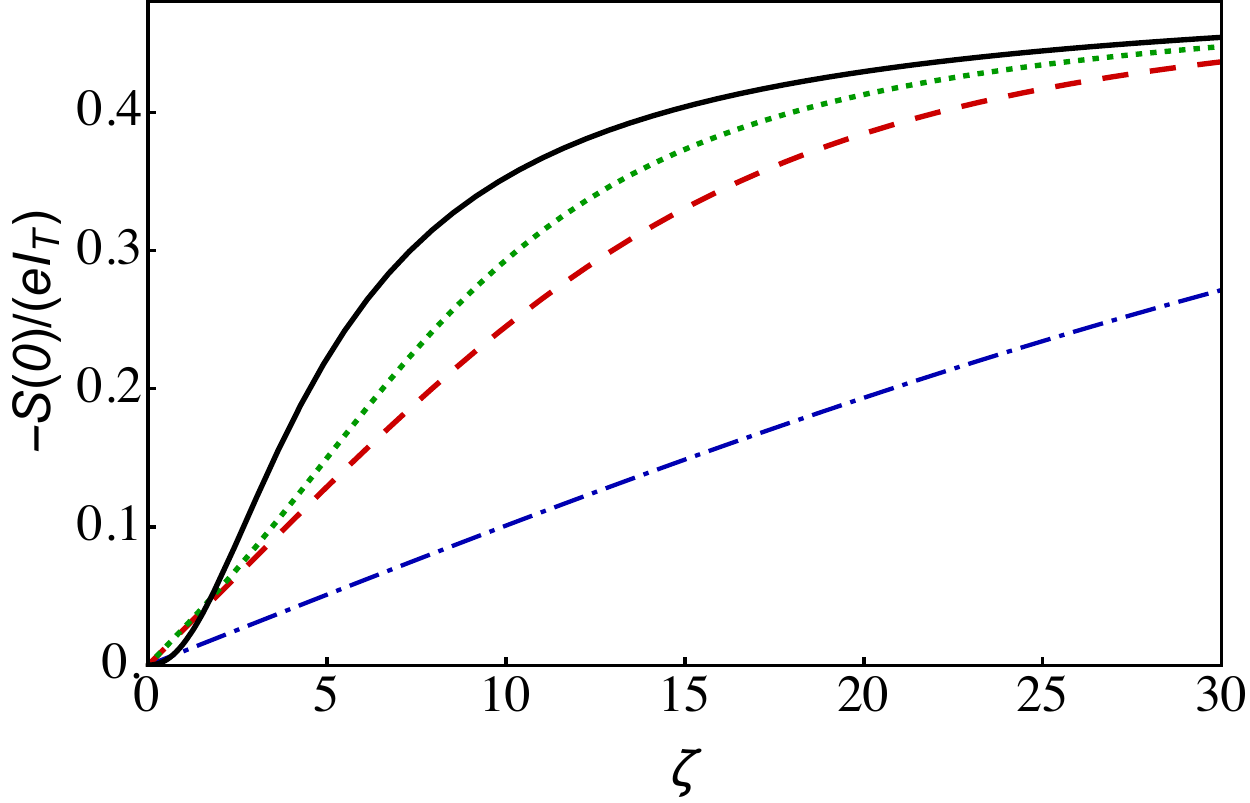}
\end{center}
\caption{Cross-correlation Fano factor $\mathcal{F}_\text{CC}=-S(0)/(e \langle I_T \rangle)$ plotted as a function of $\zeta=veV/w^2$ for several values of dimensionless inverse temperature $\tilde{\beta}=\beta w^2/v$: $\tilde{\beta}=0.1$ for blue (dot-dashed), 0.5 for red (dashed), and 1 for green (dotted) curves. The black solid curve shows the zero-temperature result, Eq.~(\ref{eq:s0_zero_temp}).}
\label{fig:fano}
\end{figure}

It is worthwhile stressing that Eq.~\eqref{eq:s0_expression} vanishes when $V \to 0$, for all values of $w^2$ and temperature. In other words, there is no thermal noise in cross-correlations for a chiral two-edge system at $\nu=1/2$. This fact is illustrated in  Fig.~\ref{fig:fano}, where noise vanishes when $\zeta = v e V/w^2 \to 0$ for all temperatures.
This result agrees with the statement of Ref.~\cite{Campagnano:2016}, where it was shown for general values of $\nu$ that the leading-order (in tunneling probability) contribution to noise $S(0)$ vanishes at $V = 0$, irrespective of temperature. Here, we have demonstrated that the vanishing of $S(\Omega \to 0)$ at $V\to 0$ is \textit{exact} in the $\nu=1/2$ case. Previously, the exact vanishing of current cross-correlations in this two-edge setup was known only for non-interacting fermions.

The analytical expression~\eqref{eq:s0_expression} for the zero-frequency \textit{current cross-correlations}, valid for arbitrary temperature, bias, and the QPC transmission amplitude, is one of the central results of our work. To the best of our knowledge, this expression is not reported by Refs.~\cite{Fendley:1995c, Fendley:1995b} or related works, where exact solutions for the tunneling current were obtained. At the same time, these references computed the tunneling-current noise $S_T$ for arbitrary $\nu$ but only in the zero-temperature limit, with the result represented by an infinite series. For the relation of the current cross-correlations to the tunneling current and the tunneling-current noise, see Sec.~\ref{sec:fd_relation} below.

In the zero temperature limit ($\beta \to \infty$), the thermal factor in the integrand of Eq.~\eqref{eq:sud} becomes proportional to the step function, 
\begin{equation}
   \lim_{\beta \to \infty} \frac{\cosh(\beta e V)-1}{ \left[\cosh(\beta eV/2)+\cosh(\beta\epsilon)\right]^2} = \left\{\begin{array}{c}
 2,\text{ if }\,\epsilon < \frac{eV}{2}      \\[0.2cm]
 0, \text{ if }\, \epsilon > \frac{eV}{2}    
\end{array}
\!\right. ,
\end{equation}
with which the integral for the current cross-correlation simplifies as
\begin{equation}
    \begin{aligned}
       & S (\Omega \to 0; x_u , x_d  )|_{\beta \to \infty} 
       \!= \! \frac{e^2 v^2}{\pi}\!\int_{0}^{\frac{e V}{2}} \! d\epsilon \frac{- w^4 \epsilon^2}{(4 w^4+v^2\epsilon^2)^2}
       \\
       &\quad  = \! - \frac{e^2 w^2}{2\pi v} \left[ -\frac{ 4v w^2  eV}{16 w^4 + (v\, eV)^2} + \arctan \left( 
\frac{v \,e V}{4 w^2} \right) \right],
    \end{aligned}
    \label{eq:s0_zero_temp}
\end{equation}
which agrees with results of Refs.~\cite{Fendley:1995b,Fendley:1995c}, except for a factor of two that comes from difference in the noise definition \cite{Note1}.
Similar to Eq.~\eqref{eq:it_small_tunneling} for the tunneling current, the zero-temperature cross-correlation functions in the two complementary asymptotic limits read as:
    \begin{align}
        &\text{weak tunneling, }\, \frac{w^2}{v} \!\ll eV:\notag \\
        &\quad S (0; x_u , x_d  )\Big|_{\beta \to \infty}\!\simeq\!-\frac{e^2 w^2}{4v} \left( 1- \frac{16}{\pi} \frac{w^2}{v\, eV} \right),
         \label{eq:st_limiting_cases-1}\\
        &\text{strong tunneling, }\, \frac{w^2}{v} \!\gg eV:\notag \\
        &\quad S (0; x_u , x_d  )\Big|_{\beta \!\to \!\infty}\!\simeq\!-\frac{e^2w^2}{24 \pi v}\! \left(\! \frac{v\,eV}{2w^2}\! \right)^3 \!\left[ 1 \!-\! \frac{3}{10}\! \left( \!\frac{v\,eV}{2w^2} \!\right)^2\!\right],
         \label{eq:st_limiting_cases}
    \end{align}   
where we replaced $S(\Omega \to 0)$ with $S(0)$, for simplicity.

Actually, at zero temperature, the cross-correlation noise $S( 0 )$ is equal to the minus value of the tunneling-current noise at the collider, $S(0)=-S_T/2$, see Sec.~\ref{sec:fd_relation}.
Thus, the cross-correlation Fano factor, $\mathcal{F}_\text{CC}$,
defined as 
\begin{equation}
   \mathcal{F}_\text{CC} = -\frac{S( 0 )}{e \langle I_T \rangle},
   \label{eq:Fg}
\end{equation} 
equals the conventionally defined tunneling Fano factor
\begin{equation}
    \mathcal{F}_T = \frac{S_T}{2 e \langle I_T \rangle},
    \label{eq:Fano}
\end{equation}
where the prefactor $1/2$, as a difference between Eqs.~\eqref{eq:Fg} and \eqref{eq:Fano}, comes from the extra factor of $1/2$ in the definition~\cite{Note1} of the cross-correlation, Eq.~\eqref{eq:sud}.
In the weak-tunneling limit, $w^2/v\ll eV$, the generalized Fano factor
equals $e/2$, which agrees with the fractional charge carried by one quasiparticle of a $\nu = 1/2$ channel. 
In the opposite strong-tunneling limit, $w^2/v\gg eV$ [cf. Eq.~\eqref{eq:it_strong_tunneling}], however, the Fano factor goes to zero, $\mathcal{F}_\text{CC} \sim \left[ veV/(2w^2) \right]^2 \to 0$.
In fact, in this limit, the generalized Fano factor does not disclose the quasiparticle charge in a single tunneling event, as the noise $S( 0 )$ and tunneling current $\langle I_T \rangle$ come from different processes. 
Indeed, $S( 0 )$ (likewise the tunneling-current noise) originates from events where particles continue to transport charge (thus do not tunnel between $u$ and $d$ at the QPC) along $u$ or $d$, while, on the contrary, $I_T$ receives contributions from tunneling events between $u$ and $d$. The latter processes clearly have a much larger probability to occur in the strong-tunneling limit.

In the strong-tunneling limit~\footnote{Note that, for ${\nu=1/2}$, the limit of infinite tunneling amplitude, ${w\to\infty}$, yields perfect transmission between the channels $u$ and $d$. This should be contrasted with the uncorrelated case ${\nu=1}$, where the transmission vanishes for ${w\to\infty}$, whereas perfect transmission is achieved at a finite value of $w$, see, e.g., Ref.~\cite{KomnikGogolinPRL03}}, the collider that connects channels $u$ and $d$ is almost transparent, see Eq.~(\ref{eq:it_strong_tunneling}). A single event of \textit{not changing the channel} at the collider then brings into small tunneling-current noise $S_T=-S (0)$ (the tunneling-current noise vanishes, when the collider is perfectly transparent) and a small deviation of tunneling current $\langle I_T \rangle$ from the perfect tunneling current, $I_0 \equiv e^2 V/(4\pi)$.
As both quantities are induced in a single event in a correlated way, the ratio $-S( 0 )/(e \langle I_T \rangle - e I_0)$ correctly quantifies the transmitted charge for such an event. 
Actually, following Eqs.~\eqref{eq:it_strong_tunneling} and \eqref{eq:st_limiting_cases}, 
$$-S( 0 )/(e \langle I_T \rangle - e I_0) \approx 1,$$ 
in agreement with the theory: only fermions with unit charge are allowed to tunnel in the strong-tunneling limit ($w^2/v\gg eV$), supported by the self-dual feature of the two-edge anyonic system (see, e.g., Refs.~[\onlinecite{Fendley:1995, WeissBook12}]).
A corresponding definition of the Fano factor in the strong-tunneling limit, i.e., 
\begin{equation}
\tilde{\mathcal{F}}_\text{CC}=-\frac{S( 0 )}{e \langle I_T \rangle - e I_0},
\end{equation} 
was introduced in studying the Kondo model~[\onlinecite{SelaOregVonOppenKochPRL06}], the charge two-channel Kondo~[\onlinecite{LandauCornfeldSelaPRL18}] model, and for the tunneling of a resonant level into a Majorana-hosted hybrid nanowire~[\onlinecite{GuHaroldPRB20}].

At finite temperature, one obtains the results shown in Fig.~\ref{fig:fano}. Here, the black curve, representing the zero-temperature result, indeed agrees with our analytical solution obtained with Eqs.~\eqref{eq:it_two_channel} and \eqref{eq:s0_zero_temp}.

\subsection{Relation between the tunneling current and noises}
\label{sec:fd_relation}

In the two-edge structure, the study of the tunneling current and the tunneling-current noise provides important information for detecting nontrivial features of quasiparticles, including, e.g., fractional charge~\cite{de-Picciotto:1997, Saminadayar:1997} and anyonic statistics (at least, indirectly), in fractional QH systems. Real experiments, however, more commonly measure the auto- or cross-correlations after the central QPC, instead of the tunneling-current noise exactly at the QPC.
It is thus important to investigate the relationship between the tunneling-current noise and cross/auto-correlations.
In this Section, we explicitly show that, for the two-edge $\nu = 1/2$ structure, the tunneling-current noise can be experimentally obtained by measuring the current cross-correlations and conductance.
This result, crucially, remains valid for all values of bias, tunneling amplitude, and temperature.

To highlight the peculiarities of the $\nu=1/2$ case, we begin by reviewing the calculation in the case of fermionic edge states ($\nu=1$), where the zero-frequency noises $S_{\rm f}(0)$, $S_{T,\text{f}}(0)$ and the current $I_{\rm T,f}$ can be expressed with the Landauer-B{\"u}ttiker method (see, e.g., Refs.~\cite{MartinLandauerPRB92,Buettiker:1992}):
\begin{align}
    S_\text{f}(0) & = -\frac{e^2}{h}\int d\epsilon \mathcal{T}_\text{f} (1 - \mathcal{T}_\text{f} ) \big[f_u(\epsilon) - f_d (\epsilon)\big]^2 ,
    \label{eq:ferimionic_expressions-S}
    \\
    S_{T,\text{f}} (0) &\! = \!\frac{e^2}{h}\!\int d\epsilon \Big\{\mathcal{T}_\text{f} f_u (\epsilon) \big[ 1\!-\!f_d (\epsilon)\big] \!+\! \mathcal{T}_\text{f} f_d (\epsilon) \big[ 1\!-\!f_u (\epsilon)\big]
    \notag
    \\
    &\qquad \qquad\quad  -  \mathcal{T}_\text{f}^2\, \big[ f_u (\epsilon) - f_d(\epsilon)\big]^2 \Big\},
    \label{eq:ferimionic_expressions-ST}\\
    I_{T,\text{f}} & = \mathcal{T}_\text{f} \frac{e}{h} \int d\epsilon \,\big[f_u(\epsilon) - f_d (\epsilon)\big].
    \label{eq:ferimionic_expressions-IT}
\end{align}
In Eqs.~\eqref{eq:ferimionic_expressions-S}-\eqref{eq:ferimionic_expressions-IT}, $f_u$ and $f_d$ are the Fermi distribution functions in the two channels $u$ and $d$, respectively, while $\mathcal{T}_\text{f}$ is the tunneling probability through the QPC.
From Eq.~\eqref{eq:ferimionic_expressions-ST}, one readily gets
\begin{equation}
    S_{T,\text{f}} (0) + S_\text{f}(0) \!=\! \frac{e^2}{h}\!\int d\epsilon\, \mathcal{T}_\text{f} \big[f_d (1 \!-\! f_d) + f_u (1 \!-\! f_u) \big] .
\end{equation}
At equilibrium, the distributions functions in the channels $f_{d,u}(\epsilon)=1/\{1+\exp[\beta(\epsilon-eV_{d,u})]\}$ satisfy the relation
\begin{equation}
    f_d (\epsilon) [1 - f_d (\epsilon)] = \frac{1}{\beta} \frac{\partial}{\partial(e V_d)} f_d (\epsilon) ,
\end{equation}
yielding the following relation between fermionic noises and tunneling current:
\begin{align}
    S_{T,\text{f}} (0) + S_\text{f}(0)   & = \frac{e}{h}\frac{1}{\beta}\int d\epsilon \mathcal{T}_\text{f} \left[ \frac{\partial}{\partial V_u} f_u (\epsilon) +  \frac{\partial}{\partial V_d}f_d (\epsilon)\right]\notag 
    \\
    & = T\frac{\partial}{\partial V} I_{T,\text{f}}.
\end{align}
Notice that this relation is only valid for the case where both channels $u$ and $d$ are in equilibrium and have the same temperature.

Now we move on to the $1/2$ case, where the tunneling current noise of the two-edge structure is given by (see Appendix~\ref{sec:tunneling_current_noise} for details)
\begin{equation}
\begin{aligned}
    &S_T  = \frac{e^2}{\pi} \int_0^\infty\!  d\epsilon\, \frac{4w^4}{\left[ \cosh\left( \frac{\beta e V}{2} \right) + \cosh (\beta\epsilon) \right]^2}\\
    & \ \times\! \left\{\!\frac{v^2\epsilon^2[\cosh(\beta eV) \!-\! 1]}{2(4w^4 + v^2\epsilon^2)^2}+\frac{1 + \cosh \left(\frac{\beta eV}{2}\right) \cosh (\beta\epsilon)}{4w^4 + v^2\epsilon^2}\!
    \right\}.
\end{aligned}
\label{eq:st_two_edge}
\end{equation}
At zero temperature, an analytical expression of the tunneling noise can be obtained by combining expressions of Refs.~\cite{Fendley:1995, Fendley:1995c}.
These references, however, do not provide the finite-temperature version of $S_T$, Eq.~\eqref{eq:st_two_edge}.

The last term in the curly brackets of Eq.~\eqref{eq:st_two_edge} provides a direct connection to the differential conductance.
Indeed, starting with the integral form of the tunneling current Eq.~\eqref{tunc.6}, we obtain
\begin{equation}
\begin{aligned}
    \frac{\partial}{\partial V} \langle I_T \rangle & = \frac{e^2}{2 \pi} \: \int_{0}^{\infty} \: d \epsilon\, \frac{4 w^4}{v^2 \epsilon^2 + 4 w^4} \\
   & \times 
\beta \frac{ 1 + \cosh \left(\frac{\beta eV}{2}\right) \cosh (\beta\epsilon)}{\left[ \cosh ( \beta \epsilon ) + \cosh 
\left( \frac{ \beta e V}{2}  \right) \right]^2 },
\end{aligned}
\label{eq:one_half_conductance}
\end{equation}
which, when multiplied by $2/\beta$, equals the contribution of the last term of Eq.~\eqref{eq:st_two_edge}.
As a consequence, we obtain the relation between differential conductance, tunneling current noise, and cross-correlation noise,
\begin{equation}
    S_T = -2S(0) + \frac{2}{\beta} \frac{\partial}{\partial V} \langle I_T \rangle,
    \label{eq:diss_fluct_half}
\end{equation}
which is valid for arbitrary temperature, voltage, and transmission amplitude [given Eqs.~\eqref{tunc.6} and \eqref{eq:s0_expression}, it also provides an explicit compact expression for $S_T$].

At this point, it is worth noting that Eq.~\eqref{eq:diss_fluct_half} can be derived \footnote{In\`es Safi, private communication} from the general finite-frequency relation between currents and noises, which was obtained \textit{non-perturbatively} in Ref.~\cite{BenaSafiPBR07} (see also Ref.~\cite{DolciniPRB05}) for arbitrary-$\nu$ chiral Luttinger liquids. By virtue of the ``generalized Kubo formula'' for setups where charge couples linearly to bias (Ref.~\cite{SafiJoyezPRB11}), the general formula of Ref.~\cite{BenaSafiPBR07} reduces, in the limit of zero frequency, to Eq.~\eqref{eq:diss_fluct_half}.

Nevertheless, to the best of our knowledge, the exactness of the current-noise relation in the form of Eq.~\eqref{eq:diss_fluct_half} for the $\nu = 1/2$ two-edge model was not
explicitly  
shown before;
neither was it demonstrated with explicit expressions of noises and conductance, cf. Eqs.~\eqref{eq:s0_expression}, \eqref{eq:st_two_edge}, and \eqref{eq:one_half_conductance}.
Actually, in general-$\nu$ cases, the current-noise relation in the concise form of Eq.~\eqref{eq:diss_fluct_half} was derived only perturbatively---to leading order in the tunneling amplitude, see, e.g., Refs.~\cite{RechJonckheereGremaudMartinPRL20,SafiPRB20}. In addition, in Refs.~\cite{RechJonckheereGremaudMartinPRL20,SafiPRB20}, the same perturbative relation was shown to be valid when the two equilibrium reservoirs have different temperatures; note also the different definition \cite{Note1} of the noise in Ref.~\cite{RechJonckheereGremaudMartinPRL20,SafiPRB20} as concerns the factor of 2 [cf. the comment below Eq.~\eqref{eq:s0_zero_temp}].
It is also worth emphasizing that Ref.~\cite{SafiPRB20} further discussed the weak-tunneling limit in the absence of thermalization (i.e., when considering time-domain braiding).

The noise-current relation, Eq.~\eqref{eq:diss_fluct_half}, can also be written through the current auto-correlation
\begin{equation}
    S_{\alpha\alpha} (0;x_\alpha,x_\alpha) \equiv \frac{1}{2}\int_{-\infty}^\infty dt\, \big\langle \big\langle \{ j_\alpha(t,x_\alpha),j_\alpha(0,x_\alpha)\}\big\rangle \big\rangle,
\end{equation}
where $\alpha = u, d$ is the label of the channel.
Indeed, following details provided by Appendix~\ref{app:auto_correlation}, Eq.~\eqref{eq:diss_fluct_half} is equivalent to
\begin{equation}
    S_T = 2 S_{\alpha\alpha}(0) + \frac{2}{\beta} \frac{\partial}{\partial V}  \langle I_T \rangle  - \frac{G_0}{\beta},
    \label{eq:auto-conductance_relation}
\end{equation}
where $G_0 \equiv e^2/4\pi$ is the conductance of a ballistic $\nu = 1/2$ channel.
Equation~\eqref{eq:auto-conductance_relation}, unlike Eq.~\eqref{eq:diss_fluct_half} for cross-correlation, has actually been explicitly demonstrated in, e.g., Refs.~\cite{WangFeldmanPRL13,FeldmanMotyPRB17}, and can also be obtained from the formulas of Refs.~\cite{BenaSafiPBR07,SafiJoyezPRB11}.
In addition, similar ``fluctuation-dissipation'' relations are known for non-chiral Luttinger liquids~\cite{DolciniPRB05,SafiBenaCrepieuxPRB80}, dissipative systems~\cite{ZamoumCrepieuxSafiPRB12}, and various other setups comprising chiral Luttinger liquids~\cite{Kane:1994b,Fendley:1995b,Fendley:1996,WangFeldmanPRL13,FeldmanMotyPRB17,FeldmanHalperinRPP21}.

Noticeably, Eq.~\eqref{eq:auto-conductance_relation} is widely used in experiments.
Indeed, at zero temperature, it provides a relation between current auto-correlations and the tunneling-current noise.
This relation, crucially, allows one to extract the Fano factor [proportional to the ratio between tunneling current noise, and the tunneling current, see Eq.~\eqref{eq:Fano}] by measuring the (experimentally accessible) auto-correlation in the current drain: a prerequisite that enables the experimental observation of the fractional charge~\cite{de-Picciotto:1997,Saminadayar:1997}. 
In real experiments, the detection of fractional charge can, however, be influenced by multiple practical issues.
For instance, following Ref.~\cite{Snizhko22}, the measurements of fractional charge, as well as the accompanied fractional scaling dimension, may be influenced by interactions between counter-propagating chiral channels.

Before closing this section, we compare our results with those of earlier works~\cite{Fendley:1995c, Kane:2003} where the $\nu=1/2$ chiral channels were addressed.
The system considered in this section has two edges and is, thus, different from the three-edge structure of Ref.~\cite{Kane:2003}, where only tunneling through the central QPC (but not the diluter) is exactly dealt with at $\nu = 1/2$ (for the analysis of noise created by diluted beams at $\nu=1/2$, see Sec.~\ref{four-edge-setup} below). 
On the other hand, Ref.~\cite{Fendley:1995c} also considered our model in the $\nu = 1/2$ situation.
However, our work goes beyond that of Ref.~\cite{Fendley:1995c}, by providing (i) compact analytical expressions, valid for all system parameters, for the tunneling-current noise and cross-correlation, as well as (ii) the exact noise-current relation for the $\nu = 1/2$ two-edge model.
In Ref.~\cite{Fendley:1995b}, another exact relation between the tunneling-current noise and the tunneling current was derived at arbitrary $\nu$ but only for \textit{zero temperature}:
\begin{equation}
  S_T=-\frac{\nu e}{2(1-\nu)} V^2 \frac{\partial}{\partial V} \frac{\langle I_T \rangle}{V}, \quad T=0. 
    \label{ST-FLS95}
\end{equation}
Our result for $S_T$, Eq.~(\ref{eq:st_two_edge}), combined with that for $\langle I_T \rangle$, Eq.~\eqref{tunc.6}, indeed satisfies Eq.~\eqref{ST-FLS95} at zero temperature and $\nu=1/2$, see Appendix~\ref{app:expansions}. However, relation \eqref{ST-FLS95} breaks down at finite temperatures, as can be explicitly checked using Eqs.~(\ref{eq:st_two_edge}) and ~\eqref{tunc.6}, in contrast to Eq.~(\ref{eq:diss_fluct_half}) that remains exact at $\nu=1/2$ for arbitrary $T$ and $V$.

\section{Four-edge setup --- tunneling between diluted edge channels}
\label{four-edge-setup}

\begin{figure} \begin{center} 
\includegraphics[width=8cm]{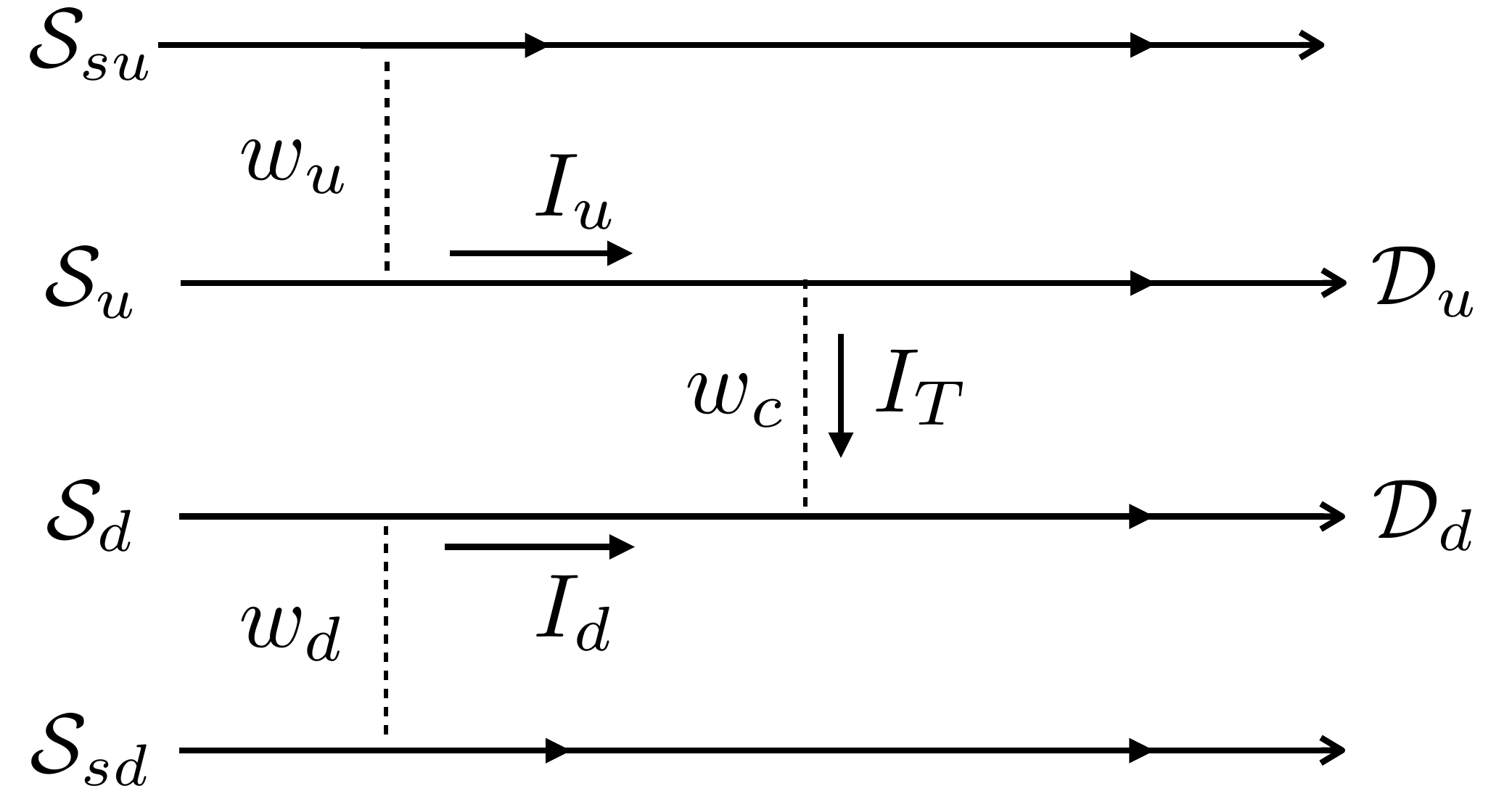}
\end{center}
\caption{Schematic representation of the four-edge-channel setup (HOM interferometer). A bias $V$ is applied at arms $\mathcal{S}_{su}$ and $\mathcal{S}_{sd}$, while $\mathcal{S}_u$ and $\mathcal{S}_d$ are kept grounded. The two QPCs located closest to these biased arms, QPC$_u$ and QPC$_d$ (diluters, indicated by the tunneling amplitudes $w_u$ and $w_d$, respectively), emit nonequilibrium quasiparticle currents $I_u$ and $I_d$ into arms $\mathcal{S}_u$ and $\mathcal{S}_d$, respectively. These dilute beams then impinge on the central QPC$_c$ (collider, indicated by the tunneling amplitude $w_c$). For all QPCs, the corresponding tunneling currents are shown (for the central QPC we use the same notation $I_T$ as in the two-edge setup). The tunneling-current noise $S_T$ studied for this setup is defined for QPC$_c$. The cross-correlations are measured for currents in drains $\mathcal{D}_u$ and $\mathcal{D}_d$. For a physical implementation of such a device in the case of the filling fraction $\nu=1/3$, see, e.g., Ref.~[\onlinecite{Bartolomei:2020}]. } 
\label{4-edge}
\end{figure}

In this Section, we analyze a four-edge setup as originally considered in Ref.~[\onlinecite{Rosenow:2016}] for the case of Laughlin filling fractions $\nu=1/(2m+1)$ with positive integer $m$. Below, we show that the case $\nu =1/2$ requires special care to obtain meaningful (finite) results. The system is illustrated in Fig.~\ref{4-edge}. The model and its analysis in  Ref.~[\onlinecite{Rosenow:2016}] (and subsequent works on the subject) are different from that of Ref.~\cite{Kane:2003} in two ways. Firstly, the setup has four edges (with two diluters), different from the three-edge system (with one diluter) of Ref.~\cite{Kane:2003}. Secondly, the model of Ref.~\cite{Kane:2003} is treated exactly for tunneling at the central QPC, but perturbatively, only to leading order in tunneling, at the first QPC (serving as a dilute).
In the analysis of Ref.~\cite{Rosenow:2016}, on the contrary, tunneling through the central QPC is treated perturbatively to the leading order. At the same time, both diluters are dealt with beyond the leading-order perturbation (albeit not exactly), by means of resummation of the series containing the most dominant terms at each order.

In the present work, we will make use of the exact solution for the central QPC (following similar steps of Ref.~\cite{Kane:2003}), to check the validity of the leading-order expansion of the tunneling current and noise in the transmission of the central QPC in the weak-tunneling limit (see Appendix~\ref{app:exact_central}).
However, we are not going to solve two diluters exactly at $\nu=1/2$, which could explicitly demonstrate the conditions of applicability of the resummation scheme for taking into account processes involving an arbitrary number of nonequilibrium anyons (for further refinements of this scheme going beyond the Poissonian statistics of duiluted beams, see Refs.~\cite{LeeNature23, IyerX2023, ThammBerndPRL24}). In short, solving the diluters exactly only provides us the knowledge of correlation functions of refermionized operators. This, however, does not facilitate solving the tunneling problem (even at the perturbative level) at the central 
QPC, located downstream of diluters. This is because the bosonic correlation functions for this QPC are not directly expressed through the fermionic operators used for solving the tunneling problem at the diluters (cf. Ref.~\cite{Vondelft:1998}). In principle, one could still expect that the four-edge model with diluters and the central QPC is exactly solvable by means of the Bethe ansatz for $\nu=1/2$. We relegate the exploration of this possibility to future work. 

Nevertheless, we account for all orders of the transmission through the diluters by the resummation of the perturbation series, similar to the approximate resummation performed in Refs.~\cite{Rosenow:2016,
Bartolomei:2020, Lee:2022, Schiller:2023}. 
However, we go beyond the approach of these works by including the subleading corrections to the correlation functions at each order of the Keldysh perturbation theory. These corrections are extremely important in the $\nu=1/2$ case, as they produce finite results for the tunneling current and the generalized Fano factor defined in Ref.~\cite{Rosenow:2016}.
Indeed, the generalized Fano factor would otherwise diverge (because of the vanishing of the tunneling current) for $\nu =1/2$, see Sec.~\ref{sec:GenFano} below.

The considered four-edge structure has the following model Hamiltonian
\begin{equation}
H=H^0_\mathrm{su\mhyphen u}+ H^0_\mathrm{d\mhyphen sd} + H^T_\mathrm{su\mhyphen u}+H^T_\mathrm{d\mhyphen sd}+H^T_\mathrm{u\mhyphen d},
\label{ham4edge}
\end{equation}
where the $H^0$ Hamiltonians describe pairs of free edge channels, whereas the $H^T$ terms describe the tunneling of quasiparticles at the three QPCs.
In Eq.~\eqref{ham4edge}, subscripts $u$, $d$, $su$ and $sd$ are labels of four edge channels, see Fig.~\ref{4-edge}, with $u$ and $d$ referring to the ``up'' and ``down'' channels connected by the central QPC and $su$, $sd$ the corresponding source channels. 
For general values of $\nu$, we have
\begin{align}
H^0_\mathrm{su\mhyphen u}&=\frac{v}{4\pi} \!\sum_{j=\mathrm{su,u}} \! \int_{-\mathcal{L}}^\mathcal{L} \: d x \: [ \partial_x \phi_j ( x )]^2,
\\
H^0_\mathrm{d\mhyphen sd}&=\frac{v}{4\pi} \!\sum_{j=\mathrm{d,sd}} \! \int_{-\mathcal{L}}^\mathcal{L} \: d x \: [ \partial_x \phi_j ( x )]^2
\end{align}
for the free Hamiltonians of the upper (combining the up channel and its source) and lower (combining the down channel and its source) edges,
\begin{align}
H^T_\mathrm{su\mhyphen u}&=\frac{2 w_u}{ (2 \pi l_c)^\nu } \: \cos \!\left\{ \sqrt{\nu}\left[\phi_{su} (0) \!-\! \phi_u (0 )\right] \right\},
\\
H^T_\mathrm{d\mhyphen sd} &= \frac{2 w_d}{ (2 \pi l_c)^\nu} \: \cos \!\left\{ \sqrt{\nu} \left[\phi_d (0) \!-\! \phi_{sd} (0 )\right] \right\},
\end{align}
for the tunneling Hamiltonian of the diluters in the upper and lower edges, and
\begin{align}
    H^T_\mathrm{u \mhyphen d}& =\frac{2 w_c}{ (2 \pi l_c)^\nu} \: \cos\! \left\{ \sqrt{\nu} \left[\phi_u (L) \!-\! \phi_d (L )\right] \right\}
\end{align}
for the tunneling Hamiltonian of the central QPC (collider) connecting the channels $u$ and $d$ of the upper and lower edges.
Here, the $\phi$ fields obey the commutation relations specified in Eq.~\eqref{commutation-rule}.
Here $w_u,w_d$, and $w_c$ are the tunneling amplitudes at the upper ($x=0$), lower ($x=0$), and central ($x=L$) QPCs, respectively.
In the present work, following Ref.~\cite{Morel:2022}, we focus on $t \ll L/v$.
As we will see, the relevant time scales for both the tunneling current and tunneling current noise satisfy $t \lesssim 2(w_u^2 + w_d^2 + w_c^2)/l_c$, which will be assumed to be smaller than $L/v$.

Like for the two-edge structure considered in Sec.~\ref{two-edge-setup}, below we focus on the $\nu = 1/2$ case.
As was done previously, we take the limit of infinite channels, $\mathcal{L} \rightarrow \infty$. 
For simplicity, we concentrate on the case when the upper and lower subsystems are biased symmetrically, i.e., the bias voltage $V_{su}$ applied to the channels $\mathcal{S}_{su}$ is equal to the voltage $V_{sd}$ applied to $\mathcal{S}_d$, i.e., $V_{su} = V_{sd} = V$; the arms $\mathcal{S}_u$ and $\mathcal{S}_d$ are assumed grounded. At the end of this Section, we will also briefly discuss the modifications introduced by the asymmetry in bias.

With $I_u$ and $I_d$ we label the tunneling current at the upper and lower QPC, respectively (see Fig.~\ref{4-edge}). 
We assume that the tunneling of quasiparticles at all three QPC's is weak (i.e., the effective transmission at each of the QPCs is small).
More specifically, for the two diluters, the smallness of their tunneling transmission can be characterized by the conditions $$I_u, I_d \ll  e^2 V.$$ 
For the central QPC, the corresponding conditions reads:
$$\frac{\partial}{\partial {(I_u - I_d)}} \langle I_T \rangle \ll 1.$$

We are interested in the tunneling current at the central QPC and in its fluctuations.
The tunneling-current operator at the central QPC is given by
\begin{equation}
I_T=\frac{i e}{2}\left(A^\dagger-A \right)    
\end{equation}
with
\begin{equation}
A(t)=\frac{w_c}{(2\pi l_c )^\frac{1}{2}}\exp\big[i(\phi_u(L,t)-\phi_d(L,t))/\sqrt{2}\big].   
\label{eq:a}
\end{equation}
In order to evaluate the tunneling current $I_T$, we move to the interaction representation with respect to the ``free Hamiltonian''
$$\tilde{H}=H^0_\mathrm{su\mhyphen u}+H^T_\mathrm{su\mhyphen u}+   H^0_\mathrm{d\mhyphen sd}+H^T_\mathrm{d\mhyphen sd},$$ 
considering $H^T_\mathrm{u-d}$ as an ``interaction''.
Assuming that the tunneling at the central QPC is adiabatically turned on at $t=-\infty$, we have 
\begin{equation}
\langle I_T(t)\rangle=\left\langle {\bf T}_K I_T(t^+)\exp\left[-i\int_K \tilde{H}(\tau)d\tau\right]\right\rangle,
\label{keldysh-current}
\end{equation}
with $K$ indicating the Keldysh contour. Here with $\tau$ we label an arbitrary time on the Keldysh contour (Fig.~\ref{keldysh-contour}). With $t^+$ we specify that the time belongs to the upper part of the Keldysh contour. For the calculation of the tunneling current, we could have equally chosen a time on the lower part of the contour (i.e.,  $t^-$).  

Following the method of Ref.~\cite{Kane:2003}, the tunneling current in our four-edge system can be obtained by evaluating the correlation function
\begin{equation}
\langle I_T \rangle =\frac{e}{2}\int_{-\infty}^{\infty} dt e^{-\frac{2 w_c^2}{v}|t|} \langle[A^\dagger(0),A(t)] \rangle_\text{neq},
\label{eq:current_exact_central}
\end{equation}
see Appendix~\ref{app:exact_central} for detailed derivation.
Here, $\langle \cdots \rangle_\text{neq}$ indicates that the expectation value is evaluated with respect to the state created by the tunneling at the biased diluters (but not affected by the tunneling at the central QPC).
Importantly, this state is a \textit{nonequilibrium} state (hence, ``neq'' in the label) with diluted quasiparticle beams for $x>0$ (downstream of diluters).
In comparison to leading-order (in tunneling through the central QPC) expressions (used in, e.g., Refs.~\cite{Rosenow:2016, Lee:2022, Morel:2022, Schiller:2023}),
\begin{equation}
\langle I_T \rangle\simeq \frac{e}{2}\int_{-\infty}^{\infty} dt \langle[A^\dagger(0),A(t)] \rangle_\text{neq},
\label{keldysh-current2}
\end{equation}
Eq.~\eqref{eq:current_exact_central} contains an extra factor $\exp(-2w_c^2 |t|/v)$ that comes from the higher-order tunneling processes at the central QPC.

Similarly, for the fluctuation of the tunneling current we have 
\begin{multline}    
S_T=\frac{1}{2}\sum_{\eta=\pm} \int_{-\infty}^{\infty} dt\, \big\langle \big\langle {\bf T}_K I_T(t^\eta)I_T(0^{-\eta}) \big\rangle \big\rangle  
 \\=\frac{e^2}{4}\int_{-\infty}^{\infty}\!\!\! dt\, e^{-\frac{2w_c^2}{v} |t|} \big\langle \{ A^\dagger(0),A(t)\} \big\rangle_\text{neq},
 \label{eq:tunneling_current_noise_4}
\end{multline}
with $\eta$ being the Keldysh index.

Importantly, in both Eqs.~\eqref{eq:current_exact_central} and \eqref{eq:tunneling_current_noise_4}, the factor $\exp(-2w_c^2 |t|/c)$, which exactly accounts for the tunneling through the central QPC, only influences the integral over time $t$. Correlation functions, marked by ``neq'', of each individual subsystem (separated by the central QPC) are not influenced by the higher-order tunneling processes at the central QPC. This is because the central QPC is in the downstream direction of both diluters.
We also note that the only difference between the integrands in Eqs.~(\ref{keldysh-current2}) and (\ref{eq:tunneling_current_noise_4}) is that the former contains a commutator, whereas the latter involves an anticommutator of $A^\dagger(0)$ and $A(t)$. Thus, for computing the tunneling current, as well as the tunneling-current noise, we need to calculate the expectation values of the product of two tunneling vertex operators of the form (\ref{eq:a}) taken at two distinct times. The averages should be done with respect to the nonequilibrium states created by biased diluters.

\subsection{Correlation function of tunneling operators at the central QPC: perturbative Keldysh approach}
\label{subsec:regions}

In order to evaluate Eqs.~\eqref{keldysh-current2} and \eqref{eq:tunneling_current_noise_4}, we need the explicit expression of correlation functions of the vertex operators involved in the Hamiltonian describing tunneling at the central QPC, such as 
\begin{multline}
\left\langle  A(t) A^\dagger(0) \right\rangle_\text{neq} =  \frac{w_c^2}{2 \pi l_c} \left\langle e^{i \phi_u(L,t)/\sqrt{2}} e^{-i \phi_u(L,0)/\sqrt{2}}\right\rangle_\text{neq}
\\ 
\quad \times \left\langle e^{-i \phi_d(L,t)/\sqrt{2}} e^{i \phi_d(L,0)/\sqrt{2}}\right\rangle_\text{neq}.
\end{multline}
Within the Keldysh formalism (cf. Ref.~\cite{Morel:2022}), we obtain
\begin{widetext}
\begin{align}
\left\langle e^{i \phi_u(L,t)/\sqrt{2}} e^{-i \phi_u(L,0)/\sqrt{2}}\right\rangle_\text{neq} 
=\left\langle {\bf T}_K \, e^{i \phi_u(L,t^-)/\sqrt{2}} e^{-i \phi_u(L,0^+)/\sqrt{2}} \,
e^{-i\int_K d\tau\,H^T_\mathrm{su\mhyphen u}(\tau)}
\right\rangle_0,
\label{correlation-function-phi-2}
\end{align}
where the subscript 0 in $\langle\cdots\rangle_0$
emphasizes that the average is now taken with respect to the state of the system without tunneling bridges. 
In order to evaluate the right-hand side of Eq.~\eqref{correlation-function-phi-2}, we expand the Keldysh evolution operator in powers of $H^T_\mathrm{su\mhyphen u}$, leading to
\begin{align}
\left\langle {\bf T}_K e^{i \phi_u(L,t^-)/\sqrt{2}} e^{-i \phi_u(L,0^+)/\sqrt{2}}  \,
e^{-i\int_K d\tau\,H^T_\mathrm{su\mhyphen u}(\tau)}\right\rangle_0 =
\sum_{n=0}^\infty 
\left\langle e^{i \phi_u(L,t)/\sqrt{2}} e^{-i \phi_u(L,0)/\sqrt{2}}\right\rangle_{2n},
\label{eq:2n-sum}
\end{align}
where we have introduced the short-hand notation for an average involving $2n$ integrals along the Keldysh contour:
\begin{align}
\left\langle O(t,t') \right\rangle_{2n}= 
\frac{(-i)^{2n}}{(2n)!} \int_K d\tau_1\ldots \int_K\tau_{2n}\,  \big\langle {\bf T}_K\, O(t,t') \, H^T_\mathrm{su\mhyphen u}(\tau_1)\ldots H^T_\mathrm{su\mhyphen u}(\tau_{2n}) \big\rangle_0.
\end{align}
Explicitly, the contribution of the second order in $H^T_\mathrm{su\mhyphen u}$ to the correlation function, i.e., the $n=1$ term in Eq.~(\ref{eq:2n-sum}), reads at zero temperature, $T=0$, as:
\begin{multline}
 \left\langle e^{i \phi_u(L,t)/\sqrt{2}}\, e^{-i \phi_u(L,0)/\sqrt{2}}\right\rangle_2
 =-\sum_{\eta_1\eta_2} \eta_1\eta_2 \int_{-\infty}^\infty \!\!\!ds_1 \int_{-\infty}^\infty \!\!\!ds_2
 \, \left\langle {\bf T}_K e^{i \phi_u(L,t^-)/\sqrt{2}}\, e^{-i \phi_u(L,0^+)/\sqrt{2}}\, H^T_\mathrm{su\mhyphen u}(s_1^{\eta_1})\,H^T_\mathrm{su\mhyphen u}(s_1^{\eta_2})\right\rangle_0 
 \\
 =
 -\frac{ w_u^2}{ 2 \pi l_c}  \sum_{\eta_1\eta_2} \eta_1\eta_2 \int_{-\infty}^\infty  ds_1 \int_{-\infty}^\infty ds_2 
 \,\: e^{-i V (s_1 - s_2)/2}
 \Big\langle {\bf T}_K e^{i\phi_{su} (0,s_1^{\eta_1})/{\sqrt{2}}}\, e^{-i \phi_{su} (0,s_2^{\eta_2})/{\sqrt{2}}}
 \Big\rangle_0\\
 \times 
 \Big\langle {\bf T}_K e^{-i \phi_u (L,t^-)/{\sqrt{2}}}\, e^{i \phi_u (L, 0^+)/{\sqrt{2}}}\, e^{-i \phi_u (0,s_1^{\eta_1})/{\sqrt{2}}}\, e^{i \phi_u (0,s_2^{\eta_2})/{\sqrt{2}}} \Big\rangle_0 \\
=-\frac{w_u^2}{ 2 \pi l_c}\, \frac{l_c^{1/2}}{(l_c + i v t)^{1/2}}\,  \sum_{\eta_1\eta_2} \eta_1\eta_2 \int_{-\infty}^\infty ds_1 \int_{-\infty}^\infty ds_2 \,\:
e^{-i V (s_1 - s_2)/2}\,
 \frac{l_c}{l_c + i v(s_1 - s_2) \chi_{\eta_1,\eta_2} (s_1 - s_2) }  \\
\times   \frac{ \sqrt{ l_c + i (vt - vs_1 - L ) \chi_{-,\eta_1 } (t - s_1) } \sqrt{ l_c + i ( - v s_2 - L ) \chi_{+,\eta_2 } ( - s_2) } }{\sqrt{ l_c + i (vt - vs_2 - L ) \chi_{-,\eta_2 } (t - s_2) } \sqrt{ l_c + i ( - vs_1 - L ) \chi_{+,\eta_1 } ( - s_1) } },
\label{second_order}
\end{multline}
\end{widetext}
where $\eta_1$ and $\eta_2$ Keldysh indexes of nonequilibrium anyonic operators and 
\begin{equation}
\chi_{\eta,\eta'} (s) = \eta' \Theta(s)-\eta \Theta(-s).
\label{chi-def}
\end{equation}

In addition, when going from the first line to the second line of Eq.~\eqref{second_order}, a factor of two, coming from different choices of creation and annihilation operators, cancels out the factor $1/2!$ that originates from the Keldysh expansion.

\begin{figure}[ht]
\begin{center} 
\includegraphics[width=8cm]{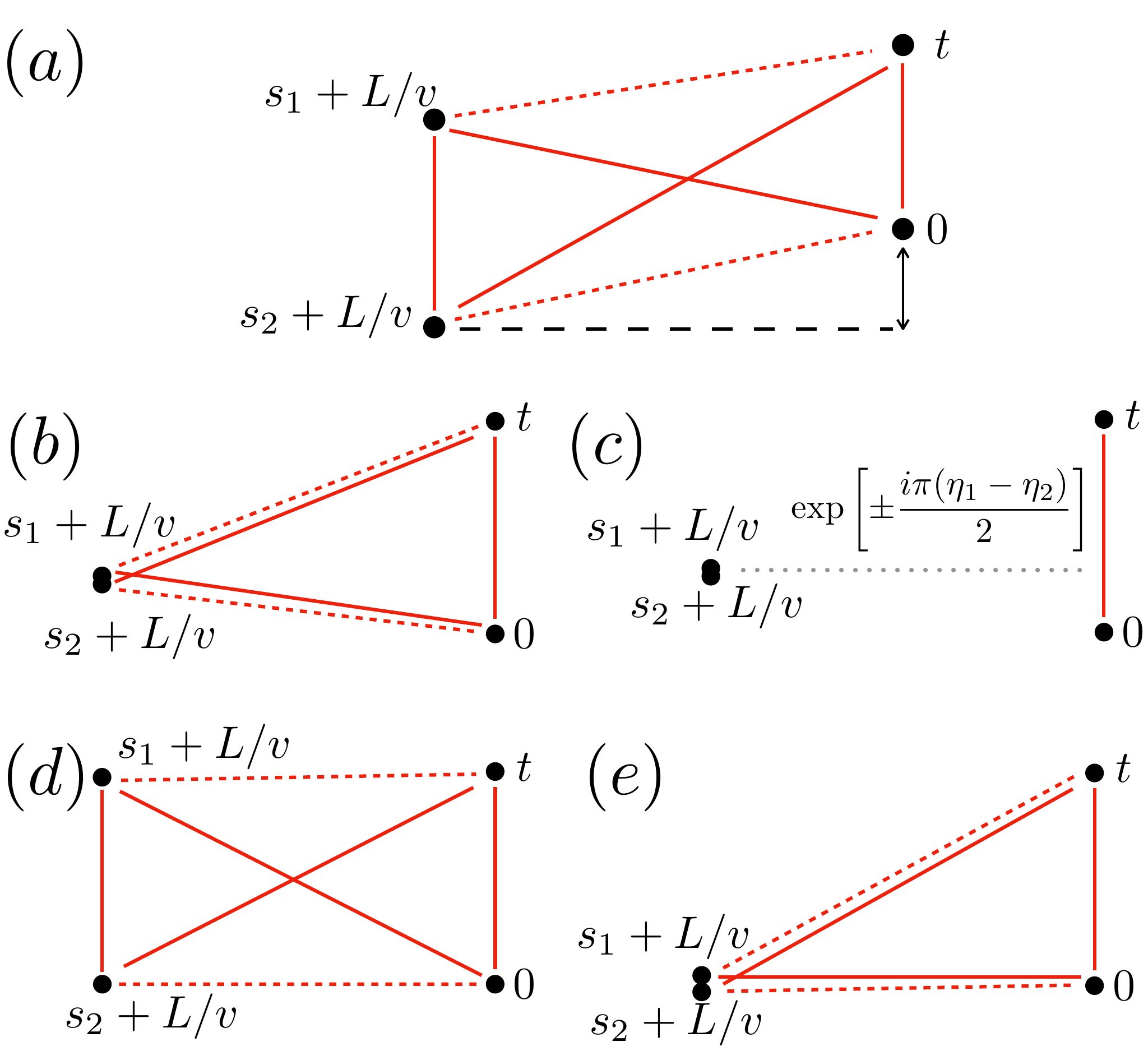}
\end{center}
\caption{Different regions contributing to the correlation function Eq.~\eqref{second_order}. (a) The general case. Here $s_1 + L/v$ and $s_2 + L/v$ refer to the times two nonequilibrium anyons arrive at the central QPC. Times 0 and $t$ mark out the time arguments of the tunneling operators at the central QPC. The red dashed and solid connecting lines refer to the correlation functions that appear in the numerator and denominator of Eq.~\eqref{second_order}, respectively. 
The time distance between a nonequilibrium anyon and a tunneling operator is represented by the projected distance along the vertical direction [for instance, the double arrow in panel (a) represents the distance $s_2 + L/v \leftrightarrow 0$, with the black dashed line being the guide for the eye for the projection]. (b) Diagram that corresponds to Region I contribution.
Here $|s_1 - s_2| \sim l_c/v$, and the projected distances for the pairs of solid and dotted lines coincide: $|s_1 + L/v |/ |s_2 + L/v|\approx |s_1 + L/v -t |/ |s_2 + L/v - t | \approx 1$. (c) These equal-length projected distances cancel out in the tangling factor in Eq.~\eqref{second_order}, leading to a ``disconnected'' diagram. Here the nonequilibrium pair becomes disconnected from tunneling operators (no red lines connect the dots at times $s_1 + L/v,\,s_2 + L/v$ with dots at times $0,t$), except for the phase shift $\exp [\pm i\pi (\eta_1 - \eta_2)/2]$ (marked out by the gray dotted line), arising from time-domain braiding. For fermions, this phase shift is absent, and the diagram becomes truly disconnected. (d) Diagram for Region II. One cannot separate the diagram into the disconnected blocks, hence the diagram is ``connected''. (e) Diagram for Region III.  
Now, $|s_1 - s_2| \sim 1/\nu e V \gg l_c /v$.
Distances $s_1 + L/v \leftrightarrow 0$ and $s_2 + L/v \leftrightarrow 0$ will not cancel exactly each other. The diagram is again ``connected''.}
\label{fig:diagrams_regions}
\end{figure}

An analysis of Eq.~(\ref{second_order}) shows that the main contributions to the integral over times $s_1,\,s_2$ come from the following regions (see Appendix \ref{app:correlation_derivations}):
\begin{itemize}
    \item Region  I: $s_1\simeq s_2$, with both times within the interval $(-L/v,\,t-L/v)$;
    \item Region  II: $s_1\simeq -L/v$ and $s_2\simeq t-L/v$; 
    \item Region  III: $s_1,s_2\simeq -L/v$ or $s_1,s_2\simeq t-L/v$.
\end{itemize}
This separation of the integration domains is not specific to the case of $\nu=1/2$; however, it becomes 
especially sharp in this case. Mathematically, these three regions  can be identified by analyzing the phase of the ``tangling factor'' [the combination of the square roots in the integrand of Eq.~\eqref{second_order}], which jumps at the boundaries between the regions (see Appendix \ref{app:correlation_derivations}). The dominant contributions to the correlation function correspond to the singularities of the integrand, which appear in the denominator. Even though the powers of the singular terms (like $1/2$ in the present case) could be insufficient for the special points to dominate the integral on their own, the oscillatory factor with the voltage-induced phase ensures that the integrals over the corresponding branch cuts are indeed dominated by the vicinity of the branching point (cf. Ref.~\cite{Morel:2022}). We illustrate the above classification in Fig.~\ref{fig:diagrams_regions}.

Following the convention of Ref.~[\onlinecite{AndreevX23}] (see also Sec.~10.C.4 of Ref.~\cite{Vondelft:1998}), in this work, we address the contribution of Region I to the correlation functions as ``disconnected''. A disconnected diagram is a notion that appears for systems with conventional (fermionic or bosonic) statistics.
For a fermionic system, an event belonging to Region I would correspond to a Feynman diagram that can be divided into disconnected sub-diagrams, hence ``disconnected''.
For a disconnected fermionic diagram (resulting from the Wick's theorem for fermions), nonequilibrium particles (whose operators appear with the time arguments $s_1$ and $s_2$) self-contract.
Fermionic tunneling operators at the central QPC (at times $0$ and $t$), resorting to thermally excited particle-hole pairs, become then decoupled from the operators at the diluter, leading to a vanishing contribution to the irreducible correlation function.

However, this concept of ``disconnected'' does not apply to anyonic systems (cf. Ref.~\cite{Han:2016}), since Wick's theorem is not valid for anyonic operators. Indeed, in great contrast to fermionic systems, processes involving anyons are inseparable even for Region I, because a product of all possible two-point correlation functions (with all possible time differences of the arguments) appear in the average of anyonic operators. This is evident in the last line of Eq.~\eqref{second_order}, which contains the tangling factor in the form of a fraction of square roots involving the time arguments of the tunneling operators at both the collider and diluters, see Fig.~\ref{fig:diagrams_regions}(a). In this situation, the anyonic operators, even well separated in time, can become correlated via the so-called ``time-domain braiding''~\cite{Schiller:2023} (corresponding to ``topological vacuum bubbles'' in the terminology of Ref.~\cite{Han:2016}). Specifically, even when the tangling factor reduces to unity by absolute value,
which happens for $s_1=s_2$ (corresponding to Region I) in Eq.~\eqref{second_order}, it produces a non-trivial (``braiding'') phase, independent of the time arguments [see Figs.~\ref{fig:diagrams_regions}(b) and \ref{fig:diagrams_regions}(c), and more detailed discussions in Appendix~\ref{app:correlation_derivations}]. Region I, thanks to this nonlocal correlation, thus generates a nonzero contribution to the irreducible correlation functions.

For convenience, in this work, we keep using ``disconnected'' to address anyonic processes corresponding to Region I, to state that their graphical representation reduces to separate entities after replacing the tangling factor by a phase factor. Physically, the appearance of disconnected blocks indicates that nonequilibrium anyons do not directly tunnel at the central QPC. 
Regions II and III, on the other hand, involve tunneling of nonequilibrium anyons at the central QPC. They are thus referred to as ``connected'', to distinguish them from processes determined by Region I. Further details on the classification of processes into connected and disconnected can be found in Appendix \ref{App:E-class}.

The difference between Region I and Region II contributions becomes especially manifest when comparing Figs.~\ref{fig:diagrams_regions}(b), and \ref{fig:diagrams_regions}(d): of the latter diagram (for Region II), nonequilibrium anyons (labelled by $s_1 + L/v$ and $s_2 + L /v$) are well separated.
In contrast (to this apparent difference), naively, Region III seems to overlap with Region I, as in both regions, $s_1$ and $s_2$ are close to each other.  Mathematically, however, there are major differences between Regions I and III.
First of all, in Region I, the times $s_1 = s_2$ are confined within the so-called causality regime, i.e., between $-L/v$ and $t-L/v$ [Fig.~\ref{fig:diagrams_regions}(b)]. In Region III, on the other hand, contributions with $s_1$ and/or $s_2$ outside of $(-L/v,t-L/v)$ are present [Fig.~\ref{fig:diagrams_regions}(e)].
For Region I, it is the delta-function term (with $|s_1 - s_2| \sim l_c /v\to 0$), which yields the contribution to the correlation function.
As will be shown shortly, Region I corresponds to the first term of Eqs.~\eqref{eq:ud_full_correlationsA} and \eqref{eq:ud_full_correlationsB} inside the curly brackets.  $\exp(-4I_{u/d} |t|/e)$, which, importantly, is a real function. The contribution of Region III instead leads to a purely imaginary contribution, $i\,\text{sgn}(t)\exp(-4I_{u/d} |t|/e) 8 I_{u/d}/e^2 V$, where the integral is given by the vicinity $\sim (eV)^{-1}$ of the point $s_{1}\approx s_{2} \sim -L/v$ [note that such a separation of time scales is only possible for $|t|,L/v>1/(e V)$].
Starting from the next section, we will explicitly address the contributions of each of the three region to the correlation functions.

\subsection{Tunneling current: Vanishing of the contributions from Region I}
\label{sec:vanishing_region_i}

Let us focus on Region I, which yields the contribution to the correlation functions considered by previous references, e.g., Refs.~[\onlinecite{Rosenow:2016, Lee:2022, Morel:2022}], and further addressed in terms of ``time-domain braiding'' by Ref.~[\onlinecite{Schiller:2023}].
Here ``braiding'' refers to the nontrivial phase that is generated by exchanging nonequilibrium anyons (that pass by the central QPC) and the anyon-hole pairs spontaneously created at the central QPC.
In Region I, the correlation function (\ref{correlation-function-phi-2}), calculated to second order in the tunneling amplitude $w_u$, reads (see Appendix~\ref{app:correlation_derivations} for details):
\begin{equation}
\langle e^{i \phi_u(L,t)/\sqrt{2}} 
e^{-i \phi_u(L,0)/\sqrt{2}}\rangle^{(\text{I})}_2 = -\frac{2w_u^2}{v} |t|\frac{l_c^{1/2}}{(l_c + i vt)^{1/2}}.
\label{eq81}
\end{equation}
Here superscript ``(I)'' indicates that we choose Region I as the integration domain in Eq.~\eqref{second_order} and subscript ``2'' indicates that we expand the correlator to second order $H^T_\mathrm{su\mhyphen u}$ [corresponding to $n = 1$ in Eq.~\eqref{eq:2n-sum}].

Summing up such contributions to all orders in $H_{su-u}^T$ [i.e., summing over $n$ of Eq.~\eqref{eq:2n-sum}]
and introducing the zero-temperature tunneling current at the upper QPC $I_u = e w_u^2/(2 v)$,
we obtain~[\onlinecite{Morel:2022}]
\begin{equation}
\begin{aligned}
 &\langle e^{i \phi_u (L,t)/\sqrt{2}} e^{-i \phi_u (L,0)/\sqrt{2}}\rangle^{(\text{I})} \\
 &\qquad\equiv 
 \sum_{n=0}^\infty \langle e^{i \phi_u (L,t)/\sqrt{2}} e^{-i \phi_u (L,0)/\sqrt{2}}\rangle_{2n}^{(\text{I})} \\
 &\qquad=  \frac{l_c^{1/2}}{(l_c+ivt)^{1/2}} \exp\left(-4\, \frac{I_{u}}{e} |t|\right).
 \end{aligned}
\label{corr-phi2-kindI} 
\end{equation}
Similarly, the resummation of perturbative contributions of Region I to the correlation function of vertex operators in subsystem $d$  
yields
\begin{align}
 &\langle e^{-i \phi_d(L,t)/\sqrt{2}} e^{i \phi_d(L,0)/\sqrt{2}}\rangle^{(\text{I})}\notag\\
 &\qquad=  \frac{l_c^{1/2}}{(l_c+ivt)^{1/2}} \exp\left(-4\, \frac{I_{d}}{e} |t|\right),
 \label{corr-phi3-kindI} 
\end{align}
where $I_d = e w_d^2/(2 v)$.
Note that the exponentials in Eqs.~(\ref{corr-phi2-kindI}) and (\ref{corr-phi3-kindI}) are purely real.

Using these results, we obtain
\begin{align}
\langle  A(t) A^\dagger(0) \rangle^{(\text{I})}&=\frac{w_c^2}{2\pi (l_c+ivt)}\exp\left[-\frac{4|t|}{e} \left(I_{u}+I_{d}\right)\right],
\label{eq:hole_correlation}
\\
\langle  A^\dagger(0) A(t)  \rangle^{(\text{I})}&=\frac{w_c^2}{2\pi ( l_c+i vt ) }\exp\left[-\frac{4|t|}{e} \left(I_{u}+I_{d}\right)\right].
\label{eq:particle_correlation}
\end{align}
Importantly, 
\begin{equation}
    \langle  A(t) A^\dagger(0) \rangle^{(\text{I})}=\langle  A^\dagger(0) A(t)  \rangle^{(\text{I})}.
    \label{AtA0=A0At}
\end{equation}
Equations~(\ref{eq:hole_correlation}) and (\ref{eq:particle_correlation}) agree with the result of Ref.~\cite{Rosenow:2016} taken at $\nu=1/2$.

By inserting the above correlation functions in Eq.~(\ref{eq:current_exact_central}), 
one immediately obtains 
\begin{equation}
    \langle I_T \rangle^{(\text{I})}=0,
    \label{IT=0}
\end{equation}
since the averaged commutator $\langle[A^\dagger(0),A(t)] \rangle_\text{neq}^{(\text{I})}$ is exactly zero.
Mathematically, this 
is related to the fact that the current difference $I_- = I_u - I_d$  does not enter Eqs.~\eqref{eq:hole_correlation} and \eqref{eq:particle_correlation} (thus, vanishing of $\langle I_T \rangle$ occurs for arbitrary biases $V_{su}$ and $V_{sd}$). From a more physical perspective, this property relies on the anyon-hole symmetry of the braiding phase: a unique feature of the $\nu=1/2$ situation. 
Indeed, for $\nu\neq 1/2$, the exponents in the expressions analogous to Eqs.~(\ref{corr-phi2-kindI}) and (\ref{corr-phi3-kindI} acquire an imaginary part, which gives a nonzero $I_-$.

\subsection{Tunneling current with Regions II and III included}
\label{sec:current_four_edge}

Above, we have shown that, if we take into account only disconnected contributions to the correlation functions of tunneling operators, 
Eqs.~\eqref{eq:hole_correlation} and \eqref{eq:particle_correlation},
the tunneling current between channels $u$ and $d$ vanishes, Eq.~\eqref{IT=0}. 
This counter-intuitive vanishing of the tunneling current is a consequence of the absence of tunneling of nonequilibrium anyons at the central QPC.
Indeed, if one allows direct tunneling of nonequilibrium anyons, then under the assumption $I_u > I_d$ (likewise, if $I_u < I_d$), the number of quasiparticles tunneling from $u$ to $d$ (which is proportional to $I_u$) is expected to become larger (likewise, smaller) than that for tunneling from $d$ to $u$ (which is proportional to $I_d$). Thus, the tunneling current is anticipated to be proportional to $I_u - I_d$.

\begin{figure} \begin{center} 
\includegraphics[width=8.6cm]{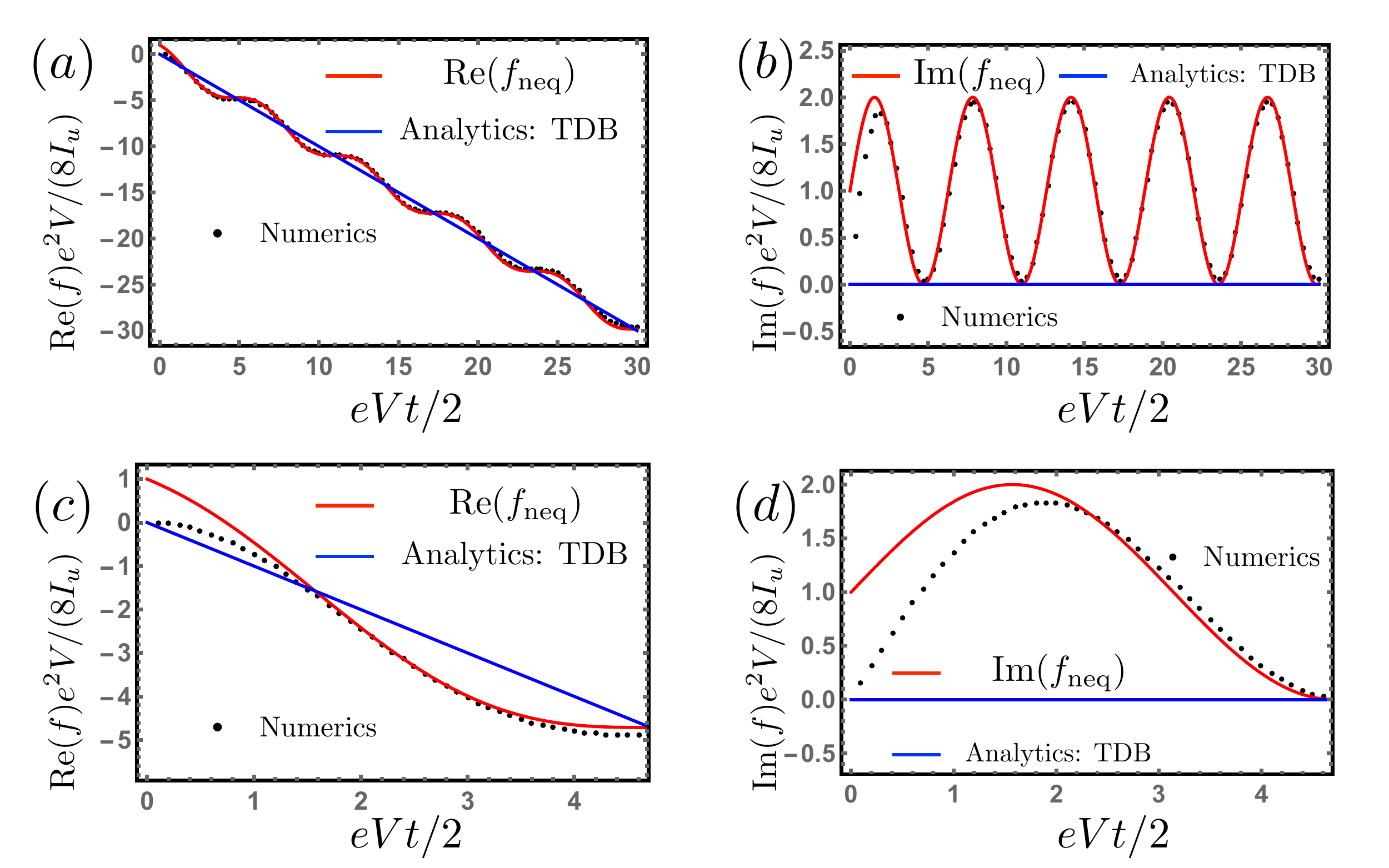}
\end{center}
\caption{
Checking the validity of Eq.~\eqref{eq:ud_full_correlationsA} to the leading order in the nonequilibrium current $I_u$. (a) and (b): Red curves show, for $t>0$, the imaginary and real parts of Eq.~\eqref{eq:fneq}, respectively, which multiplies the linear-in-$I_u$ terms in the expansion of Eq.~\eqref{eq:ud_full_correlationsA}, see Eq.~\eqref{eq-noneq}; Blue curves represent the contribution of Region I, Eq.~(\ref{corr-phi2-kindI}), corresponding to time-domain braiding (denoted as TDB in the figure); black dots show the numerical results for $f(t)$ defined in Eq.~\eqref{eq:ft}].
Negative-$t$ result can be obtained via the relations $\text{Re} f(-t) = \text{Re} f(t)$, and $\text{Im} f(-t) = -\text{Im} f(t)$.
For both real and imaginary parts of $f(t)$, red curves [Eq.~\eqref{eq:fneq}] match much better the exact numerical result [Eq.~\eqref{eq:ft}, black dots] than the TDB results (blue curves), when $eV t \gtrsim 4 $.
(c) and (d): Real and imaginary parts of $f(t)$ for not too large values of $eVt$.
When $eV t \to 0$, numerical values of function $f(t)$ drop to zero, different from the approximate analytical expression Eq.~\eqref{eq:fneq}.
} \label{fig_numerics_vs_analytics}
\end{figure}

In this section, we explicitly prove the above picture by including contributions to the correlation functions in Eq.~(\ref{keldysh-current2}) from Regions II and III. The details of derivation are provided in Appendix~\ref{app:correlation_derivations}. For $V_{su}=V_{sd}=V$, the results read:
\begin{widetext}
 
\begin{align}
\big\langle e^{i \phi_u(L,t)/\sqrt{2}} e^{-i \phi_u(L,0)/\sqrt{2}} \big\rangle_\text{neq} &\simeq 
\frac{l_c^{1/2}}{(l_c+ivt)^{1/2}}\left\{ \exp\left(-4\, \frac{I_{u}}{e} |t|\right)\left[1+i\, \text{sgn} (t) \frac{8 I_u}{e^2 V}\right] + \frac{8 I_u}{e^2 V}  
\exp\left(i\, \frac{eV}{2} t\right) \right\},
\label{eq:ud_full_correlationsA}
\\
\big\langle e^{-i \phi_d(L,t)/\sqrt{2}} e^{i \phi_d(L,0)/\sqrt{2}} \big\rangle_\text{neq} &\simeq 
\frac{l_c^{1/2}}{(l_c+ivt)^{1/2}}\left\{ \exp\left(-4\, \frac{I_{d}}{e} |t|\right)\left[1-i\, \text{sgn} (t) \frac{8 I_d}{e^2 V}\right] + \frac{8 I_d}{e^2 V}  
\exp\left(\!\!-i \frac{eV}{2} t\!\right) 
\right\}.
\label{eq:ud_full_correlationsB}
\end{align}
\end{widetext}
In correlation functions Eqs.~\eqref{eq:ud_full_correlationsA} and \eqref{eq:ud_full_correlationsB}, the equilibrium correlation function 
\begin{equation}
    \big\langle e^{i \phi_u(L,t)/\sqrt{2}} e^{-i \phi_u(L,0)/\sqrt{2}} \big\rangle_\text{eq}=\frac{l_c^{1/2}}{(l_c+ivt)^{1/2}}
\end{equation}
is multiplied by the factors in curly brackets that describe the effect of dilution.
Analogously to Eqs.~\eqref{corr-phi2-kindI} and \eqref{corr-phi3-kindI} obtained for Region I only, Eqs.~\eqref{eq:ud_full_correlationsA} and \eqref{eq:ud_full_correlationsB} also involve resummation of all orders in the transmission of diluters, see Eq.~\eqref{eq:2n-sum}.
The contributions of Regions II and III to Eqs.~\eqref{eq:ud_full_correlationsA} and \eqref{eq:ud_full_correlationsB} are those that are proportional to the factors $8 I_{u,d}/(e^2 V)$.
More specifically, Region III contributes to the term with the resummation-induced exponential factor [producing a correction to the Region-I unity in the second term $\propto\text{sgn}(t)$ in square brackets], whereas Region II yields the oscillatory voltage-dependent contribution (last term in the curly brackets).
These Region-II and Region-III terms, albeit subleading with respect to the term stemming from Region I for strong dilution, introduce imaginary parts to the content of the curly brackets in Eqs.~\eqref{eq:ud_full_correlationsA} and \eqref{eq:ud_full_correlationsB}. These imaginary parts translate into the leading nonvanishing terms in the tunneling current, since the product of real parts gives zero contribution, in view of the particle-hole symmetry at $\nu=1/2$, as discussed above. 
Importantly, correlation functions \eqref{eq:ud_full_correlationsA} and \eqref{eq:ud_full_correlationsB} are written here for $|t| \gg 1/eV$. Indeed, during derivations detailed in Appendix~\ref{app:correlation_derivations}, the condition $|t| \gg 1/eV$ was assumed. Otherwise, the separation of time scales determining the regions becomes impossible, and both correlation functions approach zero, as can be seen from Figs.~\ref{fig_numerics_vs_analytics}(c) and \ref{fig_numerics_vs_analytics}(d).

In Fig.~\ref{fig_numerics_vs_analytics}, we verify the validity of Eq.\eqref{eq:ud_full_correlationsA} by comparing its expansion to leading order in $I_{u}$ with the result of numerical integration of Eq.\eqref{second_order}.
In this figure, $f_\text{neq}$ is defined as
\begin{equation}
    f_\text{neq} (t') \equiv - |t'| + i \ \text{sgn}(t') + \exp\left( i t' \right),
    \label{eq:fneq}
\end{equation}
which equals the leading-order expansion of expressions inside the curly bracket of Eq.~\eqref{eq:ud_full_correlationsA}, after taking $t'\equiv e Vt /2$, and multiplying the result by $ e^2 V/(8 I_u)$: 
\begin{align}
   & \big\langle e^{i \phi_u(L,t)/\sqrt{2}} e^{-i \phi_u(L,0)/\sqrt{2}} \big\rangle_\text{neq} = \frac{l_c^{1/2}}{(l_c+ivt)^{1/2}}\notag \\
    &\qquad \times \left\{ 1+\frac{8I_u}{e^2 V} f_\text{neq}(eVt/2)+\mathcal{O}
    \left(I_u^2/V^2\right)\right\}.
    \label{eq-noneq}
\end{align}
Based on Fig.~\eqref{fig_numerics_vs_analytics}, $f_\text{neq}$ agrees with the numerical evaluation of the corresponding integral $f(t)$, defined by Eq.~\eqref{eq:ft} in Appendix \ref{app:correlation_derivations}, for $|t|\gtrsim 4/eV$. In the limit $t\to 0$, the function $f(t)$
goes to zero, while  $f_\text{neq}(t')$ approaches a finite value.

Though Eqs.~\eqref{eq:ud_full_correlationsA} and \eqref{eq:ud_full_correlationsB} involve resummation of the most significant higher-order tunneling processes at the diluters, they also receive corrections from other higher-order processes that involve, e.g., partial overlap between nonequilibrium anyonic pairs or simultaneous tunneling of anyons at the central collider. 
As discussed in Appendix~\ref{app:corrections}, these corrections generate the terms of the order of $O(I_u^2/V^2)$ and $O(I_d^2/V^2)$ in the prefactors of the exponential terms in the correlation functions and of order $O(I_u^2|t|/V)$ and $O(I_d^2|t|/V)$ in the arguments of the exponentials.
In the formal perturbative expansion of the correlation functions, such corrections can compete at each order with the terms already accounted for in Eqs.~\eqref{eq:ud_full_correlationsA} and \eqref{eq:ud_full_correlationsB}. However, upon resummation, such higher-order perturbative corrections turn out to be negligible, as they would only produce further subleading terms in the time integrals determining the tunneling current and its noise.

Substituting Eqs.~\eqref{eq:ud_full_correlationsA} and \eqref{eq:ud_full_correlationsB} into Eq.~\eqref{eq:current_exact_central}, we calculate 
the tunneling current at the central QPC to leading order in $I_u$ and $I_d$.
Firstly, when choosing Regime I contribution from both diluters, tunneling current equals
\begin{equation}
    \langle I_T \rangle^\text{I-I} = \frac{e w_c^2}{4\pi } \int_{-\infty}^\infty \!\! dt \left(\frac{e^{-4 \frac{I_u + I_d}{e} |t| }}{l_c + i vt} - \frac{e^{-4 \frac{I_u + I_d}{e} |t| }}{l_c + i vt}\right) \!=\! 0
\label{eq:it_i_i}
\end{equation}
In agreement with Eq.~\eqref{IT=0}, Eq.~\eqref{eq:it_i_i} indicates the vanishing of tunneling current, when including only time-domain braiding in a $\nu = 1/2$ four-edge setup. 
Thus, to obtain a finite tunneling current for $\nu=1/2$, the inclusion of Regions II and III becomes indeed necessary.

Next, we take the product of the contributions of Regions I and II to the correlation functions. The corresponding tunneling current equals
\begin{equation}
\begin{aligned}
    \langle I_T \rangle^\text{I-II}  
    &=  - \frac{4w_c^2 }{\pi e V} \!\int_{-\infty}^\infty \!\! dt\, \frac{\sin \left(\!\frac{e V t}{2}\!\right)}{vt} \left[ I_u\,  e^{-\left(\frac{4I_d}{e} + \frac{2w_c^2}{v}\right) |t| } \right.\\
     &\left. \!\!\!\!\!\! -  I_d\,  e^{-\left(\frac{4I_u}{e} + \frac{2w_c^2}{v}\right)|t| }\right] \approx \frac{4 w_c^2}{v e V} (I_u - I_d),
    \label{eq:it_i_ii}
\end{aligned}
\end{equation}
where we replaced $\arctan[e^2 V/(8 I_{u,d} + 4 w^2 e/v)]\to \pi/2$ in the weak-tunneling limit, $I_u, I_d, ew^2/v \ll e^2 V$.

Combining Regions I and III, the corresponding contribution to the tunneling current equals 
\begin{equation}
\begin{aligned}
    &\langle I_T \rangle^\text{I-III} =  - \frac{8w_c^2}{\pi eV} (I_u - I_d) 
    \\
    &\times
    \left( + \int_{\sim 1/eV}^\infty \!\! dt\  
    \frac{e^{-4\frac{I_u + I_d +e w_c^2/2v }{e}|t|}}{v t} \right)\\
    &\quad \approx \frac{8w_c^2}{\pi veV} \left[ C_2 +\ln\left( \frac{e^2V}{ I_u \!+\! I_d \!+\!e w_c^2/2v} \right) \right] (I_u \!-\! I_d),
\end{aligned}
\label{eq:it_i_iii}
\end{equation}
where the numerical constant $C_1$ comes from times $|t|\lesssim 1/eV$ and $C_2$ a number of order unity (exact evaluation of these numbers requires the knowledge of the correlation functions at $|t|\approx 1/eV$). In the weak-tunneling limit, the integral is dominated by the large logarithmic term.

Finally, the products of the terms in Eqs.~\eqref{eq:ud_full_correlationsA} and \eqref{eq:ud_full_correlationsB} that come from Region II and Region III is of higher order in $I_{u,d}/e^2V$ [a similar contribution would be produced by further corrections to Eqs.~\eqref{eq:ud_full_correlationsA} and \eqref{eq:ud_full_correlationsB}], and hence we disregard such cross-terms. It is also worth noticing that the product of the last terms in curly brackets of Eqs.~\eqref{eq:ud_full_correlationsA} and \eqref{eq:ud_full_correlationsB} (a product of two Region-II terms) yields an exactly zero contribution to the tunneling current. Interestingly, in contrast to other contributions to $\langle[A^\dagger(0),A(t)] \rangle_\text{neq}$, this product is independent of $I_u, I_d$, or bias $V$. Therefore, the time integral for this product could have only been cut off by the factor $\exp(-2 w_c^2 |t|/v)$ in Eq.~\eqref{eq:current_exact_central}, which stems from the resummation of the tunneling processes at the central QPC.

Adding up Eqs.~\eqref{eq:it_i_ii} and \eqref{eq:it_i_iii}, we arrive at the leading-order (with the logarithmic accuracy) expression for the tunneling current in the weak-tunneling limit,
\begin{align}
    \langle I_T \rangle
    \approx \frac{8w_c^2}{\pi v e V} 
    \ln\left( \frac{e^2V}{ I_u \!+\! I_d \!+\!e w_c^2/2v} \right) 
    (I_u \!-\! I_d),
    \label{eq:it_four_arm}
\end{align}
which is dominated by the product of the Region-I and Region-III contributions to the correlation functions,
Eq.~\eqref{eq:it_i_iii}.
As expected, the leading-order result for the tunneling current is proportional to the difference of nonequilibrium currents, $I_-=I_u - I_d$. 
In Eqs.~\eqref{eq:it_i_ii}-\eqref{eq:it_four_arm}, we have used the assumption of weak tunneling at all three QPCs, i.e., $w_c^2/v,\, I_u/e,\, I_d/e \ll e V$.
In the weak-tunneling limit, Eq.~\eqref{eq:it_four_arm} is dominated by the logarithmic term.
As discussed in Appendix~\ref{app:exact_central}, when all three QPCs are in the weak-tunneling limit, the relation $w_c^2 \ll w_u^2 + w_d^2 = 2v (I_u + I_d)/e$ holds. Therefore, it is safe to neglect $w_c^2$ in the argument of the logarithm in Eq.~\eqref{eq:it_four_arm} in this limit.

When two source biases are different, i.e., $V_{su} \neq V_{sd}$, the bias $V$ in Eq.~\eqref{eq:ud_full_correlationsA} and Eq.~\eqref{eq:ud_full_correlationsB} is replaced by $V_{su}$ and $V_{sd}$, respectively. 
In Eq.~\eqref{eq:it_four_arm}, one then simply replaces the common bias $V$, by different source biases:
$$(I_u-I_d)/V\, \to \, I_u/V_{su}-I_d/V_{sd}.$$

We would like to emphasize that our result, Eq.~\eqref{eq:it_four_arm}, will not reproduce the results of Ref.~\cite{Kane:2003} if the transmission of one of the diluters is set to zero, reducing our four-edge setup to the three-edge setup of Ref.~\cite{Kane:2003}).
The reason is that Ref.~\cite{Kane:2003} did not perform a resummation of processes of multiple tunneling events at the diluter. 
Firstly, in our work, the nonequilibrium state is manifested (in the form of the voltage or diluted currents) in all three terms of correlation functions Eq.~\eqref{eq:ud_full_correlationsA} and Eq.~\eqref{eq:ud_full_correlationsB}.
In great contrast, only the contribution of Region II (without the resummation-induced exponential factor) was taken into account in Ref.~\cite{Kane:2003}.
Secondly, even if we keep only the Region-II terms, it is the nonequilibrium current which serves as a long-time cutoff in the time integral for the tunneling current upon the resummation.
In Ref.~\cite{Kane:2003}, in great contrast, the cutoff by the nonequilibrium current did not appear, because the resummation over the train of quasiparticles generated by the diluter was not performed. In this case, the time integral for the tunneling is cut off by a finite temperature Ref.~\cite{Kane:2003} or by the transmission of the central QPC.

\subsection{Tunneling-current noise}

With the correlation functions Eqs.~\eqref{eq:ud_full_correlationsA} and \eqref{eq:ud_full_correlationsB}, we can further analyze the tunneling-current noise in the four-edge setup.
Similar to the evaluation of the tunneling current, here we also include the exponential factor, $\exp ( -2w_c^2 |t|/v )$, obtained after exactly solving the central QPC.
When multiplying the two Region-I contributions to the correlation functions Eqs.~\eqref{eq:ud_full_correlationsA} and \eqref{eq:ud_full_correlationsB}, the integration over time in Eq.~\eqref{eq:tunneling_current_noise_4} yields:
\begin{equation}
    \big\langle \delta I_T^2 \big\rangle^\text{I-I}= \frac{e^2 w_c^2}{4\pi} \int_{-\infty}^{\infty} dt \frac{e^{-\left(4\frac{I_u + I_d}{e} + \frac{2w_c^2}{v}\right)|t|}}{l_c + i v t} = \frac{e^2 w_c^2}{4v},
    \label{eq:st_leading}
\end{equation}
Note that this result does not depend on the inverse cutoff $-[4(I_u + I_d )/e + 2w_c^2 /v ]$.
Crucially, the Region-I contribution to the tunneling-current noise, Eq.~\eqref{eq:st_leading}, is induced by the time-domain braiding. 
Now, we can combine the contributions of Region I and Region II, which leads to 
\begin{align}
    & \big\langle \delta I_T^2 \big\rangle^\text{I-II}  = \frac{2 w_c^2}{\pi V} \int_{-\infty}^{\infty}\!dt\,  e^{-\frac{2w_c^2}{v} |t|}  \, \frac{\cos\left(eVt/2\right)}{l_c + i v t}
     \notag
     \\
     &\quad\times \left(I_u e^{- 4 I_u |t | }
     + I_d e^{- 4 I_d |t | }\right)\,= \frac{2  w_c^2}{v} \frac{I_u + I_d}{ V},
     \label{eq:st_next_leading}
\end{align}
which is proportional to the sum of diluted currents, thus being much smaller than the Region-I result, Eq.~\eqref{eq:st_leading}, in the strongly-diluted limit.
Similar to Eq.~\eqref{eq:st_leading} for the leading-order contribution, the integral in Eq.~\eqref{eq:st_next_leading} also does not depend on the inverse cutoffs, $eV/2$ or $-[4(I_u + I_d )/e + 2w_c^2 /v ]$.

When combining the Region-I and Region-III contributions, the integral vanishes because of an extra minus sign between particle-like and hole-like correlations, Eqs.~\eqref{eq:ud_full_correlationsA} and \eqref{eq:ud_full_correlationsB}.
Equation~\eqref{eq:st_next_leading} gives then the only finite correction that is linear in $I_u/V$ or $I_d/V$.
Indeed, when combining Regions II and III, two Region II terms, or two Region III terms, the results are quadratic in diluted currents and thus negligible in the strongly diluted limit. 
Adding up Eqs.~\eqref{eq:st_leading} and \eqref{eq:st_next_leading}, we obtain the expression for the tunneling-current noise in a compact form,
\begin{equation}
    \big\langle \delta I_T^2 \big\rangle \approx \frac{e^2 w_c^2}{4v}
    \left(1+ 8\, \frac{I_u + I_d}{ e^2 V}\right),
    \label{eq:st_four_arm}
\end{equation}
which is exact with respect to $w_c^2$ (tunneling through the central QPC) at this order of expansion in $I_u/e^2V$ and $I_d/e^2V$ (diluters' transmissions).

In fact, to the lowest order in $w_c^2$ but at higher order in diluted currents, the tunneling-current noise suffers from the logarithmic divergence of the time integral, when the two Region-II terms are multiplied [last terms in the curly brackets in Eqs.~\eqref{eq:ud_full_correlationsA} and \eqref{eq:ud_full_correlationsB}], yielding a contribution to the noise proportional to $I_u I_d/(e^2 V)^2$.
However, the exact solution for the central QPC, which introduces the exponential decay of the integrand in Eq.~\eqref{eq:tunneling_current_noise_4}, cuts off this divergence. At finite temperatures, the time integral is additionally bounded by the inverse temperature. Thus, this contribution can be safely ignored in the weak-tunneling limit. Its smallness is further accentuated by the removal of zero-bias noise from the measured data---a conventional procedure in real experiments.

\subsection{Generalized Fano factor (cf. Ref.~\cite{Rosenow:2016})}
\label{sec:GenFano}

The relevant result here is that, in comparison to the tunneling current $\langle I_T \rangle$, Eq.~\eqref{eq:it_four_arm}, which is proportional to the difference of nonequilibrium currents $I_u - I_d$, the tunneling-current noise $\langle \delta I_T^2 \rangle$, Eq.~\eqref{eq:st_four_arm}, contains a constant term, $e^2 w_c^2/4v$.
As discussed after Eq.~\eqref{eq:st_leading}, this constant term comes from the Region-I terms in the correlation functions (associated with time-domain braiding), which do not contribute to the tunneling current $I_T$ at $\nu=1/2$. As a consequence, in the diluted limit, $I_u, I_d \ll e^2 V$, the tunneling current becomes much smaller than the corresponding noise.
This leads to the divergence of the generalized Fano factor (defined in Ref.~\cite{Rosenow:2016}; here, $I_\pm=I_u\pm I_d$), 
\begin{equation}
    P\equiv \frac{-\langle \delta I_T^2\rangle + \nu e \left( I_u \frac{\partial}{\partial I_u} - 
I_d \frac{\partial}{\partial I_d} \right) \langle I_T\rangle}{\nu e\, I_+ \frac{\partial}{\partial I_- } \langle I_T \rangle },
\label{eq:generalized_fano}
\end{equation}
for $\nu = 1/2$, when only Region-I processes (time-domain braiding) are taken into account. Specifically, the calculation of Ref.~\cite{Rosenow:2016} involved only the contributions from Region I, with which the denominator of the generalized Fano factor exactly vanished at $\nu = 1/2$, leading to a divergent $P$. This denominator becomes finite, yet much smaller than the numerator determined by the noise, after including Regions II and III into consideration.

Our analysis yields a finite result for $P$. Using Eq.~\eqref{eq:it_four_arm} for $\langle I_T \rangle$ and Eq.~\eqref{eq:st_four_arm} for $\langle \delta I_T^2 \rangle$ at $\nu=1/2$ in the diluted limit, $I_u, I_d \ll e^2 V$, we obtain the following expressions for the numerator,
\begin{align*}
& -\langle \delta I_T^2\rangle + \frac{e}{2} \left( I_u \frac{\partial}{\partial I_u} - 
I_d \frac{\partial}{\partial I_d} \right) \langle I_T\rangle \\
&\ \simeq -\frac{e^2 w_c^2}{4v}
    \left(1+ 8\, \frac{I_u + I_d}{ e^2 V}\right) -\frac{4w_c^2}{\pi v V}  \\
&\ \times \left[\! \frac{(I_u - I_d)^2}{I_u \!+\! I_d \!+\! w_c^2 e/2 v} \!-\! (I_u + I_d) \ln\left( \frac{e^2V}{ I_u \!+\! I_d \!+\!e w_c^2/2v} \right) \! \right],
\end{align*}
and the denominator,
\begin{align*}
    &\frac{e}{2}\, I_+ \frac{\partial}{\partial I_- } \langle I_T\rangle  \simeq \frac{4w_c^2}{\pi v  V} (I_u + I_d) \ln\left( \frac{e^2V}{ I_u + I_d +e w_c^2/2v} \right),
\end{align*}
of the generalized Fano factor \eqref{eq:generalized_fano}.
After keeping only the leading-order contributions in both the numerator and denominator, the generalized Fano factor, Eq.~\eqref{eq:generalized_fano}, takes a finite (yet large) value: 
\begin{equation}
    P \simeq -\frac{\pi e^2}{16} \frac{V}{I_u + I_d} \left[\ln\left(\!\frac{e^2V }{I_u + I_d + e w_c^2/2v}\right)\right]^{-1}\gg 1. 
\end{equation}

\section{Discussion and Summary}
\label{sec:summary}

In this work, we have studied anyonic quasiparticles with charge $e/2$ described by the $\nu=1/2$ chiral Luttinger liquid in a Hong–Ou–Mandel-like interferometer.
We have computed the tunneling current and the noise in two different devices realized with one-dimensional edge states hosting anyons described by the chiral $\nu=1/2$ Luttinger liquid.
Specifically, we have analyzed two geometries: (i) a two-edge-channel setup where anyons originate from equilibrium reservoirs (Fig.~\ref{2-edge}) and (ii) a four-edge-channel setup where nonequilibrium anyons arrive at the collider as diluted beams (Fig.~\ref{4-edge}).

The two-edge model can be refermionized for $\nu=1/2$ anyons, leading to a quadratic Hamiltonian, whose eigenmodes can be readily obtained.
This has allowed us to derive exact expressions for the tunneling current and noise, for all relevant parameters.
Interestingly, when the transmission is weak, the tunneling current is approximately voltage-independent, in agreement with the scaling feature predicted for $\nu = 1/2$ Luttinger liquids (see, e.g., Ref.~\cite{Fendley:1996}).

The four-edge model has been analyzed in the weak-tunneling limit.
We have shown that ``time-domain braiding'' contributions (that correspond to Region I defined in  Sec.~\ref{four-edge-setup}) provide a \textit{vanishing} net tunneling current, even for non-equal nonequilibrium currents, i.e., $I_u \neq I_d$.
The cancellation of the tunneling current is a consequence of the particle-hole symmetry featured by  $\nu = 1/2$ quasiparticles when evaluating their exchange phase. 
Noteworthily, the vanishing of the tunneling current induced by the time-domain braiding leads to the divergence of the generalized Fano factor introduced in Ref.~[\onlinecite{Rosenow:2016}], to capture the anyonic fingerprint.

However, the vanishing of tunneling current for nonequal values of $I_u$ and $I_d$ is unphysical. Indeed, if $I_u > I_d$, one would intuitively anticipate a net tunneling current from channel $u$ to channel $d$.
To resolve this issue, we include processes where nonequilibrium anyons directly tunnel at the central QPC.
These processes, neglected in Refs.~\cite{Rosenow:2016,Lee:2022,Schiller:2023} for other filling factors, stand out at $\nu = 1/2$.
When considering the direct tunneling of nonequilibrium anyons, the rate of their tunneling from channel $u$ to channel $d$ (or from $d$ to $u$) is proportional to the particle density in $u$ (or $d$), and thus the nonequilibrium current $I_u$ (or $I_d$).
The inclusion of nonequilibrium anyon tunneling thus produces a nonequilibrium tunneling current proportional to $I_u - I_d$, to leading order in the diluter transmissions [see Eq.~\eqref{eq:it_four_arm}].

This result, in combination with the vanishing of time-domain braiding contribution to the tunneling current, indicates that for $\nu=1/2$ anyonic systems (in contrast to systems with Laughlin filling factors), it is the direct tunneling of anyons, rather than time-domain braiding, which becomes the dominant process when evaluating the tunneling current, no matter how strong currents are diluted.
In contrast to the tunneling current, its noise receives major contributions from the time-domain-braiding processes. The difference between the sources of leading contributions to the tunneling current and noise is the key reason that induces the divergence of generalized Fano factor of Ref.~[\onlinecite{Rosenow:2016}] for $\nu = 1/2$ chiral Luttinger liquids.

In this work, we have focused on the situation where all three QPCs are in the weak-tunneling limit.
It is then interesting to ask what happens when the system is away from this limit.
What is the role of the ``connected'' contributions to the correlation functions, which are small under the assumption of a strongly diluted limit, when one or some of three QPCs (especially one or two diluters) are away from the weak-tunneling limit? Does the distinction between the ``disconnected'' and ``connected'' processes remain meaningful when reducing the degree of dilution of nonequilibrium anyonic beams? Is the situation for $\nu=1/2$ different in this regard from that for Laughlin filling fractions? 
Corrections to the results obtained in the weak-tunneling limit were recently studied in Refs.~\cite{LeeNature23, ThammBerndPRL24, IyerX2023}.
More specifically, Ref.~\cite{LeeNature23} discussed the crossover between two regimes: (i) The \textit{extremely} diluted regime, where nonequilibrium anyons can be considered as uncorrelated entities governed by the Poissonian statistics (cf. Ref.~\cite{Rosenow:2016}) and (ii) a \textit{moderately} diluted regime, where nonequilibrium anyons follow instead the binomial distribution. As another example, Refs.~\cite{ThammBerndPRL24, IyerX2023} investigated the influence of a ``finite width'' of nonequilibrium anyons on their braiding with anyon-hole pairs described by the tunneling operators at the central QPC.
Corrections discussed in these references can, in principle, be understood with the Green's function method
by systematically analyzing the higher-order terms in the perturbative expansions in powers of the diluter transmissions.

As another option to deal with the four-edge structure, one can try to solve the scattering problem for the two diluters exactly, while treating the central QPC perturbatively. 
By doing so, one (i) would deal with only one QPC 
perturbatively and (ii) could explicitly derive the range of applicability of resummation over higher-order tunneling events at both diluters, thus shedding light on the general-$\nu$ case.
This option was, however, not taken by relevant references, e.g., Ref.~\cite{Kane:2003} (dealing with one diluter and the central QPC).
To understand the complications associated with this approach, we remind the reader that by solving the central QPC (collider) exactly, one can express the operators appearing downstream of the collider with those upstream of it.
After that, evaluation of the correlation function can be carried out by neglecting the central QPC by simply including $w_c$ at proper places [e.g., by introducing the $\exp (-2 w_c^2|t|/v)$ factor in Eq.~\eqref{eq:current_exact_central}].
As a consequence, after exactly solving the central QPC and performing an approximate resummation of the series for the diluter(s), the correlation functions are simply given by products of correlation functions of different \textit{independent} anyonic channels.
This is, however, not the case if we solve two diluters exactly. Since the central QPC is placed downstream of the central QPC, expressing operators downstream of the diluter through those upstream of it does not simplify the evaluation of correlation functions of operators at the central QPC.
In addition, by solving diluters exactly, one only knows correlation functions of the refermionizable bosonic vertexes, e.g., $\exp[\pm i (\phi_u - \phi_{su})/\sqrt{2} ]$ and $\exp[\pm i (\phi_u + \phi_{su})/\sqrt{2} ]$. Correlation functions of operators encountered at the central QPC, e.g., $\exp[\pm i (\phi_u - \phi_d)/\sqrt{2}]$, remain unknown.
Consequently, when trying to solve diluters exactly, other methods, e.g., Bethe Ansatz, are required instead of refermionization for obtaining the correlation functions at the central QPC.

Treating the central QPC exactly leads to the exponential factor, $\exp (-2w_c^2 |t|/v)$, when evaluating, e.g., the tunneling-current correlations,  Eq.~\eqref{eq:current_exact_central}.
It is interesting to note that this factor has the same form as those obtained at two diluters -- after resummation over corresponding higher-order tunneling processes [$\exp (-2w_u^2 |t|/v)$ and $\exp (-2w_d^2 |t|/v)$, see the first exponential term of Eqs.~\eqref{eq:ud_full_correlationsA} and \eqref{eq:ud_full_correlationsB}].
Based on this fact, it would be interesting to check whether this factor indeed originates from an (approximate) resummation over higher-order tunnelings through the central QPC, between two $\nu=1/2$ channels.
If yes, can this conclusion be extended to systems with general values of $\nu$?

To summarize, our work has studied the tunneling current and its noise, for both two-edge and four-edge structures in chiral Luttinger liquids at $\nu=1/2$. For the two-edge structure, we have provided exact analytical expressions for both the tunneling current and tunneling-current noise, as well as an exact general current-noise relation between them. For the four-edge structure, our work has taken into consideration the direct tunneling and collisions of nonequilibrium anyons.
Importantly, inclusion of these processes resolves the problem of vanishing of the tunneling current, which was the reason for the divergence of the generalized Fano factor~\cite{Rosenow:2016} in the $\nu=1/2$ case.
The obtained results suggest that the effect of direct tunneling and collisions of diluted anyons in the Hong-Ou-Mandel interferometer can be important for various observables in physical quantum-Hall edges also at Laughlin filling fractions. 

\section*{Acknowledgments}

We are grateful to Y. Gefen, A. Mirlin, J. Park, B. Rosenow, and K. Snizhko for discussions. We are particularly thankful to I. Safi for the comments on the manuscript, especially concerning Eq.~\eqref{eq:diss_fluct_half}. I.G. acknowledges the support by the Deutsche Forschungsgemeinschaft (DFG) through grant No. MI\,$658/10$-$2$. G.Z. acknowledges the support from National Natural Science
Foundation of China (Grant No.~12374158) and Innovation Program for Quantum Science and Technology
(Grant No.~2021ZD0302400).

\appendix

\begin{widetext}

\section{Integrals for tunneling current and noise in the two-edge structure}
\label{app:tunneling current and noise}

In this Appendix, we review the calculations of the relevant integrals required to derive the results of Sec.~\ref{two-edge-setup}. To begin with, we note the following identity \cite{SelaAffleckPRB09}:
\begin{equation}
    \int_{-\infty}^\infty\!\!\!\! d\omega\, 
    \frac{f_p (\omega - eV) - f_p (\omega + e V)}{-i\omega+\Gamma} 
        \!=\!  
        \text{Im}\, 
        \psi\!\left( \frac{1}{2} + \frac{\Gamma + i e V}{2\pi T}\! \right),
\end{equation}
from which it is straightforward to obtain Eq.~\eqref{tunc.6}, after replacing $\Gamma \to 2w^2/v$.
To compute the integral in Eq.~\eqref{eq:sud}, we define the function
\begin{equation}
\begin{aligned}
    &g_1(\beta,V,\Gamma)\equiv \int_0^\infty \frac{d\epsilon}{\epsilon^2 + \Gamma^2} \, \frac{1}{\cosh (\beta\epsilon) + \cosh (\beta e V/2)} = \frac{1}{\Gamma \sinh \left( 
\beta e V/2 \right)}\, \text{Im}\, \psi\left( \frac{1}{2} + \frac{\Gamma \beta}{2\pi} + \frac{i e V \beta}{4\pi} \right) .
\end{aligned}
\end{equation}
We further derive another integral, $g_2$, by differentiating $g_1$ over $eV$:
\begin{equation}
\begin{aligned}
 g_2(\beta,V,\Gamma) & \equiv  \int_0^\infty \frac{d\epsilon}{\epsilon^2 + \Gamma^2} \frac{1}{[\cosh (\beta\epsilon) + \cosh (\beta e V/2)]^2}
 =-\frac{2}{\beta  \sinh (\beta e V/2) }\, \frac{\partial g_1}{\partial (eV)}\\
 & = -\frac{1}{\Gamma \sinh^2(\beta eV/2)} \left[\frac{1}{2\pi}  \text{Re}\, \psi' \left( \frac{1}{2} + \frac{\Gamma \beta}{2\pi} + \frac{i e V \beta}{4\pi} \right) -  \coth (\beta e V/2) \, \text{Im}\, \psi \left( \frac{1}{2} + \frac{\Gamma \beta}{2\pi} + \frac{i e V \beta}{4\pi} \right)\right].
     \label{eq:g2}
\end{aligned}
\end{equation}

Finally, we evaluate the third relevant integral, $g_3$, by taking a derivative of $g_2$ with respect to $\Gamma$:
\begin{equation}
\begin{aligned}
g_3(\beta,V,\Gamma) & \equiv  \int_0^\infty d\epsilon \frac{1}{(\epsilon^2 + \Gamma^2)^2} \frac{1}{[\cosh (\beta\epsilon) + \cosh (\beta e V/2)]^2}=-\frac{1}{2\Gamma}\, \frac{\partial}{\partial \Gamma} g_2\\
    &= \frac{1}{2\Gamma^3 \sinh^2 (\beta e V /2 )} 
    \left\{ 
   \frac{\beta \Gamma}{2\pi}\, \left[\frac{1}{2\pi}\, \text{Re}\, \psi'' \left( \frac{1}{2} + \frac{\Gamma \beta}{2\pi} + \frac{i e V \beta}{4\pi} \right)  - \coth (\beta e V/ 2)\, \text{Im}\, \psi' \left( \frac{1}{2} + \frac{\Gamma \beta}{2\pi} + \frac{i e V \beta}{4\pi} \right)  \right]  \right. \\
    & \left. - \left[  \frac{1}{2\pi}\, \text{Re}\, \psi'\left( \frac{1}{2} + \frac{\Gamma \beta}{2\pi} + \frac{i e V \beta}{4\pi} \right) - \coth (\beta e V /2) \, \text{Im}\, \psi \left( \frac{1}{2} + \frac{\Gamma \beta}{2\pi} + \frac{i e V \beta}{4\pi} \right)  \right] \right\}.
    \label{eq:g3}
\end{aligned}
\end{equation}
With Eqs.~\eqref{eq:g2} and \eqref{eq:g3}, we are ready to compute the integral in Eq.~\eqref{eq:sud}.
Indeed, Eq.~\eqref{eq:sud} can be reorganized into
\begin{equation}
\begin{aligned}
    S(0) = & \frac{e^2 \Gamma^2}{4\pi}\int_{0}^{\infty} d\epsilon \frac{ \epsilon^2}{(\Gamma^2+\epsilon^2)^2}
    \frac{1-\cosh(\beta e V)}{\left[\cosh(\beta eV/2)+\cosh(\beta\epsilon)\right]^2}\\
    = & \frac{e^2 \Gamma^2}{4\pi} [1-\cosh(\beta e V)] \int_{0}^{\infty} d\epsilon \frac{ \epsilon^2 + \Gamma^2 - \Gamma^2}{(\Gamma^2+\epsilon^2)^2} \frac{1}{\left[\cosh(\beta eV/2)+\cosh(\beta\epsilon)\right]^2}\\
    = & \frac{e^2 \Gamma^2}{4\pi} [1 - \cosh(\beta e V)] \left[ g_2 (\beta ,V,\Gamma ) - \Gamma^2 g_3 (\beta ,V,\Gamma )\right].
\end{aligned}
\label{eq:s0_general}
\end{equation}
Equation~\eqref{eq:s0_general} leads to Eq.~\eqref{eq:s0_expression} of the main text, after substituting Eqs.~\eqref{eq:g2} and \eqref{eq:g3} into Eq.~\eqref{eq:s0_general} and taking $\Gamma=2w^2/v$.

\section{Tunneling-current noise in the two-edge structure}
\label{sec:tunneling_current_noise}

In this Appendix, we provide details of the derivation of the tunneling-current noise, which is used in the relation between tunneling current, tunneling-current noise, and cross-correlation noise reported in Sec.~\ref{sec:fd_relation}.
We start with Eq.~\eqref{tunc.3} for the tunneling current operator, to obtain the tunneling-current noise,
\begin{equation}
\begin{aligned}
    S_T  =&\frac{e^2 v^2}{4} \int d (t_1 - t_2) \Big[  \big\langle \psi^\dagger_\sigma (x,t_1) \psi_\sigma (x,t_1) \psi^\dagger_\sigma (x,t_2) \psi_\sigma (x,t_2) \big\rangle - \big\langle \psi^\dagger_\sigma (x,t_1) \psi_\sigma (x,t_1) \big\rangle\big\langle \psi^\dagger_\sigma (x,t_2) \psi_\sigma (x,t_2) \big\rangle  \\
    +  & \big\langle \psi^\dagger_\sigma (-x,t_1) \psi_\sigma (-x,t_1) \psi^\dagger_\sigma (-x,t_2) \psi_\sigma (-x,t_2) \big\rangle - \big\langle \psi^\dagger_\sigma (-x,t_1) \psi_\sigma (-x,t_1) \big\rangle\big\langle \psi^\dagger_\sigma (-x,t_2) \psi_\sigma (-x,t_2) \big\rangle\\
    -  & \big\langle \psi^\dagger_\sigma (x,t_1) \psi_\sigma (x,t_1) \psi^\dagger_\sigma (-x,t_2) \psi_\sigma (-x,t_2) \big\rangle + \big\langle \psi^\dagger_\sigma (x,t_1) \psi_\sigma (x,t_1) \big\rangle\big\langle \psi^\dagger_\sigma (-x,t_2) \psi_\sigma (-x,t_2) \big\rangle\\
    -  &  \big\langle \psi^\dagger_\sigma (-x,t_1) \psi_\sigma (-x,t_1) \psi^\dagger_\sigma (x,t_2) \psi_\sigma (x,t_2) \big\rangle + \big\langle \psi^\dagger_\sigma (-x,t_1) \psi_\sigma (-x,t_1) \big\rangle\big\langle \psi^\dagger_\sigma (x,t_2) \psi_\sigma (x,t_2) \big\rangle \Big],
\end{aligned}
\label{eq:st_start}
\end{equation}
where $x > 0$ is a position after the central QPC. Each line in Eq.~\eqref{eq:st_start} can be expressed in terms of scattering-wave amplitudes, Eqs.~\eqref{ef.6} and \eqref{ef.7}, as well as particle and hole distribution functions, Eq.~\eqref{tunc.4} 
\begin{align}
    &\int d (t_1 - t_2) \left[ \big\langle \psi^\dagger_\sigma (x_1,t_1) \psi_\sigma (x_1,t_1) \psi^\dagger_\sigma (x_2,t_2) \psi_\sigma (x_2,t_2) \big\rangle
    - \big\langle \psi^\dagger_\sigma (x_1,t_1) \psi_\sigma (x_1,t_1) \big\rangle\big\langle \psi^\dagger_\sigma (x_2,t_2) \psi_\sigma (x_2,t_2) \big\rangle \right]
    \notag
    \\
    &=  \int d(t_1 - t_2)  
    \sum_{\epsilon_1 > 0,\epsilon_2 > 0} \sum_{\lambda_1,\lambda_2 = p,h}     
    \Bigg( e^{i(\epsilon_1 - \epsilon_2) (t_1 - t_2)} 
    f_{\lambda_1} (\epsilon_1) [ 1 - f_{\lambda_2} (\epsilon_2) ]
    \notag\\
    &\qquad\qquad \times \Big\{
    \left[ \mathcal{P}^{\lambda_1}_{\epsilon_1} (x_1) \right]^* \mathcal{P}^{\lambda_1}_{\epsilon_1} (x_2) \mathcal{P}^{\lambda_2}_{\epsilon_2} (x_1) \left[ \mathcal{P}^{\lambda_2}_{\epsilon_2} (x_2) \right]^*
   -      \left[ \mathcal{P}^{\lambda_1}_{\epsilon_1} (x_1) \right]^* \mathcal{H}^{\lambda_1}_{\epsilon_1} (x_2) \mathcal{P}^{\lambda_2}_{\epsilon_2} (x_1) \left[ \mathcal{H}^{\lambda_2}_{\epsilon_2} (x_2) \right]^* 
     \Big\}
    \notag\\
    & \qquad\qquad +  e^{-i (\epsilon_1 - \epsilon_2) (t_1 - t_2)} 
    [ 1 - f_{\lambda_1} (\epsilon_1) ]f_{\lambda_2} (\epsilon_2)
     \notag\\
    &
    \qquad\qquad \times \Big\{\mathcal{H}^{\lambda_1}_{\epsilon_1} (x_1) \left[ \mathcal{H}^{\lambda_1}_{\epsilon_1} (x_2) \right]^* \left[ \mathcal{H}^{\lambda_2}_{\epsilon_2} (x_1) \right]^* \mathcal{H}^{\lambda_2}_{\epsilon_2} (x_2) 
    -       \mathcal{H}^{\lambda_1}_{\epsilon_1} (x_1) \left[ \mathcal{P}^{\lambda_1}_{\epsilon_1} (x_2) \right]^* \left[ \mathcal{H}^{\lambda_2}_{\epsilon_2} (x_1) \right]^* \mathcal{P}^{\lambda_2}_{\epsilon_2} (x_2)
      \Big\}
      \notag
      \\
     &\qquad\qquad+ e^{i (\epsilon_1 + \epsilon_2) (t_1 - t_2)}
      f_{\lambda_1} (\epsilon_1) f_{\lambda_2} (\epsilon_2)
      \notag\\
    &
    \qquad\qquad  \times
    \left\{ \left[ \mathcal{P}^{\lambda_1}_{\epsilon_1} (x_1) \right]^* \mathcal{P}^{\lambda_1}_{\epsilon_1} (x_2) \left[ \mathcal{H}^{\lambda_2}_{\epsilon_2} (x_1) \right]^* \mathcal{H}^{\lambda_2}_{\epsilon_2} (x_2) \!-\! \left[ \mathcal{P}^{\lambda_1}_{\epsilon_1} (x_1) \right]^* \mathcal{H}^{\lambda_1}_{\epsilon_1} (x_2) \left[ \mathcal{H}^{\lambda_2}_{\epsilon_2} (x_1) \right]^* \mathcal{P}^{\lambda_2}_{\epsilon_2} (x_2) \right\}\!
     \notag
     \\
     &\qquad\qquad+  e^{-i (\epsilon_1 + \epsilon_2) (t_1 - t_2)} 
     [1-f_{\lambda_1} (\epsilon_1) ][1-f_{\lambda_2} (\epsilon_2)]
     \notag\\
    &
    \qquad\qquad \times\left\{ \mathcal{H}^{\lambda_1}_{\epsilon_1} (x_1) \left[ \mathcal{H}^{\lambda_1}_{\epsilon_1} (x_2) \right]^* \mathcal{P}^{\lambda_2}_{\epsilon_2} (x_1) \left[ \mathcal{P}^{\lambda_2}_{\epsilon_2} (x_2) \right]^*  - \mathcal{H}^{\lambda_1}_{\epsilon_1} (x_1) \left[ \mathcal{P}^{\lambda_1}_{\epsilon_1} (x_2) \right]^* \mathcal{P}^{\lambda_2}_{\epsilon_2} (x_1) \left[ \mathcal{H}^{\lambda_2}_{\epsilon_2} (x_2) \right]^* \right\} 
      \Bigg).
     \label{eq:sigma_correlations}
\end{align}

Different choices of $x_1$ and $x_2$ in Eq.~\eqref{eq:sigma_correlations} correspond to different terms of Eq.~\eqref{eq:st_start}.
The integral over time $t_1 - t_2$ in Eq.~\eqref{eq:sigma_correlations} produces delta-functions that describe energy conservation,
greatly simplifying further evaluation.
To evaluate the first line of Eq.~\eqref{eq:st_start}, we choose $x_1=x_2=x>0$ in Eq.~\eqref{eq:sigma_correlations}, leading to
\begin{equation}
\begin{aligned}
     & \int d (t_1 - t_2)  \Big[  \big\langle \psi^\dagger_\sigma (x,t_1) \psi_\sigma (x,t_1) \psi^\dagger_\sigma (x,t_2) \psi_\sigma (x,t_2) \big\rangle  - \big\langle \psi^\dagger_\sigma (x,t_1) \psi_\sigma (x,t_1) \big\rangle\big\langle \psi^\dagger_\sigma (x,t_2) \psi_\sigma (x,t_2) \big\rangle \Big]
     \\
     & = \int d\epsilon \Big( \left[  (|t_\epsilon|^4 + |r_\epsilon|^4) - 2 |t_\epsilon|^2 |r_\epsilon|^2\right] f_p (\epsilon) [1 - f_p (\epsilon)] 
     + \left[ (|t_\epsilon|^4 + |r_\epsilon|^4)- 2 |t_\epsilon|^2 |r_\epsilon|^2 \right] f_h (\epsilon) [1 \!-\! f_h (\epsilon)] 
          \\
     &\qquad +  2 |t_\epsilon|^2 |r_\epsilon|^2 \left\{ f_p(\epsilon) f_p (-\epsilon) + [1 - f_p(\epsilon)] [1 - f_p (-\epsilon) ]\right\}
     + 2 |t_\epsilon|^2 |r_\epsilon|^2 \left\{ f_h(\epsilon) f_h (-\epsilon) + [1 - f_h(\epsilon)] [1 - f_h (-\epsilon) ]\right\}
     \\
     &\qquad + 2 |t_\epsilon|^2 |r_\epsilon|^2\left\{ f_p (\epsilon) [1 - f_h (\epsilon)] + f_h (\epsilon) [1 - f_p (\epsilon)]  \right\}
     + 2 |t_\epsilon|^2 |r_\epsilon|^2\left\{ f_p (\epsilon) [1 - f_h (\epsilon)] + f_h (\epsilon) [1 - f_p (\epsilon)]  \right\}
          \\
     &\qquad + (|r_\epsilon|^4 - |t_\epsilon|^2 |r_\epsilon|^2)[1- f_p(\epsilon)] [1- f_h (-\epsilon)] 
      + (|t_\epsilon|^4 - |t_\epsilon|^2 |r_\epsilon|^2)[1- f_h(\epsilon)] [1- f_p (-\epsilon)] 
          \\
     &\qquad + (|r_\epsilon|^4 - |t_\epsilon|^2 |r_\epsilon|^2) f_h(\epsilon) f_p (-\epsilon)+ (|t_\epsilon|^4 - |t_\epsilon|^2 |r_\epsilon|^2) f_p(\epsilon) f_h (-\epsilon) \Big).
\end{aligned}
\label{eq:x1_x2_positive}
\end{equation}
Now we move to the second case, $x_1=x_2=-x < 0$, corresponding to the second line of Eq.~\eqref{eq:st_start}. In this case, Eq.~\eqref{eq:sigma_correlations} produces a much simpler expression [indeed, the terms with $\Theta(-x)$ in Eqs.~\eqref{ef.6-1} and \eqref{ef.7} are not multiplied by $t_\epsilon$ and $r_\epsilon$]:
\begin{equation}
\begin{aligned}
&\int dt \Big[ \big\langle \psi^\dagger_\sigma (-x,t) \psi_\sigma (-x,t) \psi^\dagger_\sigma (-x,0) \psi_\sigma (-x,0) \big\rangle - \big\langle \psi^\dagger_\sigma (-x,t) \psi_\sigma (-x,t) \big\rangle\big\langle \psi^\dagger_\sigma (-x,0) \psi_\sigma (-x,0) \big\rangle \Big]\\
&=  \int d\epsilon\Big\{ f_p (\epsilon) [ 1- f_p (\epsilon)] + f_h (\epsilon) [ 1- f_h (\epsilon)] + f_p(\epsilon) f_h (-\epsilon) + [1 - f_p(-\epsilon)] [1 - f_h(\epsilon)] \Big\}.
\end{aligned}
\label{eq:x1_x2_neg}
\end{equation}
Next, we consider the case $x_1=x> 0$ and $x_2=-x < 0$, corresponding to the third line of Eq.~\eqref{eq:st_start}, for which we obtain:
\begin{equation}
\begin{aligned}
    &\int dt \Big[ \big\langle \psi^\dagger_\sigma (x,t) \psi_\sigma (x,t) \psi^\dagger_\sigma (-x,0) \psi_\sigma (-x,0) \big\rangle  -  \big\langle \psi^\dagger_\sigma (x,t) \psi_\sigma (x,t) \big\rangle\big\langle \psi^\dagger_\sigma (-x,0) \psi_\sigma (-x,0) \big\rangle \Big] \\
    &= \int \!d\epsilon  (|t_\epsilon|^2 \!-\! |r_\epsilon|^2)\left\{ f_p(\epsilon) [1 \!-\! f_p(\epsilon)] \!+\! f_h(\epsilon) [1 \!-\! f_h(\epsilon)] \right\}
    \\
    &+ \int d\epsilon\Big(\left\{ t_\epsilon^* t_{-\epsilon}^* f_p(\epsilon) f_h(-\epsilon) \!-\! r_\epsilon r_{-\epsilon} [1 \!-\! f_p (\epsilon)] [1 \!-\! f_h (-\epsilon)] \right\}-\left\{r_\epsilon^* r_{-\epsilon}^* f_h(\epsilon) f_p(-\epsilon) \!-\! t_\epsilon t_{-\epsilon} [1 \!-\! f_h (\epsilon)] [1 \!-\! f_p (-\epsilon)]\right\}\Big).
\end{aligned}
\label{eq:x1_pos_x2_neg}
\end{equation}
Finally, the case $x_1=-x <0$ and $x_2=x > 0$ corresponds to the last line of Eq.~\eqref{eq:st_start}:
\begin{equation}
\begin{aligned}
    & \int dt \Big[ \big\langle \psi^\dagger_\sigma (-x,t) \psi_\sigma (-x,t) \psi^\dagger_\sigma (x,0) \psi_\sigma (x,0) \big\rangle -  \big\langle \psi^\dagger_\sigma (-x,t) \psi_\sigma (-x,t) \big\rangle\big\langle \psi^\dagger_\sigma (x,0) \psi_\sigma (x,0) \big\rangle\Big] \\
    &= \int d\epsilon (|t_\epsilon|^2 - |r_\epsilon|^2)\left\{ f_p(\epsilon) [1 - f_p(\epsilon)] + f_h(\epsilon) [1 - f_h(\epsilon)] \right\}\\
    &+ \int d\epsilon\left\{ (t_\epsilon t_{-\epsilon} - r_\epsilon r_{-\epsilon}) f_p(\epsilon) f_h(-\epsilon)+ (t_\epsilon^* t_{-\epsilon}^* - r_\epsilon^* r^*_{-\epsilon}) [1 - f_h (\epsilon)] [1 - f_p (-\epsilon)] \right\}.
\end{aligned}
\label{eq:x1_neg_x2_pos}
\end{equation}
With Eqs.~\eqref{ef.8} and \eqref{tunc.4}, we evaluate Eqs.~\eqref{eq:x1_x2_positive}-\eqref{eq:x1_neg_x2_pos}, leading to
\begin{equation}
\begin{aligned}
    &S_T  = \frac{2e^2w^4}{\pi} \left\{\!\int_0^\infty\!  d\epsilon\ \frac{ v^2\epsilon^2\,[\cosh(\beta eV) \!-\! 1]}{(4w^4 \!+\! v^2\epsilon^2)^2 \!\left[ \cosh\!\left( \frac{\beta e V}
{2} \right)\! +\! \cosh (\beta\epsilon) \right]^2}
     +\int_0^\infty\! d\epsilon \frac{2 \left[ 1 + \cosh \left(\frac{\beta eV}{2}\right) \cosh 
(\beta\epsilon)\right]}{(4w^4 + v^2\epsilon^2) \left[ 
\cosh\left( \frac{\beta e V}{2} \right) + \cosh 
(\beta\epsilon) \right]^2}\right\},
\end{aligned}
\end{equation}
which is 
Eq.~\eqref{eq:st_two_edge} of the main text.

We further go beyond the integral form of Eq.~\eqref{eq:st_two_edge}, to provide a compact analytical form of $S_T$, leading too
\begin{equation}
\begin{aligned}
    S_T & = -\frac{e^2 \Gamma^2}{2\pi} [1 - \cosh(\beta e V)] \left[ g_2 (\beta ,V,\Gamma ) - \Gamma^2 g_3 (\beta ,V,\Gamma )\right] + \frac{e^2 \Gamma}{4 \pi^2 } \text{Re}\, \psi' \!\left( \frac{1}{2} + \frac{\Gamma}{2\pi }\beta + i\frac{ e V \beta}{4\pi} \right).
\end{aligned}
\end{equation}

In the limit $\Gamma \ll 1/\beta$, the expressions for functions $g_2$ and $g_3$, Eqs.~(\ref{eq:g2}) and (\ref{eq:g3}), can be approximated by neglecting the term $\Gamma \beta/2\pi$ in the arguments of the digamma functions:
\begin{equation}
\begin{aligned}
 g_2(\beta,V,\Gamma) 
 & \approx -\frac{1}{\Gamma \sinh^2(\beta eV/2)} \left[\frac{1}{2\pi}  \text{Re}\, \psi' \left( \frac{1}{2} + i\frac{e V \beta}{4\pi} \right) -  \coth (\beta e V/2) \, \text{Im}\, \psi \left( \frac{1}{2} + i\frac{e V \beta}{4\pi} \right)\right],\\
 g_3(\beta,V,\Gamma) 
    & \approx -\frac{1}{2\Gamma^3 \sinh^2 (\beta e V /2 )}  \left[  \frac{1}{2\pi}\, \text{Re}\, \psi'\left( \frac{1}{2}  + i\frac{e V \beta}{4\pi} \right) - \coth (\beta e V /2) \, \text{Im}\, \psi \left( \frac{1}{2}  + i\frac{e V \beta}{4\pi} \right)  \right], 
\end{aligned}
\end{equation}
leading to 
\begin{equation}
\begin{aligned}
    S_T & \approx \frac{e^2 \Gamma}{2\pi}  \coth (\beta e V /2) \, \text{Im}\, \psi \left( \frac{1}{2}  + i\frac{e V \beta}{4\pi} \right) .
\end{aligned}
\end{equation}
Neglecting the term with $w$ in the expression for the tunneling current, Eq.~\eqref{tunc.6}, the Fano factor  acquires a very simple form in the limit $w^2/v\ll T$:
\begin{equation}
    \mathcal{F}_T = \frac{\coth(\beta eV/2)}{2}.
\end{equation}
Similarly, in this limit, the cross-correlation and the associated cross-correlation Fano factor become 
\begin{equation}
\begin{aligned}
    S (0) & \approx \frac{e^2 \Gamma^2}{4\pi} \left[\frac{1}{2\pi}  \text{Re}\, \psi' \left( \frac{1}{2} + i\frac{e V \beta}{4\pi} \right) -  \coth (\beta e V/2) \, \text{Im}\, \psi \left( \frac{1}{2} + i\frac{e V \beta}{4 \pi} \right)\right] 
    = -  \frac{e^2 \Gamma^2}{8} \tanh^2\left( \frac{e V\beta}{4} \right) ,\\
    \mathcal{F}_\text{CC} & \approx \tanh \left( \frac{e V\beta}{4} \right) = \frac{1}{\pi} \text{Im}\, \psi \left( \frac{1}{2}  + i\frac{ e V \beta}{4\pi} \right).
\end{aligned}
\end{equation}
The result for $ \mathcal{F}_\text{CC} $ agrees with the general expression~\cite{Snizhko22,SnizhkoX24} for the cross-correlation Fano factor in the weak-tunneling limit
(when taken at $\nu=1/2$, i.e., for the scaling dimension $1/4$ and the quasiparticle charge $e/2$).

\section{Auto-correlations}
\label{app:auto_correlation}

Equation~\eqref{eq:auto-conductance_relation} provides the exact relation between auto-correlation, tunneling current noise, conductance, and the thermal noise.
Here we prove this relation for the $\nu=1/2$ case.
Derivation of auto-correlations can be simply performed by writing current-current correlations in terms of the $\sigma$ and $\rho$ modes:
\begin{equation}
\begin{aligned}
    s_{uu} ( t_1\! -\! t_2 ; x_u , x_u )  &= \frac{1}{2} \sum_{\eta = \pm } \{ \langle {\bf T}_K  j_u ( x_u , t_1 , \eta ) j_u ( x_u , t_2 , - \eta ) \rangle
 -\langle j_u ( x_u , t_1 , \eta ) \rangle \langle  j_u ( x_u , t_2 , - \eta ) \rangle \} \\
  &= s_\rho( t_1 - t_2 ; x_u , x_d )+s_\sigma( t_1 - t_2 ; x_u , x_d )\\
 s_{dd} ( t_1\! -\! t_2 ; x_d , x_d ) &= \frac{1}{2} \sum_{\eta = \pm } \{ \langle {\bf T}_K  j_d ( x_d , t_1 , \eta ) j_d ( x_d , t_2 , - \eta ) \rangle
  -\langle j_d ( x_d , t_1 , \eta ) \rangle \langle  j_d ( x_d , t_2 , - \eta ) \rangle \}\\
 &= s_\rho( t_1 - t_2 ; x_u , x_d )+s_\sigma( t_1 - t_2 ; x_u , x_d ).
\end{aligned}
\end{equation} 
\end{widetext}
With Eqs.~\eqref{tunc.x11} and \eqref{tunc.13}, we thus obtain the following expression for the auto-correlations (assuming $x > 0$, downstream of the collider):
\begin{align}
    &S_{uu} (0;x,x)= S_{dd} (0;x,x) = 2 S_\rho (0;x,x) - S (0;x,x)\notag\\
    &\quad = \frac{e^2}{8\pi} \int_0^\infty d\epsilon \frac{1}{1 + \cosh (\beta\epsilon)} + \frac{1}{2} S_T - \frac{1}{\beta} \frac{\partial}{\partial V} \langle I_T \rangle\notag \\
&\quad = \frac{1}{2} \frac{G_0}{\beta} + \frac{1}{2} S_T - \frac{1}{\beta} \frac{\partial}{\partial V} \langle I_T \rangle,
\label{eq:suu_sdd}
\end{align}
where $G_0 \equiv e^2/4\pi$ is the conductance of a clean $\nu = 1/2$ channel.
The first term of Eq.~\eqref{eq:suu_sdd} is the thermal Nyquist-Johnson noise, which comes from the $\rho$ mode that decouples from the collider.
Equation~\eqref{eq:suu_sdd} is equivalent to Eq.~\eqref{eq:auto-conductance_relation} of the main text.

\medskip

\section{Zero-temperature results of Refs.~\cite{Fendley:1995b,Fendley:1995c} for general-$\nu$ two-edge systems}
\label{app:expansions}

As stated in the main text, at zero temperature, analytical expressions for the tunneling current and tunneling-current noise for the $\nu = 1/2$ two-edge structure can be found in Refs.~\cite{Fendley:1995b, Fendley:1995c}.
Moreover, these references also provided analytical expressions for the zero-temperature tunneling current and tunneling-current noise, for general values of $\nu$.
In contrast to the $\nu = 1/2$ case, for a system with general $\nu$, the tunneling current and its noise are not described by compact formulas, but are expressed in terms of infinite series in the regimes of small bias (for the strong-tunneling regime) or small transmission through the QPC (for the weak-tunneling regime). 
In this Appendix, we quote these formulas from Refs.~\cite{Fendley:1995b,Fendley:1995c}, rewrite them using the parameters of our model, and compare with our results. 
For later convenience, following Ref.~\cite{Fendley:1995c}, we can define an energy scale
\begin{equation}
    W \equiv \frac{4\sqrt{\pi}\, \Gamma \left( \frac{1}{2  - 2\nu} \right) }{\nu \Gamma \left( \frac{\nu}{2 - 2\nu } \right)}\, \frac{w^2}{v} ,
\end{equation}
which actually corresponds to $T_B'$ of Ref.~\cite{Fendley:1995c}.

We begin by providing expressions for the tunneling current:
\begin{equation}
\begin{aligned}
    & e V \ll W:\\
    &\tilde{I}_T  = \frac{e^2 V}{2\pi} \sum_{n=1}^\infty  \frac{(-1)^{n + 1}\sqrt{\pi} \Gamma (n/\nu) \left( \frac{eV}{W} \right)^{2n (1/\nu - 1)}}{2 \Gamma (n) \Gamma[3/2 + (1/\nu - 1)n]},  \\
    & e V \gg W:\\
    & \tilde{I}_T = \frac{e^2 V \nu}{2\pi} \left\{1-  \sum_{n=1}^\infty \frac{(-1)^{n + 1} \nu\sqrt{\pi} \Gamma (n\nu) \left( \frac{eV}{W} \right)^{2n (\nu - 1)}}{2 \Gamma(n) \Gamma [3/2 + (\nu - 1)n]}\right\},
\end{aligned}
\label{eq:it_resummations}
\end{equation}
In the equations above, $\nu < 1$ is assumed to guarantee convergence of the sums in both limits.
Note that $\tilde{I}_T \to e^2 \nu V/(2\pi)$ in the $e V \gg W$ limit, in contrast to our model, where tunneling is perfect in the opposite limit. This difference originates from the definition of the ``tunneling current'' $\tilde{I}_T$, which is different from our $I_T$. 
Indeed, in Ref.~\cite{Fendley:1995c}, the tunneling current is defined as the one continuing along the same channel after the QPC.

For $\nu = 1/2$, Eq.~\eqref{eq:it_resummations} becomes
\begin{equation}
    \tilde{I}_T = \frac{e^2 V}{4\pi} - \frac{e w^2}{\pi v}\!
\arctan \!\left( \!\frac{v e V }{4w^2}\!\right),
\end{equation}
which yields our Eq.~\eqref{eq:it_two_channel} after a shift of the perfect tunneling current, $e^2 V/(4\pi)$, and the change of sign of the correction term.
These two modifications again originate from a different definition of tunneling current in Ref.~\cite{Fendley:1995c}:
$$\langle I_T \rangle=-(\tilde{I}_T-I_0).$$

To obtain the noise, we take the zero-temperature noise-current relation of Ref.~\cite{Fendley:1995b} [equivalent to Eq.~\eqref{ST-FLS95}],
\begin{equation}
\begin{aligned}
    S_T 
    & = -\frac{\nu e}{2 (1 - \nu)} W \partial_W I_T,
\end{aligned}
\end{equation}
leading to zero-temperature tunneling-current noise, $S_T(0)$, in the form of the series:
\begin{align}
    & e V \ll W: \notag
    \\
    &S_T(0) \! =\! \frac{n e^3 V}{2\pi} \sum_{n=1}^\infty  \frac{(-1)^{n + 1}\sqrt{\pi} \Gamma (n/\nu) \left( \frac{eV}{W} \right)^{2n (1/\nu - 1)}}{2 \Gamma (n) \Gamma[3/2 + (1/\nu - 1) n ]},  
    \end{align}
    \begin{align}
    & e V \gg W:\notag
    \\
    &S_T(0)\! =\! \frac{n e^3 V \nu^2}{2\pi}  \sum_{n=1}^\infty \frac{(-1)^{n + 1} \nu\sqrt{\pi} \Gamma (n\nu) \left( \frac{eV}{W} \right)^{2n (\nu - 1)}}{2 \Gamma(n) \Gamma [3/2 +  (\nu - 1) n]}.
\end{align}
\label{eq:st_resummations}
For $\nu = 1/2$, $S_T(0)$ becomes
\begin{equation}
    S_T(0) = \! - \frac{e^2 w^2}{2\pi v} \left[ -\frac{ 4v w^2  eV}{16 w^4 + (v\, eV)^2} + \arctan \left( 
\frac{v \,e V}{4 w^2} \right) \right],
\end{equation}
which perfectly agrees with our Eq.~\eqref{eq:s0_zero_temp}.

\section{Central QPC in the four-edge setup: Exact treatment}
\label{app:exact_central}

In this Appendix, following the method of Ref.~\cite{Kane:2003}, we provide details on exactly solving the central QPC. Our starting point is the refermionized Hamiltonian, Eq.~\eqref{bos.7}, which is quadratic in fermionic operators $\psi_\sigma$ and, thus, is exactly diagonalizable. 
Here, we use the notation of Ref.~\cite{Kane:2003} for the scattering-state operators (the correspondence to the notation of Sec.~\ref{sec:referm} is obvious):
\begin{equation}
\psi_{k,\text{out}} = t_k \psi_{k,\text{in}} - r_k \psi^\dagger_{-k,\text{in}},
\label{eq:scattering_wf}
\end{equation}
with coefficients [equivalent to those given by Eq.~\eqref{ef.8}]
\begin{equation}
t_k = \frac{v^2 k}{v^2k + 2 i w_c^2}, \ \ r_k = \frac{2iw^2_c}{v^2 k + 2 i w^2_c}.
\end{equation}

In Eq.~\eqref{eq:scattering_wf}, subscripts ``in'' and ``out'' refer to operators before and after scattering off the central QPC, respectively.
In comparison to the scattering-wave operators in Ref.~\cite{Kane:2003}, Eq.~\eqref{eq:scattering_wf} contains an extra minus sign for the Andreev-scattering term (the term proportional to $r_k$), which is due to a different definition of the current operator, of our work.
The scattering operators (\ref{eq:scattering_wf}) in the $k$-space are related to the operator in the coordinate space by the following Fourier transformations:  
\begin{equation}
\begin{aligned}
&\psi^\dagger_{k,\text{in}}\equiv \frac{1}{2\mathcal{L}} \int_{-\mathcal{L}}^\mathcal{L} dx\, \psi^\dagger_{\sigma,\text{in}} (x) e^{i \pi n_k x/\mathcal{L}}, \\
&\psi^\dagger_{k,\text{out}}\equiv \frac{1}{2\mathcal{L}} \int_{-\mathcal{L}}^\mathcal{L} dx\, \psi^\dagger_{\sigma,\text{out}} (x) e^{i \pi n_k x/\mathcal{L}},\\
&\psi^\dagger_{\sigma,\text{in}}(x) \equiv \sum_{n_k} \psi^\dagger_{k,\text{in}} e^{-i \pi n_k x/\mathcal{L}}, \\
& \psi^\dagger_{\sigma,\text{out}}(x) \equiv \sum_{n_k} \psi^\dagger_{k,\text{out}} e^{-i \pi n_k x/\mathcal{L}},
\end{aligned}
\label{eq:op_fourier}
\end{equation}
with $k=2\pi n_k/\mathcal{L}$ on the finite-size system and $n_k$ being a relative integer number. 

To proceed, we choose the sign of $I_T$ to be positive, if the net current is directed from channel $u$ to channel $d$ (see Fig.~\ref{2-edge}). With this convention, the tunneling current is given by
\begin{equation}
\begin{aligned}
I_T  =& -\frac{1}{2} [ j_u (x > L) - j_u (x < L) ] \\
& + \frac{1}{2} [ j_d (x > L ) - j_d (x < L) ] \\
 = &-\frac{1}{2} [I_{\sigma,\text{out}} - I_{\sigma,\text{in}} ].
\end{aligned}
\end{equation}
Here, we have defined the ``out-going'' and ``incoming'' current operators as
\begin{equation}
\begin{aligned}
&I_{\sigma,\text{out}}  = j_u  (x>L)- j_d (x > L),\\
&I_{\sigma,\text{in}}  = j_u  (x<L)- j_d (x < L),
\end{aligned}
\end{equation}
referring to the current operators for particles leaving and approaching the QPC, respectively.

Now, we continue to express the current operator in terms of $k$-space operators, Eq.~\eqref{eq:scattering_wf}, following Eq.~\eqref{eq:op_fourier}, which leads to
\begin{equation}
\begin{aligned}
I_{\sigma,s} (x) & = ev\, \psi^\dagger_{\sigma,s} (x) \psi_{\sigma,s} (x) \\
& = ev \sum_{k_1, k_2} \psi^\dagger_{k_1,s} \psi_{k_2,s} e^{-i \pi(n_{k_1} - n_{k_2} ) x/D},
\end{aligned}
\end{equation}
for $s=\{$in,\,out$\}$.
We follow Ref.~\cite{Kane:2003} to get rid of the phase factors, by averaging over the entire space, yielding
\begin{equation}
\begin{aligned}
\bar{I}_{\sigma,s} & \equiv \frac{1}{2D}\int_{-\mathcal{L}}^\mathcal{L} dx\, I_{\sigma,s} (x) \\
& =  \frac{ev}{2\mathcal{L}}  \sum_{n_{k_1},n_{k_2}} \psi^\dagger_{k_1,s} \psi_{k_2,s} \int_{-\mathcal{L}}^\mathcal{L} dx e^{-i \pi(n_{k_1} - n_{k_2} ) x/\mathcal{L}}\\
& = ev\sum_{n_{k_1},n_{k_2}} \psi^\dagger_{k_1,s} \psi_{k_2,s} \delta_{n_{k_1},n_{k_2}} = ev \sum_{n_k} \psi_{k}^\dagger \psi_{k}.
\end{aligned}
\label{eq:momentum_summation}
\end{equation}
Note that the averaging performed in Eq.~\eqref{eq:momentum_summation}, relies on the fact that the currents and noise are independent of the position in the contact where the current operator is evaluated~\cite{Kane:2003}.

Next, we express the outgoing modes in terms of incoming ones, following Eq.~\eqref{eq:scattering_wf}:
\begin{align}
I_T & = -\frac{1}{2} [\bar{I}_{\sigma,\text{out}} - \bar{I}_{\sigma,\text{in}}] =
\frac{ ev}{2} \sum_{n_k}  \big[ \psi^\dagger_{k,\text{in}} \psi_{k,\text{in}} 
\notag
\\
&-( t_k^* \psi^\dagger_{k,\text{in}} - r^*_k \psi_{-k,\text{in}})( t_k \psi_{k,\text{in}} - r_k 
\psi^\dagger_{-k,\text{in}})\big]
\notag
\\
& = \frac{ev}{2} \sum_{n_k} \left(\psi^\dagger_{k,\text{in}} \psi_{k,\text{in}} (1 - |t_k|^2) - |r_k|^2 \psi_{-k,\text{in}}  \psi^\dagger_{-k,\text{in}} \right)
\notag
\\
& + \frac{ev}{2} \sum_{n_k} \left( t_k^* r_k \psi^\dagger_{k,\text{in}} \psi^\dagger_{-k,\text{in}} + t_k r^*_k \psi_{-k,\text{in}} \psi_{k,\text{in}} \right)
\notag
\\
&= \frac{ev}{2} \sum_{n_k} |r_k|^2 \left( \psi^\dagger_{k,\text{in}} \psi_{k,\text{in}}  -  \psi_{-k,\text{in}}  \psi^\dagger_{-k,\text{in}} \right)
\notag\\
& + \frac{ev}{2} \sum_{n_k} |t_k| |r_k| \left(  \psi^\dagger_{k,\text{in}} \psi^\dagger_{-k,\text{in}} - \psi_{-k,\text{in}} \psi_{k,\text{in}} \right).
\label{IT-referm}
\end{align}
Compared to the expression of the tunneling current in Ref.~\cite{Kane:2003}, Eq.~\ref{IT-referm} differs by the replacement $|r_k|\leftrightarrow|t_k|$, because of different definitions of the current operator.

The final expression of the tunneling current can be written as
\begin{equation}
\begin{aligned}
I =&\ \frac{e w_c^2}{4\pi l_c v} \frac{1}{2\mathcal{L}} \int_{-\mathcal{L}}^\mathcal{L} dx_1 \int_{-\mathcal{L}}^\mathcal{L} dx_2\,  e^{-\frac{2w_c^2}{v^2} |x_1 - x_2|} \\
& \times \left\langle e^{-i\phi_\sigma (x_1,t_1)} e^{i\phi_\sigma (x_2,t_2)} - e^{i\phi_\sigma (x_2,t_2)} e^{-i\phi_\sigma (x_1,t_1)} \right\rangle_\text{neq} \\
= &\ \frac{e w_c^2}{4\pi l_c }  \int_{-\infty}^\infty d t  \, e^{-\frac{2w_c^2}{v} |t|}\\
& \times \Bigg\langle \mathbf{T}_K \big[ e^{-i\phi_\sigma (L,t)} e^{i\phi_\sigma (L,0)} - e^{i\phi_\sigma (L,0)} e^{-i\phi_\sigma (L,t)}\big]\\
& \times \exp\left\{-i\int_K d\tau \left[ H^T_\mathrm{su\mhyphen u} + H^T_\mathrm{d\mhyphen sd} \right] (\tau) \right\}\Bigg\rangle_0.
\end{aligned}
\label{eq:exact_central}
\end{equation}
Based on Eq.~\eqref{eq:exact_central}, after dealing with the central QPC exactly, one can still address each subsystem separately. In comparison to the leading-order expansion, Eq.~\eqref{keldysh-current2}, exactly solving the central QPC brings in an extra factor, $\exp(-2w_c^2|t|/v)$ (which comes from the Fourier transform of $|r_k|^2$), after including higher-order tunneling processes at the central QPC.
This fact, in combination with exponential terms in Eqs.~\eqref{eq:ud_full_correlationsA} and \eqref{eq:ud_full_correlationsB}, indicates that when performing the integration over $t$, one encounters terms of the form $\propto \exp[-2(w_u^2 + w_d^2 + w_c^2)|t|/v]$, meaning that squares of tunneling amplitudes at each QPC sum up to bound integral over time.

In the previous literature devoted to HOM colliders with diluted beams (see, e.g., Refs.~\cite{Rosenow:2016,
Lee:2022,
Morel:2022,
Schiller:2023}), resummation over tunneling through the central QPC was not considered (since for $\nu\neq 1/2$ exact calculation is not feasible).
Before ending this section, we establish the limits of validity for such a consideration that neglects higher-order tunneling processes at the central QPC.
When all three QPCs are in the weak-tunneling limit (assumptions taken by Refs.~\cite{Rosenow:2016, Lee:2022, Morel:2022, Schiller:2023}), we can write down the currents through two diluters and the central QPC for a general value of $\nu$ as 
\begin{equation}
\begin{aligned}
I_u & = \nu \frac{e^2}{2\pi} \mathcal{T}_u V_{su}\,   ( \nu e V_{su} \tau_c)^{2 \nu - 2} ,\\
I_d & = \nu \frac{e^2}{2\pi} \mathcal{T}_d  V_{sd}\,  ( \nu e V_{sd} \tau_c)^{2 \nu - 2},\\
I_T & = \mathcal{T}_c I_-\,  ( I_+ \tau_c/e)^{2 \nu - 2} ,
\label{noneq-currents}
\end{aligned}
\end{equation}
where $I_\pm \equiv I_u \pm I_d$ and $\tau_c=l_c/v$.
Here we keep $I_u$ and $I_d$ to the leading order in $\mathcal{T}_u$ and $\mathcal{T}_d$, respectively, and keep $I_T$ to the leading order of $\mathcal{T}_c$.
Actually, Eq.~\eqref{noneq-currents} defines the experimentally measurable transmission probabilities $\tilde{\mathcal T}_u \equiv \mathcal{T}_u  ( \nu e V_{su} \tau_c)^{2 \nu - 2}$, $\tilde{\mathcal T}_d \equiv \mathcal{T}_d  ( \nu e V_{sd} \tau_c)^{2 \nu - 2}$ for diluters, and $\tilde{\mathcal T}_c \equiv ( I_+ \tau_c/e)^{2 \nu - 2}$ for the central QPC.
Importantly, the Luttinger renormalization of the transmission through the central QPC is cut off by the total non-equilibrium current $I_+$ rather than by the voltages $V_{su}$ and $V_{sd}$. This effect was not taken into account in Ref.~\cite{Kane:2003}, where no resummation of higher-order diluter contributions to the nonequilibrium currents was performed. 
The cutoff by the nonequilibrium current $I_+$ in the present calculation is akin to the cutoff introduced by the ``nonequilibrium dephasing'' in Luttinger liquids out of equilibrium~\cite{Gutman2008, Gutman2010}.

Bare transmission probabilities are given by
$$\mathcal{T}_u \equiv a (\nu) w_u^2\tau_c/v, \quad \mathcal{T}_d \equiv a (\nu) w_d^2\tau_c/v $$ at two diluters, and 
$$\mathcal{T}_c \equiv b (\nu) w_c^2\tau_c/v, $$ 
where 
\begin{equation}
\begin{aligned}
    a (\nu) & \equiv \frac{\sin(2\pi\nu) \Gamma (1-2\nu)}{4\pi^3},\\
   b (\nu) & \equiv \frac{2}{\pi} \left[ \frac{2\pi\sin^2 (\pi\nu)}{\nu }\right]^{2\nu- 1} \Gamma(1 - 2\nu) (2 \nu -1 )\cos(\pi\nu),
\end{aligned}
\end{equation}
are the $\nu$-dependent numbers that are of the same order,
which are finite for $\nu=1/2$.
Interestingly, when $\nu=1/2$, the nonequilibrium currents ($I_u$ and $I_d$) given by Eq.~(\ref{noneq-currents}), corresponding to the weak-tunneling limit, is bias ($V_{su}$ or $V_{sd}$)-independent: a feature of $\nu = 1/2$ chiral Luttinger liquid systems~\cite{Fendley:1995c}.

Since all three QPCs are in the weak-tunneling limit, the experimentally measured transmissions are small  ($\tilde{\mathcal T}_{u,d,c}\ll 1$), meaning that the bare transmission probabilities satisfy the conditions:
\begin{equation}
\begin{aligned}
&\mathcal{T}_u \ll (\nu e V_{su} \tau_c)^{2-2\nu}, \quad \mathcal{T}_d \ll (\nu e V_{sd} \tau_c)^{2-2\nu},\\
&\mathcal{T}_c \ll [(I_u + I_d) \tau_c/e]^{2 - 2\nu}.
\end{aligned}
\label{eq:small_relations}
\end{equation}
Plugging in $I_u$ and $I_d$ into the last inequality, we 
\begin{equation}
\begin{aligned}
 & \mathcal{T}_c \, \ll \big[\mathcal{T}_u (\nu e V_{su} \tau_c)^{2\nu - 1}  + \mathcal{T}_d (\nu e V_{sd} \tau_c)^{2\nu - 1}\big]^{2 -2\nu}.
\end{aligned}
\label{eq:inequality_1}
\end{equation}
For $\nu  = 1/2$, Eq.~\eqref{eq:inequality_1} immediately leads to
\begin{equation}
    \mathcal{T}_c \ll (\mathcal{T}_u + \mathcal{T}_d)\ \text{ or } \ w_c^2 \ll (w_u^2 + w_d^2),
    \label{eq:inequality_2}
\end{equation}
when all three QPCs are in the weak-tunneling limit.

Importantly, Eq.~\eqref{eq:inequality_2} remains valid also for $\nu < 1/2$, where Eq.~\eqref{eq:inequality_1} can be further simplified by noting that
$$\mathcal{T}_u \ll (\nu e V_{su} \tau_c)^{2 - 2\nu}\ \Rightarrow\ (\nu e V _u\tau_c)^{ (2\nu -1)} \ll \mathcal{T}_u^{\frac{2\nu -1}{2 - 2\nu}},$$
leading to
\begin{equation}
    \mathcal{T}_c \ll \left( \mathcal{T}_u^{\frac{1}{2-2\nu}} + \mathcal{T}_d^{\frac{1}{2-2\nu}} \right)^{2 - 2\nu} \sim \mathcal{T}_u + \mathcal{T}_d.
\end{equation}
As a consequence, when $\nu \le 1/2$, and when all three QPCs are in the weak-tunneling limit (small transmission probabilities obtained experimentally), 
we get
$$w_c^2 \ll w_u^2 + w_d^2.$$ 
Thus, in this limit, the exponential factor $\exp(-2w_c^2/v)$ in Eq.~\eqref{eq:exact_central} is perfectly negligible (the integral over $t$ is cut off at large $t$ by the total nonequilibrium current $I_+$), justifying the perturbative approach utilized in Refs.~\cite{Rosenow:2016, Lee:2022, Morel:2022, Schiller:2023} that focused on the weak-tunneling limit.

In the opposite limit, $w_c^2 \gg w_u^2 + w_d^2$, the integral over time $t$ in expressions for the tunneling current and noise is bounded by $v/(2 w_c^2)$.
In addition, transmission through the central QPC becomes ballistic: $\tilde{\mathcal{T}}_c\sim 1$. 
As a consequence, any nonequilibrium current difference, $I_-$, will tunnel through the central QPC.

Importantly, we arrive at the conclusions above after assuming zero temperature. With large enough temperature, Eqs.~\eqref{eq:small_relations} will be modified, after replacing $V_{su}$, $V_{sd}$ and/or $I_+$ by the ambient temperature. Consequently, the relation $w_c^2 \ll w_u^2 + w_d^2$, corresponding to the weak-tunneling limit at zero temperature, can be violated at a finite temperature.

\section{Derivation of the correlation functions in the four-edge structure}
\label{app:correlation_derivations}

In this Appendix, we provide the details of the derivation of the various contributions to the nonequilibrium correlation function of the tunneling operators for the channel $u$, Eq.~\eqref{eq:ud_full_correlationsA}, together with a discussion of their physical meaning.
The correlation function \eqref{eq:ud_full_correlationsB} for channel $d$ is obtained in exactly the same way.
Below, we omit the Keldysh ordering symbol, ${\bf T}_K$, for brevity.
Actually, Keldysh correlation functions below all contain Keldysh indexes, and are clearly Keldysh-ordered.

\subsection{Classification of processes into three regions: Connected and disconnected diagrams}
\label{App:E-class}

Before moving to the calculation of the correlation functions in each of the Regions introduced in Sec.~\ref{subsec:regions}, we explain, with diagrams in Fig.~\ref{fig:diagrams_regions}, what we mean by ``disconnected'' and ``connected'' when dealing with the integral of Eq.~\eqref{second_order}. These ideas, briefly discussed at the end of Sec.~\ref{subsec:regions}, actually apply to the general-$\nu$ case; therefore, we formulate this approach here keeping arbitrary $\nu$ and later set $\nu=1/2$.
To begin with, Eq.~\eqref{second_order} is the integral that one deals with, after expanding to the leading order in the diluter transmission.
This equation involves correlations of different types of operators, the product of which is averaged with respect to the ground state. Since Wick's theorem does not apply to anyon operators, the result of the averaging is a product of pairwise correlations, where each of the operators is simultaneously involved in several correlation functions.
When a creation operator and an annihilation operator correlate, the time/coordinate dependence of the corresponding correlation function appears in the denominator of Eq.~\eqref{second_order}; otherwise, the time/coordinate dependence appears in the numerator of Eq.~\eqref{second_order}.

In Fig.~\ref{fig:diagrams_regions}, each of the involved operators is represented by a dot (vertex of a graph). 
Each vertex is connected to other vertices by solid and dashed red lines, indicating the correlation functions of the ``denominator'' and ``numerator'' types, respectively.
The distance between two diluter tunneling operators (with arguments $s_1 + L/v$ and $s_2 + L/v$) is represented by the length of a vertical line connecting them (similarly to the distance between the two collider tunneling operators with arguments 0 and $t$). 
The distance between one diluter operator and one collider operator, on the other hand, is represented by the projected distance, i.e., the distance between two points along the vertical direction [for instance, the length of the black double arrow of Fig.~\ref{fig:diagrams_regions}(a) refers to the (projected) distance between the operators with arguments $s_2 + L/v$ and 0].  
For later convenience, we denote the distance between two points by $t_1\leftrightarrow t_2$, with $t_1$, $t_2 \in \{0, t ,s_1 + L/v, s_2 + L/v\}$, labeling the time argument of the corresponding point.
Figure~\ref{fig:diagrams_regions}(a) shows the general situation with arbitrary distances between any two points.
We note in passing that the integrand in Eq.~\eqref{second_order} also contains a factor coming from the correlation of tunneling operators at the diluter in channel su; this factor is analogous to the diluter factor in Fig.~\ref{fig:diagrams_regions}(a). As a result, this correlation function (vertical line depending on $s_1-s_2$) appears squared in Eq.~\eqref{second_order}, in contrast to the correlation function (vertical line depending on $t$) for the tunneling operators at the collider.

Now we move to Region I, with its diagram shown in Fig.~\ref{fig:diagrams_regions}(b).
As the major feature, in Fig.~\ref{fig:diagrams_regions}(b), the two diluter's anyon operators are close to each other, meaning that two upper red lines (solid, $s_2 + L/v\leftrightarrow t$ and dashed, $s_1 + L/v \leftrightarrow t$) have the same projected distance; and so do the lower red lines (solid, $s_1 + L/v \leftrightarrow 0$ and dashed, $s_2 + L/v \leftrightarrow 0$). Since solid and dashed red lines appear in the denominator and numerator of Eq.~\eqref{second_order}, their norm will cancel out.
Consequently, as shown by Fig.~\ref{fig:diagrams_regions}(c), lines connecting any diluter operator with any collider operator disappear, leading to a ``disconnected'' diagram comprising two subgraphs (one for diluter operators and one for the tunneling operators).
Notice that this cancellation even works when $s_1 + L/v \leftrightarrow 0\approx s_2 + L/v \leftrightarrow 0 \sim 1/\nu e V$, i.e., close to one tunneling operator. This results from the fact that $|s_1 - s_2| \sim l_c/v \ll 1/\nu e V$. We emphasize that, different from the fermionic case, for the anyonic situation, these two ``disconnected'' parts are actually related by an extra phase factor $\exp [\pm i\pi (\eta_1 - \eta_2)/2]$ (indicated by the gray dotted line) that is generated from so-called time-domain braiding.
As mentioned after Eq.~\eqref{second_order}, when dividing the integral into three dominant regions, we keep using this terminology ``disconnected'', to refer to its analogy to disconnected diagrams in fermionic systems.

In contrast, in Fig.~\ref{fig:diagrams_regions}(d) illustrating Region II, both $|(s_2 + L/v\leftrightarrow t)/(s_1 + L/v \leftrightarrow t)|$ and $|(s_2 + L/v\leftrightarrow 0)/(s_1 + L/v \leftrightarrow 0)|$ are generically quite different from unity. The solid and dashed lines in Fig.~\ref{fig:diagrams_regions}(d) do not cancel each other, leading to a ``connected'' diagram with all vertices connected by correlators. Therefore, when evaluating Eq.~\eqref{second_order}, the norm of the corresponding tangling factor is not equal to one.
Finally, in Fig.~\ref{fig:diagrams_regions}(d) corresponding to Region III, we choose, without loss of generality, the case where $s_1 + L/v $ and $ s_2 + L/v $ are close to zero, the moment where tunneling at the central QPC occurs.
In this case, $|s_1 - s_2| \sim 1/\nu e V$.
As $1/\nu e V \ll t\sim e/4 I_{u/d} $, 
$s_1 + L/v \leftrightarrow t \sim t\pm 1/\nu e V\sim t $ and $s_2 + L/v \leftrightarrow t \sim t\pm 1/\nu e V\sim t$ have similar projected distances, and will thus cancel out in Eq.~\eqref{second_order}.
The difference between Regions III and I occurs when considering distances $s_1 + L/v \leftrightarrow 0$ and $s_2 + L/v \leftrightarrow 0$.
Indeed, although these distances are small (as those in Region I), they are both of the order of $1/\nu e V$, the same as that of $|s_1 - s_2|$.
As a consequence, $s_1 + L/v \leftrightarrow 0$ and $s_2 + L/v \leftrightarrow 0$ are not necessarily close.
For instance, when $s_1 + L/v = l_c/v$ and $s_2 + L /v = l_c / 2v$, although both distances are small, $\sim l_c/v$, their ratio is equal to two, a value significantly different from one. The corresponding factors appearing in the denominator and numerator of the tangling part in Eq.~\eqref{second_order} will thus not cancel out.
As a result, the corresponding diagram for Region III remains ``connected'', as that of Region II.

\subsection{``Disconnected'' diagram (Region I)}

\begin{figure}[ht!] 
\begin{center} 
\vspace*{0.5cm}
\includegraphics[width=7cm]{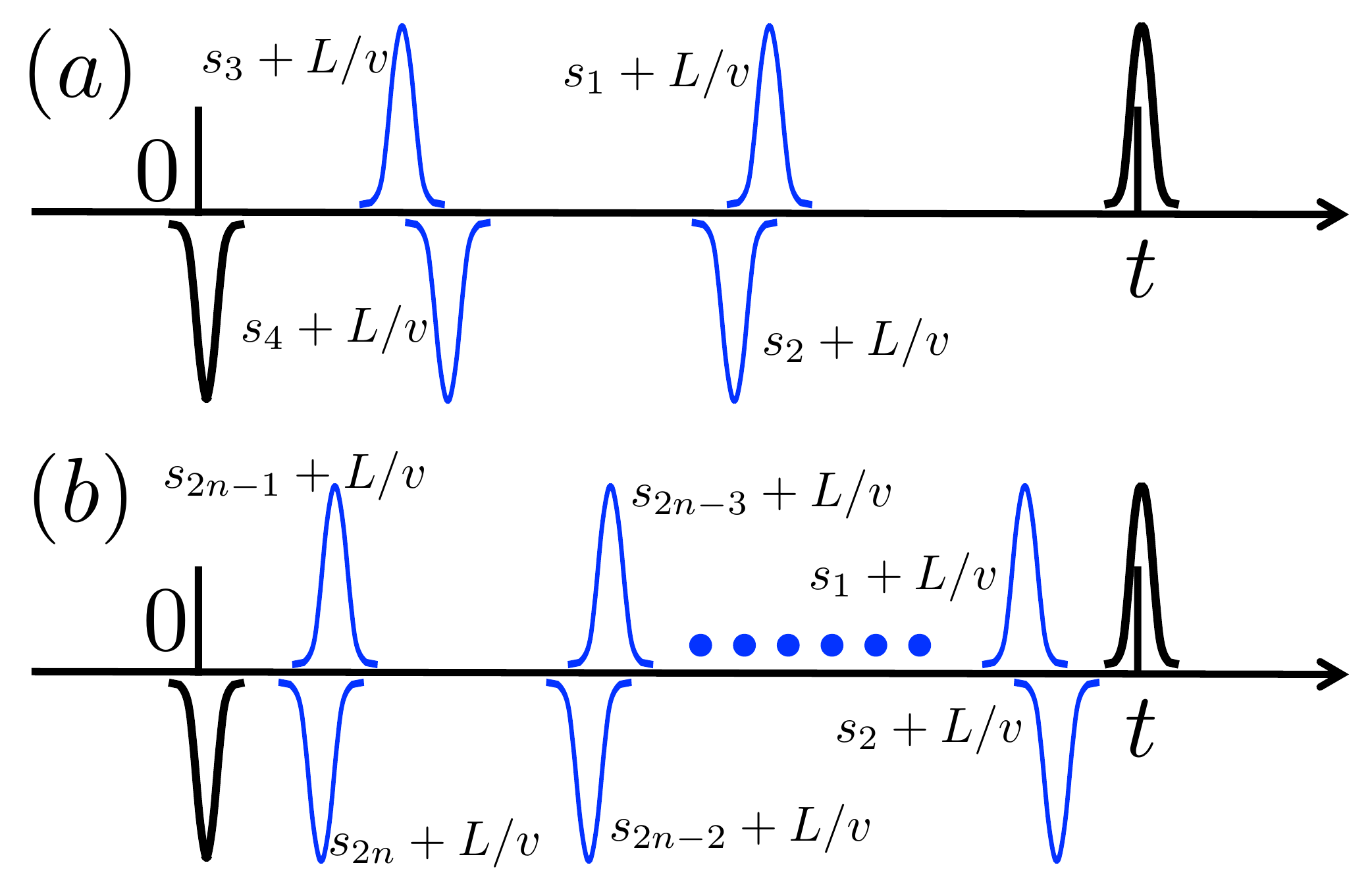}
\end{center}
\caption{Resummation over tunneling events at the diluters for the Region-I contribution to the correlation functions.
The horizontal (time) axis shows the time moments when anyons arrive at the central QPC.
Operators representing tunneling events at the central QPC (with transmission $\propto |w_c|^2$) and those referring to nonequilibrium anyons from the upper diluter (with transmission $\propto |w_u|^2$) are shown by black and blue pulses, respectively. (a) The case with two disconnected pairs. (b) The situation with $n$  disconnected pairs. All disconnected pairs are distant from each other and the corresponding integrals can thus be performed independently.} \label{fig:dis_resum}
\end{figure}

We begin by deriving the first term in Eq.~\eqref{eq:ud_full_correlationsA}.
This term, which agrees with the expression that was obtained in Refs.~[\onlinecite{Rosenow:2016, Lee:2022, Morel:2022, Schiller:2023}], after taking $\nu = 1/2$, results from the so-called time-domain braiding between an anyon-hole excitation at the central QPC and by-passing (Fig.~\ref{fig:dis_resum}) nonequilibrium anyons.
Crucially, in these processes, nonequilibrium anyons do not \textit{directly} participate in the tunneling at the central QPC; instead, they \textit{indirectly} influence the tunneling via time-domain braiding. 

To leading order in the tunneling transparency of the upper diluter, we follow the method described in 
Ref.~[\onlinecite{Morel:2022}] to compute the integral of the form 
\begin{widetext}
\begin{align}
&\left\langle e^{i \phi_u(L,t)/\sqrt{2}}\, e^{-i \phi_u(L,0)/\sqrt{2}}\right\rangle_2=
- \frac{|w_u|^2}{ 2 \pi l_c}
\sum_{\eta_1\eta_2} \eta_1\eta_2 
\iint  ds_1 ds_2 \, e^{-i eV(s_1 - s_2)/2}
\Big\langle e^{\frac{i}{\sqrt{2}} \phi_{su} (0,s_1^{\eta_1})} e^{-\frac{i}{\sqrt{2}} \phi_{su} (0,s_2^{\eta_2})} \Big\rangle 
\notag
\\
&\qquad \times 
\Big\langle e^{-\frac{i}{\sqrt{2}} \phi_u (L,t^-)} e^{\frac{i}{\sqrt{2}} \phi_u (L, 0^+)} e^{-\frac{i}{\sqrt{2}} \phi_u (0,s_1^{\eta_1})} e^{\frac{i}{\sqrt{2}} \phi_u (0,s_2^{\eta_2})} \Big\rangle 
\equiv - \frac{|w_u|^2}{ 2 \pi v} \,\frac{l_c^{1/2}}{(l_c + i vt)^{1/2}}\, f( t) ,
\label{eq:psi_a_correlation}
\end{align}
which appears in Eq.~\eqref{second_order}.
Here, the integration over $s_{1,2}$ is performed in infinite limits, $\eta_{1,2}$ are the Keldysh indices, and $\chi_{\eta,\eta'} (x) = \eta'$ if $x>0$ and equals $-\eta$ otherwise, see Eq.~\eqref{chi-def}.
The function $f(t)$ introduced in Eq.~\eqref{eq:psi_a_correlation} reads as
\begin{equation}
f(t) \!\equiv\! \sum_{\eta_1\eta_2} \eta_1\eta_2 \!\iint ds_1 ds_2\, \frac{v\, e^{-i eV (s_1 - s_2)/2}}{l_c \!+\! i v(s_1 \!-\! s_2) \chi_{\eta_1,\eta_2} (s_1 \!-\! s_2) }\, 
\frac{ \sqrt{ l_c\! +\! i (vt \!-\! vs_1 \!-\! L ) \chi_{\!-,\eta_1 } (t \!-\! s_1) }\, \sqrt{ l_c \!+\! i ( \!- vs_2 \!-\! L ) \chi_{+,\eta_2 } ( \!- s_2) } }{\sqrt{ l_c \!+\! i (vt \!-\! vs_2 \!-\! L ) \chi_{-,\eta_2 } (t \!-\! s_2) } \,\sqrt{ l_c \!+\! i ( \!-v s_1 \!-\! L ) \chi_{+,\eta_1 } ( \!- s_1) } }.
\label{eq:ft}
\end{equation}
For later convenience, we will denote by $\mathcal{R}_{\eta_1,\eta_2}(s_1,s_2,t)$ the ``tangling part'' of the integrand (the fraction containing the square roots) that arises from the correlations between the diluter and collider tunneling operators, as indicated by the non-vertical solid and dashed lines in Fig.~\ref{fig:diagrams_regions}(a): 
\begin{align}
\mathcal{R}_{\eta_1,\eta_2}(s_1,s_2,t)\equiv
\frac{ \sqrt{ l_c + i (vt - vs_1 - L ) \chi_{-,\eta_1 } (t - s_1) }\, \sqrt{ l_c - i ( vs_2 + L ) \chi_{+,\eta_2 } ( - s_2) } }{\sqrt{ l_c + i (vt - vs_2 - L ) \chi_{-,\eta_2 } (t - s_2) } \,\sqrt{ l_c - i ( v s_1 + L ) \chi_{+,\eta_1 } ( - s_1) } }.
\label{R-general}
\end{align}

To evaluate the disconnected contribution to the correlation function, we focus on the singular point $s_1 \to s_2$ (Region I), where the two operators describing the non-equilibrium anyons at $x=0$ have close-time arguments. This point corresponds to the pole-like singularity of the first factor in the integrand of Eq.~\eqref{eq:ft}.
We remind the reader that here we consider the integration intervals 
\begin{equation}
-L/v<s_{1,2}<t-L/v<0,
\label{RegionI-domain}
\end{equation}
(which, in particular, implies that $L > vt >0$ is assumed). For negative values of $s_{1,2}$, all the arguments of the $\chi$ functions involved in the tangling factor $\mathcal{R}_{\eta_1,\eta_2}$ are all positive, which yields
\begin{align}
    &\mathcal{R}_{\eta_1,\eta_2}(s_1,s_2,t)
    =\frac{ \sqrt{ l_c + i (vt - vs_1 - L ) \eta_1 } \, \sqrt{ l_c - i ( vs_2 + L ) \eta_2 } }{\sqrt{ l_c + i (vt - vs_2 - L ) \eta_2  } \, \sqrt{ l_c - i (v s_1 + L ) \eta_1  } },
    \qquad s_{1,2}<0.
    \label{quadrant}
\end{align}  
Moreover, all the length combinations in brackets multiplying the $\chi$ functions in Eq.~\eqref{R-general} are also positive in the domain (\ref{RegionI-domain}).  
This allows us to further simplify the tangling factor: 
\begin{align}
    &\mathcal{R}_{\eta_1,\eta_2}(s_1,s_2,t)
    \simeq  \frac{\left(\sqrt{|vt - vs_1 - L|}\,e^{i\pi \eta_1/4}\right) \, \left(\sqrt{|vs_2 + L|}\,e^{-i\pi \eta_2/4}\right)}{\left(\sqrt{|vt - vs_2 - L|}\,e^{i\pi \eta_2/4}\right)\, \left(\sqrt{| vs_1 + L|}\,e^{-i\pi \eta_1/4}\right)} = \sqrt{\frac{|vt - vs_1 - L|\,|vs_2 + L| }{|vt - vs_2 - L| \, | vs_1 + L|}} \,  e^{i\pi (\eta_1-\eta_2)/2}.
    \label{eqtangling}
\end{align}    
In Eq.~\eqref{eqtangling}, we have set $l_c\to 0$.
Importantly, the phase factor of the tangling factor (\ref{eqtangling}) does not depend on the time arguments.

In the strongly-diluted limit, we further set $s_1 = s_2$ in the tangling factor \eqref{eqtangling}, which then reduces to an exponential of a complex phase:
    \begin{align}
  -L/v<s_{1,2}<t-L/v<0: \qquad  \left.\mathcal{R}_{\eta_1,\eta_2}(s_1,s_2,t)\right|_{s_1\to s_2} = e^{i\pi (\eta_1-\eta_2)/2}.
    \label{eqE3}
\end{align}
The integral over $s=s_1-s_2$ in $f(t)$, however, should be treated carefully, to correctly grasp the feature of the branch cut introduced by $\chi_{\eta_1,\eta_2}(s)$. By assuming $t$ positive, extending the integration limits to infinity since (for $eVt\gg 1$) the integral at large $|s|$ is cut off by the oscillatory phase factor, and writing explicitly the terms with $s<0$ and $s>0$ for $(\eta_1,\eta_2)=(+,+),\,(+,-),\,(-,+),\,(-,-)$,  Eq.~\eqref{eq:psi_a_correlation} becomes:
\begin{equation}
    \begin{aligned}
  & \big\langle e^{-i\phi_u (t^-,L)/\sqrt{2}}  e^{i\phi_u (0^+, L)/\sqrt{2}} \big\rangle_2^\text{(I)} = 
  -\frac{ |w_u|^2}{ 2 \pi}  t\, \frac{l_c^{1/2} }{(l_c + i vt)^{1/2}} \sum_{\eta_1\eta_2} \eta_1\eta_2 \int_{-\infty}^\infty ds\, e^{-i V s/2}  \, \frac{e^{i\pi (\eta_1-\eta_2)/2}}{l_c + i v s \chi_{\eta_1,\eta_2} (s) }   
  \\
&=-\frac{ |w_u|^2}{ 2 \pi}  t \frac{l_c^{1/2} }{(l_c + i vt)^{1/2}} 
\Bigg[\int_{-\infty}^0 ds\, \frac{e^{-i V s/2}}{l_c - i v s  }+\int_{0}^\infty ds\, \frac{e^{-i V s/2}}{l_c + i v s } - \int_{-\infty}^0 ds\, \frac{e^{-i V s/2}\, e^{i\pi}}{l_c + i v s  }-\int_{0}^\infty ds\, \frac{e^{-i V s/2} \, e^{i\pi}}{l_c - i v s } 
\\
&\qquad \qquad \qquad \qquad  -\int_{-\infty}^0 ds\, \frac{e^{-i V s/2}\, e^{-i\pi}}{l_c - i v s  }-\int_{0}^\infty ds\, \frac{e^{-i V s/2} \, e^{-i\pi}}{l_c + i v s }+ 
\int_{-\infty}^0 ds\, \frac{e^{-i V s/2}}{l_c + i v s  }+\int_{0}^\infty ds\, \frac{e^{-i V s/2}}{l_c - i v s }
\Bigg]
  \\
  &
  = -\frac{ |w_u|^2}{\pi}\,t\,  \frac{l_c^{1/2}}{(l_c + i vt)^{1/2}} \int_{-\infty}^\infty d s\, e^{-i V s/2} \left( \frac{1}{l_c + i v s} + \frac{1}{l_c - i v s} \right)
   \simeq -\frac{ 2|w_u|^2}{ v}\,t\,  \frac{ l_c^{1/2}}{(l_c + i vt)^{1/2}}.
    \end{aligned}
    \label{eq:disconnected_leading}
\end{equation}
Using Eq.~\eqref{eq:it_small_tunneling}, i.e., $I_u = e |w_u|^2/(2 v)$, this yields Eq.~\eqref{eq81} of the main text
\begin{equation}
\langle e^{i \phi_u(L,t)/\sqrt{2}} 
e^{-i \phi_u(L,0)/\sqrt{2}}\rangle^{(\text{I})}_2 = -\frac{4 I_u |t|}{e} \frac{l_c^{1/2}}{(l_c + i vt)^{1/2}}.
\label{App:eq81}
\end{equation}
\end{widetext}

It is worth noticing that the last line of Eq.~\eqref{eq:disconnected_leading} contains an integral of a delta-function $\delta(s)$, which is fully consistent with setting $s_1=s_2$ in the tangling factor \eqref{eqE3}. In fact, for $V>0$, the evaluation of Eq.~\eqref{eq:disconnected_leading} would yield the same result, if we replaced $\chi_{\eta_1,\eta_2} $ with $-1$ at the very beginning. Similarly, for $V<0$, one could set $\chi_{\eta_1,\eta_2}=1$ in Eq.~\eqref{eq:disconnected_leading}.
This feature, being unique for $\nu = 1/2$, demonstrates that in Region I, the integral over $s_1 - s_2$ captures the pole contribution, $s_1 \to s_2$, of $1/[l_c \mp i v(s_1 - s_2)]$. 
This suggests the following prescription for evaluation of the integrals in Region I for $\nu=1/2$ (also in higher-order terms in the expansion in diluter transmission): 
(i) Replace $\chi_{\eta_1,\eta_2}(s_1-s_2) $ with $-\text{sgn}(V)$ in the denominator corresponding to the correlator of diluter's operators, and replace this correlator with a delta-function of $s_1-s_2$.
(ii) Set $s_1=s_2$ in the tangling part (as determined by the pole of the above correlator of diluter's operators), this replaces the tangling factor with a phase factor $\exp[ i\pi (\eta_1-\eta_2)/2]$, Eq.~(\ref{eqE3}).
(iii) Use $\eta_1\eta_2 \exp[ i\pi (\eta_1-\eta_2)/2]\equiv 1$ for all combinations of $\eta_1,\, \eta_2$.

The contribution to the correlation function given by  Eq.~\eqref{eq:disconnected_leading} is proportional to $I_u t$, in agreement with expressions derived in Refs.~\cite{Rosenow:2016, Lee:2022, Morel:2022}. Note that Eq.~\eqref{eq:disconnected_leading} contains an imaginary part only through $(l_c+ivt)^{1/2}$ in the denominator. Indeed, the factor $1 - \exp (2i\pi \nu)$ that reflects time-domain braiding~\cite{Rosenow:2016, Lee:2022, Morel:2022} for general $\nu$, becomes a real number for $\nu = 1/2$. Substituting Eq.~(\ref{eq:disconnected_leading}) into Eq.~(\ref{second_order}), we arrive at Eq.~(\ref{eq81}) of the main text.

Now, following Refs.~\cite{Lee:2022,Morel:2022}, we move to consider higher-order tunneling processes at diluters.
The next-to-leading-order ($\propto |w_u|^4$) contribution to the correlation function is explicitly written as follows:  
\begin{widetext}
\begin{equation}
    \begin{aligned}
 \Big\langle e^{i \phi_u(L,t)/\sqrt{2}}\, & e^{-i \phi_u(L,0)/\sqrt{2}}\Big\rangle_4 = 
   \frac{|w_u|^4}{ (2 \pi l_c)^2}\!\!\!\!
\sum_{\eta_1\eta_2 \eta_3\eta_4} \!\!\!\!\!\!\eta_1\eta_2 \eta_3\eta_4\!\!
\iiiint \! ds_1 ds_2 ds_3 ds_4\,
e^{-i eV(s_1 - s_2 + s_3 - s_4)/{2} }\\
&\times
\Big\langle  e^{{i} \phi_{su} (0,s_1^{\eta_1})/{\sqrt{2}}} \,e^{-i\phi_{su} (0,s_2^{\eta_2})/{\sqrt{2}}}\, e^{{i}\phi_{su} (0,s_3^{\eta_3})/{\sqrt{2}}}\, e^{-i \phi_{su} (0,s_4^{\eta_4})/{\sqrt{2}}} \Big\rangle_0
\\
&\times
\!\Big\langle e^{-{i} \phi_u (L,t^-)/{\sqrt{2}}}\, e^{i \phi_u (L, 0^+)/{\sqrt{2}}}\, e^{-{i}\phi_u (0,s_1^{\eta_1})/{\sqrt{2}} }\, e^{{i} \phi_u (0,s_2^{\eta_2})/{\sqrt{2}}}\, e^{-{i} \phi_u (0,s_3^{\eta_3})/{\sqrt{2}}}\, e^{{i}\phi_u (0,s_4^{\eta_4})/{\sqrt{2}} } \Big\rangle_0 \\
 &= \frac{{l_c}^{1/2}}{(l_c + i v t)^{1/2}}   \frac{ |w_u|^2}{ 2 \pi }  
\sum_{\eta_1\eta_2} \eta_1\eta_2 \iint ds_1 ds_2 \frac{ e^{-i{e V} (s_1 - s_2)/{2}}}{l_c + i v(s_1 - s_2) \chi_{\eta_1,\eta_2} (s_1 - s_2) } 
\mathcal{R}_{\eta_1,\eta_2}(s_1,s_2,t)\\
&\qquad \qquad \quad\ \times  \frac{  |w_u|^2}{ 2\pi }  \sum_{\eta_3\eta_4} \eta_3\eta_4 \iint ds_3 ds_4  \frac{ e^{-i{eV}(s_3 - s_4)/{2} } }{l_c + i v(s_3 - s_4) \chi_{\eta_3,\eta_4} (s_3 - s_4) } \mathcal{R}_{\eta_3,\eta_4}(s_3,s_4,t)\\
&\qquad \qquad \qquad\qquad \times  \frac{[(s_1-s_3) \chi_{\eta_1,\eta_3} (s_1 - s_3) ] [(s_2-s_4) \chi_{\eta_2,\eta_4} (s_2 - s_4)]}{[(s_1-s_4) \chi_{\eta_1,\eta_4} (s_1 - s_4)] [(s_2-s_3) \chi_{\eta_2,\eta_3} (s_2 - s_3)]}.
    \end{aligned}
    \label{eq:disconnected_4th}
\end{equation}
\end{widetext}

In Eq.~\eqref{eq:disconnected_4th}, one observes the same structures as in Eq.~\eqref{eq:ft}, now appearing for the variables labeled by $(1,2)$ and $(3,4)$.
These structures are not independent: they are connected by the factor in the last line of Eq.~\eqref{eq:disconnected_4th}.
However, when setting $s_1=s_2$ and $s_3=s_4$ in this common factor, i.e., taking the effective ``pole contribution'' in the integrals over $s_1,s_2$ and $s_3,s_4$, as was done when evaluating Eq.~\eqref{eq:ft} above, the common factor turns to unity. As a result, the ``pole contribution'' factorizes into a product of two integrals, each yielding the Region-I contribution to $f(t)$ (within each of these integrals, the fractions comprising the square roots again become the phase factors). We refer to the resulting term in the full correlation function as ``disconnected contraction'',
where operators with time arguments $s_1$ and $s_2$ and those with time arguments $s_3$ and $s_4$ are ``contracted''. To obtain the full correlation function, other contraction options should also be considered; these options will be discussed later.

When considering Region I, we can describe the above contraction (i.e., contracting the diluter tunneling operators with time arguments $s_1$ and $s_2$ and those with arguments $s_3$ and $s_4$) as illustrated in Fig.~\ref{fig:dis_resum}(a). In this figure, the two nonequilibrium anyonic pairs (indicated by blue pulses) are fully separated in time, when they arrive at the central QPC.
It is in this case that the last line of Eq.~\eqref{eq:disconnected_4th}, describing the ``interaction'' between the disconnected pairs, becomes unity, indicating that integrals over $s_1$ and $s_2$, for the first nonequilibrium anyonic pair and integrals over $s_3$, $s_4$ for the second pair can be evaluated independently.
With this observation, Eq.~\eqref{eq:disconnected_4th} becomes
\begin{equation}
   \Big\langle e^{i \phi_u(L,t)/\sqrt{2}}\, e^{-i \phi_u(L,0)/\sqrt{2}}\Big\rangle_4\!=\! \frac{l_c^{1/2}}{(l_c + ivt)^{1/2}} \left(\!\frac{-4 I_u |t|}{e}\!\right)^2\!,
\end{equation}
where the factor $(-4 I_u |t|/e)^2$ is the squared leading-order disconnected contraction [which equals $-4 I_u |t|/e$, cf. Eq.~\eqref{eq:disconnected_leading}].

This observation actually can be extended to arbitrary higher orders in the expansion in diluter transmission, as long as anyonic pairs of the nonequilibrium beam are well-separated in time (forming diluted beams), and distant from the two operators describing the tunneling at the central QPC (which is, actually, the definition of Region I). In this case [cf. Fig.~\ref{fig:dis_resum}(b)], nonequilibrium anyonic pairs are independent of each other, meaning that the nonequilibrium averaging is simply given by the product of factors coming from each of the single pairs.
Thus, when considering the contribution of Region I to the nonequilibrium correlation function, $n$ injected anyonic pairs produce the correlation $(-4 I_u |t| /e)^n$.

The above consideration involves only one contraction for $n$ ``disconnected'' non-equilibrium anyon pairs in Region I. One needs to know the total number of contractions, to carry out the resummation in the correlation functions in a general expression that contains $2n$ nonequilibrium operators. Again, we take the configuration of Fig.~\ref{fig:dis_resum}(b), where all $n$ anyonic pairs are well separated from each other--the requirement of Region I.
Each contraction leads to the nonequilibrium correlation $(-4 I_u t/e)^n$. The number of contraction is given by
\begin{equation}
    \frac{1}{2n!} \frac{A_{2n}^2 A_{2n-2}^2 ... 2 \times 1}{n!} = \frac{1}{n!}.
    \label{eq:disconnected_2n}
\end{equation}
Here, $1/(2n!)$ comes from the Keldysh expansion to the $2n^\text{th}$ order, $A_{2n} = 2n (2n-1)$ is the number of options to choose one creation and one annihilation operator out of $2n$ operators. The factorial $n!$ in the denominator gives the number of permutations among these nonequilibrium operator pairs.
With Eq.~\eqref{eq:disconnected_2n},
we sum over the number, $n$, of disconnected pairs, to obtain the correlation function of Region I:
\begin{equation}
\begin{aligned}
&\big\langle e^{-i\phi_u (t^-,L)/\sqrt{2}} e^{i\phi_u (0^+, L)/\sqrt{2}} \big\rangle^\text{(I)}\\
=& \frac{l_c^{1/2}}{(l_c\!+\!ivt)^{1/2}} \sum_n\frac{(-4 I_u t/e)^n}{n!}\! \\
= & \frac{l_c^{1/2}}{(l_c\!+\!ivt)^{1/2}} e^{-4 I_u t/e},
\end{aligned}
\label{eq:a_disconnected}
\end{equation}
which is the first term of Eq.~\eqref{eq:ud_full_correlationsA} for $t > 0$.
For negative $t$, the exponential factor becomes instead  $\exp (4 I_u t/e )$, leading to $|t|$ in Eq.~\eqref{eq:ud_full_correlationsA}.
The correlation function for channel $d$ is obtained in the same way.

\begin{widetext}

\subsection{Refined calculation of the second-order correlation function: subleading terms}

The above result for the disconnected contraction (obtained by setting $s_1=s_2$ in the tangling factor, which yields its unit norm) agrees with the general-$\nu$ calculation reported in Refs.~\cite{Rosenow:2016, Morel:2022}. However, as explained in the main text, taking into account only this leading correlation function is insufficient for the calculation of the tunneling current and the generalized Fano factor at $\nu=1/2$. Therefore, we need to refine the above derivation by including the contributions of the integration regions in the $s_1$-$s_2$ plane, which do not yield the unit modulus of the tangling factor unlike Eq.~(\ref{eqE3}).
For this purpose, we return to the second-order correlation function, Eq.~\eqref{eq:psi_a_correlation}, and change the integration variables to $s=s_1-s_2$ and $\tilde{s}_1=s_1+L/v$:
\begin{align}
\big\langle e^{-i\phi_u (t^-,L)/\sqrt{2}} e^{i\phi_u (0^+, L)/\sqrt{2}} \big\rangle_2
= &  -\frac{|w_u|^2\, l_c^{1/2}}{2\pi (l_c + i vt)^{1/2}}\sum_{\eta_1\eta_2} 
\int_{-\infty}^\infty ds
\int_{-\infty}^\infty d\tilde{s}_1 \,  \frac{ \eta_1\eta_2   \, e^{-i eV s/2}}{ l_c + i v  s\chi_{\eta_1,\eta_2} (s)}
\notag
\\
\times &
\frac{[l_c + i v ( t  -\tilde{s}_1 ) \chi_{-,\eta_1}(t+L/v-\tilde{s}_1)]^{1/2}\, [l_c + i v(s - \tilde{s}_1) \chi_{+,\eta_2}(L/v+s-\tilde{s}_1)]^{1/2}}
{[l_c + i v( t + s -\tilde{s}_1 ) \chi_{-,\eta_2}(t+L/v+s-\tilde{s}_1)]^{1/2}\,
[l_c - i v \tilde{s}_1 \chi_{+,\eta_1}(L/v-\tilde{s}_1)]^{1/2} } ,
\end{align}
Let us start with the case of $t>0$.
For the range of original variables corresponding to Region I,
Eq.~(\ref{RegionI-domain}),
we have the following domain of integration:
$$0<\tilde{s}_1<t, \qquad \tilde{s}_1-t<s<\tilde{s}_1.$$

In this range, all the $\chi$-functions in the tangling factor have positive arguments. Moreover, all combinations of times in brackets multiplying the $\chi$-functions are also positive.
This greatly simplifies the tangling factor to [cf. Eq.~\eqref{eqtangling}]
\begin{align}
    \mathcal{R}_{\eta_1,\eta_2}
    =\sqrt{\frac{(t - \tilde{s}_1)\,(\tilde{s}_1-s) }
    {(t + s - \tilde{s}_1) \, \tilde{s_1}}} \,  e^{i\pi (\eta_1-\eta_2)/2}.
    \label{eqtangl-new}
\end{align}    
The correlation function for Region I can then be written as
\begin{align}
&\quad\quad\big\langle e^{-i\phi_u (t^-,L)/\sqrt{2}} e^{i\phi_u (0^+, L)/\sqrt{2}} \big\rangle_2^{\text{(I)}}
\\
&=   -\frac{|w_u|^2\, l_c^{1/2}}{2\pi (l_c + i vt)^{1/2}}\sum_{\eta_1\eta_2}  \eta_1\eta_2 e^{i\pi (\eta_1-\eta_2)/2}\,
\int_{0}^t d\tilde{s}_1 
\int_{\tilde{s}_1 -t}^{\tilde{s}_1 }\!\! ds \,
\frac{   \, e^{-i eV s/2}}
{ l_c + i v  s\chi_{\eta_1,\eta_2} (s)}
\, 
\sqrt{\frac{(t - \tilde{s}_1)\,(\tilde{s}_1-s) }
    {(t + s - \tilde{s}_1) \, \tilde{s_1}}} \notag
\\
& =   -\frac{|w_u|^2\, l_c^{1/2}}{2\pi (l_c + i vt)^{1/2}}
\sum_{\eta_1\eta_2} 
\int_{0}^t d\tilde{s}_1 
\left(\int_{0}^{\tilde{s}_1} ds  
\frac{   \, e^{-i eV s/2}}{ l_c + i v  s\, \eta_2}
+ \int_{\tilde{s}_1-t}^0 ds 
\frac{   \, e^{-i eV s/2}}{ l_c - i v  s\, \eta_1}
\right) 
\sqrt{\frac{(t - \tilde{s}_1)\,(\tilde{s}_1-s) }
    {(t + s - \tilde{s}_1) \, \tilde{s_1}}} \notag
    \\
&  = -\frac{2|w_u|^2\, l_c^{1/2}}{\pi (l_c + i vt)^{1/2}} 
\int_{0}^t d\tilde{s}_1 
\left[ \int_{0}^{\tilde{s}_1} ds \, e^{-i eV s/2} 
\left(\frac{1}{ l_c + i v  s}+\frac{1}{ l_c - i v  s}\right)
\sqrt{\frac{(t - \tilde{s}_1)\,(\tilde{s}_1-s) }
    {(t + s - \tilde{s}_1) \, \tilde{s_1}}}\right. \notag\\
&\quad + \left.\int_{\tilde{s}_1-t}^0 ds 
\, e^{-i eV s/2} 
\left(\frac{1}{ l_c - i v  s}+\frac{1}{ l_c + i v  s}\right)
\,\sqrt{\frac{(t - \tilde{s}_1)\,(\tilde{s}_1-s) }
    {(t + s - \tilde{s}_1) \, \tilde{s_1}}}\ \right] 
\notag
\\
& =    -\frac{2|w_u|^2\, l_c^{1/2}}{\pi (l_c + i vt)^{1/2}} \int_{0}^t d\tilde{s}_1 
\int_{\tilde{s}_1 -t}^{\tilde{s}_1 }\!\! ds \,e^{-i eV s/2}
\underbrace{\,\frac{  l_c }{ l_c^2 +  v^2  s^2}\,}_{=(\pi/v) \delta(s)\ \text{for} \ l_c\to 0}  \sqrt{\frac{(t - \tilde{s}_1)\,(\tilde{s}_1-s) }
    {(t + s - \tilde{s}_1) \, \tilde{s_1}}}   =   -\frac{2|w_u|^2\, l_c^{1/2}}{v(l_c + i vt)^{1/2}} t.
\end{align}
This exact calculation yields the same result as Eq.~(\ref{eq:disconnected_leading}), justifying the prescription formulated above for dealing with Region I.
Thus, the definition of Region I is rigorously captured by the integration domain (\ref{RegionI-domain}), where the integral in the correlation function is automatically (after the summation over $\eta_1$ and $\eta_2$) determined by the delta-function term at $s_1=s_2$ rendering the ``disconnected'' character of this diagram. 
Importantly, the defining property of Region I is the phase of the tangling factor, which appears to be the same [see Eq.~\eqref{eqtangling}] for the entire domain given by Eq.~\eqref{RegionI-domain}, i.e.,  even without setting $s_1=s_2$. Once this phase jumps at the boundary of Region I, one gets another integration region.

Above, we have considered the Region-I domain, Eq.~\eqref{RegionI-domain}, where both $s_1$ and $s_2$ are restricted to a square in the $s_1$-$s_2$ plane. 
Let us now consider the contributions to the correlation function from the domain around this square for negative $s_1$ and $s_2$. In the whole quadrant $s_1<0,\ s_2<0$, the arguments of the $\chi$-functions in the tangling factor (\ref{R-general}) are positive, so that it reduces to Eq.~(\ref{quadrant}). Now, the phase factor in $\mathcal{R}_{\eta_1,\eta_2}(s_1,s_2,t)$ jumps when going around the square domain \eqref{RegionI-domain} of Region I
(the cyan square of Fig.~\ref{fig:diff_range_integrals}):
\begin{align}
    \mathcal{R}_{\eta_1,\eta_2}(s_1,s_2,t)
      & = \sqrt{\frac{|vt - vs_1 - L|\,|vs_2 + L| }{|vt - vs_2 - L| \, | vs_1 + L|}}\! \times\!
        \left\{ 
       \begin{array}{cc}
      e^{i\pi \eta_1/2},& \quad -\frac{L}{v}<s_1<t-\frac{L}{v}\ \ \text{and}\ \, \left(t-\frac{L}{v}<s_2<0 \ \text{or}  \ s_2<-\frac{L}{v}\right),\\[0.1cm]
      e^{-i\pi \eta_2/2},& \quad -\frac{L}{v}<s_2<t-\frac{L}{v}\ \ \text{and}\ \, \left(t-\frac{L}{v}<s_1<0 \ \text{or}  \ s_1<-\frac{L}{v}\right),\\[0.1cm]
      1, &\quad \text{otherwise}.
      \end{array}
      \right.
    \label{tangling-around}
\end{align}    

The structure of the expression for $f(t)$, Eq.~(\ref{eq:ft}), in the Keldysh space is such that when the phase factor is equal to unity, the summation over $\eta_1$ or $\eta_2$ will produce zero contribution of the corresponding integration domain to the correlation function. Therefore we focus to the subdomains that are attached to the side boundaries of Region I: the first two lines of Eq.~\eqref{tangling-around} 

\begin{figure}[h] \begin{center} 
\includegraphics[width=9cm]{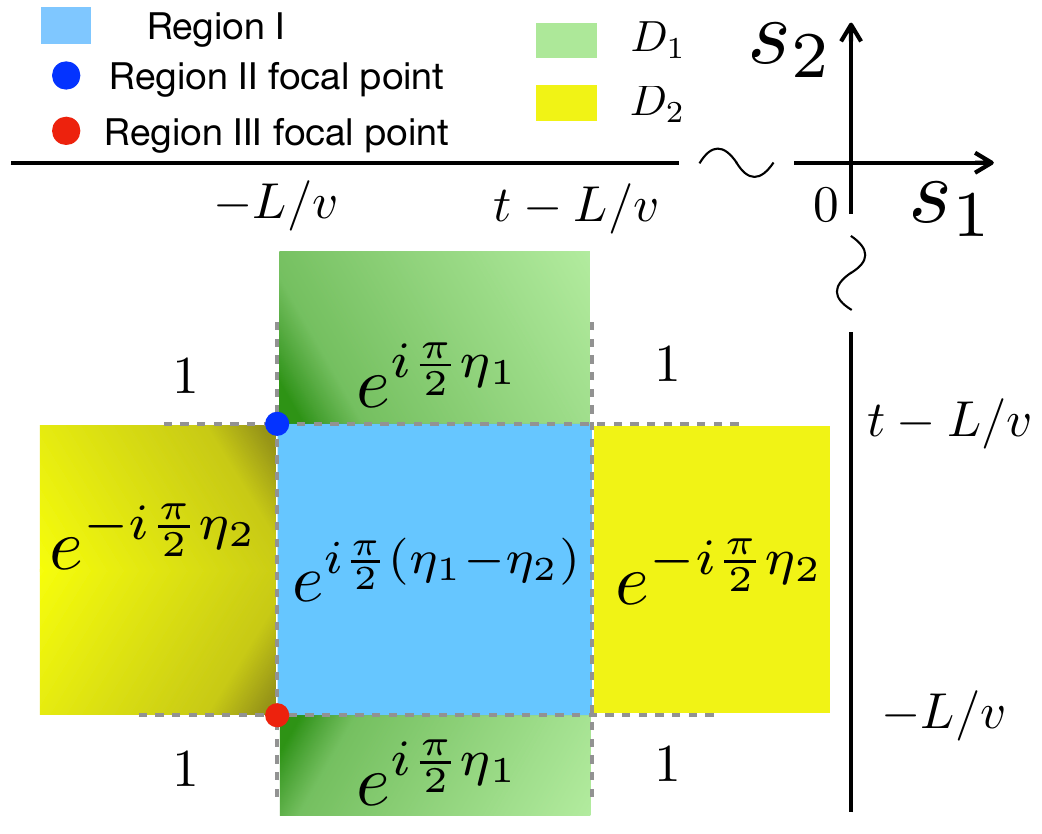}
\end{center}
\caption{Phase factor of the tangling term, Eq.~\eqref{quadrant}, for different values of $s_1$ and $s_2$ and $t > 0$. We show only negative values of $s_1$ and $s_2$, since we have assumed $L$ to be much larger than other relevant time scales.
In this figure, the cyan square area corresponds to Region I
(where only the diagonal $s_1=s_2$, associated with time-domain braiding, contributes to the correlation functions).
The blue and red dots instead indicate the focal points corresponding to contributions from Region II and Region III, respectively.
We also mark out integration subdomain ${D_1}$ [Eq.~(\ref{subdomain-D1})] and ${D_2}$ [Eq.~(\ref{subdomain-D1})], with light green and yellow, respectively. Here, areas located closer to the two focal points [within the distance of the order of $(eV)^{-1}$] have darker colors: these areas dominate the corresponding Region-II and Region-III contributions to the correlation functions.
} \label{fig:diff_range_integrals}
\end{figure}

Let us start with the subdomain in the first line of Eq.~\eqref{tangling-around}, corresponding to light green area of Fig.~\eqref{fig:diff_range_integrals}:
\begin{equation}
{D_1}: \quad -\frac{L}{v}<s_1<t-\frac{L}{v}\ \ \text{and}\ \, \left(t-\frac{L}{v}<s_2<0 \ \text{or}  \ s_2<-\frac{L}{v}\right).
\label{subdomain-D1}
\end{equation}
Using the tangling factor in this domain, we write the correlation function as
\begin{align}
&\left\langle e^{i \phi_u(L,t)/\sqrt{2}}\, e^{-i \phi_u(L,0)/\sqrt{2}}\right\rangle_2^{({D_1})}=
- \frac{|w_u|^2}{ 2 \pi } \,\frac{l_c^{1/2}}{(l_c + i vt)^{1/2}}\, \sum_{\eta_1\eta_2} \eta_1\eta_2\,  e^{i\pi \eta_1/2} \, \int_{-L/v}^{t-L/v} ds_1 \, \sqrt{\frac{vt - vs_1 - L}{vs_1 + L}}
\notag
\\
&\qquad \times
\left(\int_{-\infty}^{-L/v} ds_2+\int_{t-L/v}^{0} ds_2\right)\, 
\frac{ e^{-i eV (s_1 - s_2)/2}}{l_c \!+\! i v(s_1 \!-\! s_2) \chi_{\eta_1,\eta_2} (s_1 \!-\! s_2) }\,
\sqrt{\frac{|vs_2 + L| }{|vt - vs_2 - L|}}
\notag 
\\
&=
- \frac{|w_u|^2}{ 2 \pi } \,\frac{l_c^{1/2}}{(l_c + i vt)^{1/2}}\,  \int_{-L/v}^{t-L/v} ds_1 \, \sqrt{\frac{vt - vs_1 - L}{vs_1 + L}}  e^{-i eV s_1/2}\, \sum_{\eta_1\eta_2} \eta_1\eta_2\,  e^{i\pi \eta_1/2} \,
\notag 
\\
&\qquad \times
\left[\int_{-\infty}^{-L/v}\!\! ds_2
\frac{ e^{i eV  s_2/2}}{l_c \!+\! i v(s_1 \!-\! s_2) \,\eta_2 }\,
\sqrt{\frac{-vs_2 -L }{vt - vs_2 - L}}
+\int_{t-L/v}^{0}\!\! ds_2
\frac{ e^{i eV s_2/2}}{l_c \!-\! i v(s_1 \!-\! s_2) \,\eta_1}\,
\sqrt{\frac{vs_2 + L }{vs_2-v t + L}}
\right]\,.
\end{align}
Shifting the integration variables by introducing \begin{equation}
    \tilde{s}_1\equiv s_1+\frac{L}{v},  \quad
\tilde{s}_2\equiv s_2+\frac{L}{v}, 
\label{shift-stilde}
\end{equation}
we get
\begin{align}
&\left\langle e^{i \phi_u(L,t)/\sqrt{2}}\, e^{-i \phi_u(L,0)/\sqrt{2}}\right\rangle_2^{({D_1})}
=
- \frac{|w_u|^2}{ 2 \pi } \,\frac{l_c^{1/2}}{(l_c + i vt)^{1/2}}\,  \int_{0}^{t} d\tilde{s}_1 \, \sqrt{\frac{vt - v\tilde{s}_1}{v\tilde{s}_1}}  e^{-i eV \tilde{s}_1/2}\, \sum_{\eta_1\eta_2} \eta_1\eta_2\,  e^{i\pi \eta_1/2} \,
\notag 
\\
&\qquad \qquad \times
\left[\int_{-\infty}^{0}\!\! d\tilde{s}_2
\frac{ e^{i eV  \tilde{s}_2/2}}{l_c \!+\! i v(\tilde{s}_1 \!-\! \tilde{s}_2) \,\eta_2 }\,
\sqrt{\frac{-v\tilde{s}_2 }{vt - v\tilde{s}_2}}
+\,\underbrace{\int_{t}^{L/v}\!\! d\tilde{s}_2
\frac{ e^{i eV \tilde{s}_2/2}}{l_c \!-\! i v(\tilde{s}_1 \!-\! \tilde{s}_2) \,\eta_1}\,
\sqrt{\frac{v\tilde{s}_2}{v\tilde{s}_2-v t}}}_{\text{gives zero after summation over}\ \eta_2}\,
\right].
\end{align}
The second term in the second brackets in the last line is independent of $\eta_2$ and hence goes away after summation over $\eta_2$. The summation over $\eta_1$ in the remaining part of the correlation function produces a non-zero result because of the nontrivial (dependent on $\eta_1$) phase of the tangling factor. Changing the variable $\tilde{s}_2\to -\tilde{s}_2$, we have 
\begin{align}
&\left\langle e^{i \phi_u(L,t)/\sqrt{2}}\, e^{-i \phi_u(L,0)/\sqrt{2}}\right\rangle_2^{({D_1})}
=
- \frac{|w_u|^2}{ 2 \pi } \,\frac{l_c^{1/2}}{(l_c + i vt)^{1/2}}\,  \int_{0}^{t} d\tilde{s}_1 \, \sqrt{\frac{t - \tilde{s}_1}{\tilde{s}_1}}  e^{-i eV \tilde{s}_1/2}\, \underbrace{\left(e^{i\pi/2}-e^{-i\pi/2}\right)}_{\text{sum over}\ \eta_1} \,
\notag 
\\
&\qquad \times
\int_{0}^{\infty}\!\! d\tilde{s}_2 e^{-i eV  \tilde{s}_2/2}
\underbrace{\left[\frac{ 1}{l_c \!+\! i v(\tilde{s}_1 \!+\! \tilde{s}_2) }\,-\frac{ 1}{l_c \!-\! i v(\tilde{s}_1 \!+\! \tilde{s}_2)  } \right]}_{\text{sum over } \eta_2}\,
\sqrt{\frac{\tilde{s}_2 }{t + \tilde{s}_2}}
\notag
\\
&\qquad= - \frac{2|w_u|^2}{v \pi } \,\frac{l_c^{1/2}}{(l_c + i vt)^{1/2}}\,  \int_{0}^{t} d\tilde{s}_1 \, \sqrt{\frac{t - \tilde{s}_1}{\tilde{s}_1}}\,  e^{-i eV \tilde{s}_1/2}
\int_{0}^{\infty}\!\! d\tilde{s}_2\, e^{-i eV  \tilde{s}_2/2}
\frac{1}{\tilde{s}_1 \!+\! \tilde{s}_2} 
\sqrt{\frac{\tilde{s}_2 }{t + \tilde{s}_2}}\,.
\,
\notag 
\end{align}
The integral over $\tilde{s}_2$ here is cut off by the oscillatory factor at $\tilde{s}_2\sim (e V)^{-1}$. Assuming
$(e V)^{-1}\ll t$, we can neglect $\tilde{s}_2$ in the sum $t+\tilde{s}_2$; keeping it there only produces subleading correction in powers $(e V t)^{-1}$. 
The resulting integrals can be evaluated exactly in a closed form.
Note that the integral over $\tilde{s}_1$ has two special points:  $\tilde{s}_1=0$ and
$\tilde{s}_1=t$, which correspond to the branching points in the integrand coinciding with the integral limits. 
The point $\tilde{s}_1=t$ produces an oscillatory contribution to the integral, proportional to $e^{-i e V t/2}$, which is, however, suppressed by the factor $(e V t)^{-2}$, and hence can be ignored with our accuracy. We thus neglect $\tilde{s}_1$ in $t-\tilde{s}_1$ and extend the integral to infinity:
\begin{align}
\left\langle e^{i \phi_u(L,t)/\sqrt{2}}\, e^{-i \phi_u(L,0)/\sqrt{2}}\right\rangle_2^{({D_1})}&
\approx
- \frac{2|w_u|^2}{ \pi v} \,\frac{l_c^{1/2}}{(l_c + i vt)^{1/2}}\,  
\int_{0}^{\infty} d\tilde{s}_1 \, \frac{e^{-i eV \tilde{s}_1/2}}{\sqrt{{\tilde{s}_1}}}
\int_{0}^{\infty}\!\! d\tilde{s}_2\, 
\frac{e^{-i eV  \tilde{s}_2/2}}{\tilde{s}_1 \!+\! \tilde{s}_2 } 
\sqrt{\tilde{s}_2}
\notag 
\\
&=i \frac{2|w_u|^2}{ v \, eV} \,\frac{l_c^{1/2}}{(l_c + i vt)^{1/2}}
=i \frac{4 I_u}{e^2V} \,\frac{l_c^{1/2}}{(l_c + i vt)^{1/2}}
.
\label{D1-result}
\end{align}

Now, we turn to the second subdomain in Eq.~\eqref{tangling-around}, corresponding to the yellow area of Fig.~\eqref{fig:diff_range_integrals}:
\begin{equation}
{D_2}:\quad -\frac{L}{v}<s_2<t-\frac{L}{v}\ \ \text{and}\ \, \left(t-\frac{L}{v}<s_1<0 \ \text{or}  \ s_1<-\frac{L}{v}\right).
\label{subdomain-D2}
\end{equation}
The calculation of the correlation function in this domain follows the same lines as the calculation in ${D_1}$:
\begin{align}
&\left\langle e^{i \phi_u(L,t)/\sqrt{2}}\, e^{-i \phi_u(L,0)/\sqrt{2}}\right\rangle_2^{{D_2}}=
- \frac{|w_u|^2}{ 2 \pi } \,\frac{l_c^{1/2}}{(l_c + i vt)^{1/2}}\, \sum_{\eta_1\eta_2} \eta_1\eta_2\,  e^{-i\pi \eta_2/2} \, \int_{-L/v}^{t-L/v} ds_2 \, 
\sqrt{\frac{|vs_2 + L| }{|vt - vs_2 - L|}}
\notag
\\
&\qquad \times
\left(\int_{-\infty}^{-L/v} ds_1+\int_{t-L/v}^{0} ds_1\right)\, 
\frac{ e^{-i eV (s_1 - s_2)/2}}{l_c \!+\! i v(s_1 \!-\! s_2) \chi_{\eta_1,\eta_2} (s_1 \!-\! s_2) }\,
\sqrt{\frac{|vt - vs_1 - L|}{|vs_1 + L|}}.
\end{align}
It is again convenient to introduce the shifted variables, Eq.~\eqref{shift-stilde}, yielding
\begin{align}   
&\left\langle e^{i \phi_u(L,t)/\sqrt{2}}\, e^{-i \phi_u(L,0)/\sqrt{2}}\right\rangle_2^{{D_2}}
=- \frac{|w_u|^2}{ 2 \pi } \,
\frac{l_c^{1/2}}{(l_c + i vt)^{1/2}}\,  
\int_{0}^{t} d\tilde{s}_2 \, 
\sqrt{\frac{v\tilde{s}_2}{vt - v\tilde{s}_2 }} e^{i eV \tilde{s}_2/2}\, \sum_{\eta_1\eta_2} \eta_1\eta_2\,  e^{-i\pi \eta_2/2} \,
\notag 
\\
&\qquad \qquad\times
\left[\int_{-\infty}^{0}\!\! d\tilde{s}_1
\frac{ e^{-i eV  \tilde{s}_1/2}}{l_c \!-\! i v(\tilde{s}_1 \!-\! \tilde{s}_2) \,\eta_1}\,
\sqrt{\frac{vt - v\tilde{s}_1}{-v\tilde{s}_1}}
+\,\underbrace{\int_{t}^{L/v}\!\! d\tilde{s}_1
\frac{ e^{-i eV \tilde{s}_1/2}}{l_c \!+\! i v(\tilde{s}_1 \!-\! \tilde{s}_2) \,\eta_2}\,
\sqrt{\frac{-vt +v\tilde{s}_1}{v\tilde{s}_1}}}_{\text{gives zero after summation over}\ \eta_1}\,
\right].
\end{align}
Like in the case of subdomain $D_1$, one of the terms in the square brackets yields zero after the summation over $\eta_1$. Changing the sign of $\tilde{s}_1$, we then have
\begin{align}
&\left\langle e^{i \phi_u(L,t)/\sqrt{2}}\, e^{-i \phi_u(L,0)/\sqrt{2}}\right\rangle_2^{{(D_2)}}
=
- \frac{|w_u|^2}{ 2 \pi } \,
\frac{l_c^{1/2}}{(l_c + i vt)^{1/2}}\,  
\int_{0}^{t} d\tilde{s}_2 \, 
\sqrt{\frac{\tilde{s}_2}{t - \tilde{s}_2 }} e^{i eV \tilde{s}_2/2}\, \underbrace{\left(e^{-i\pi/2}-e^{i\pi/2}\right)}_{\text{sum over}\ \eta_2} \,
\notag 
\\
&\qquad \times
\int_{0}^{\infty}\!\! d\tilde{s}_1 e^{i eV  \tilde{s}_1/2}
\underbrace{\,\left[\frac{ 1}{l_c \!+\! i v(\tilde{s}_1 \!+\! \tilde{s}_2) }\,-\frac{ 1}{l_c \!-\! i v(\tilde{s}_1 \!+\! \tilde{s}_2)  } \right]\,}_{\text{sum over }\eta_1}
\sqrt{\frac{t + \tilde{s}_1}{\tilde{s}_1}}
\,
\notag 
\\
&\qquad =
\frac{2|w_u|^2}{ \pi v } \,\frac{l_c^{1/2}}{(l_c + i vt)^{1/2}}\,  
\int_{0}^{t} d\tilde{s}_2 \, 
\sqrt{\frac{\tilde{s}_2}{t - \tilde{s}_2 }} e^{i eV \tilde{s}_2/2}\,
\int_{0}^{\infty}\!\! d\tilde{s}_1 e^{i eV  \tilde{s}_1/2}
\frac{1}{\tilde{s}_1 \!+\! \tilde{s}_2} 
\sqrt{\frac{t + \tilde{s}_1}{\tilde{s}_1}}.
\end{align}
Utilizing the large parameter $eVt\gg 1$, we can neglect $\tilde{s}_1$ near $t$ in the last integral. However, 
since, in contrast to the correlator in domain ${D_1}$, the branching point of the square-root branch cut for the other integral is now in the denominator, we cannot neglect the resulting oscillatory term stemming from upper limit $\tilde{s}_2=t$. 
We proceed by taking the remaining integrals exactly and expanding the result to the leading order in the prefactor:
\begin{align}
&\left\langle e^{i \phi_u(L,t)/\sqrt{2}}\, e^{-i \phi_u(L,0)/\sqrt{2}}\right\rangle_2^{(D_2)}
\approx
\frac{2|w_u|^2}{ \pi v} \,\frac{l_c^{1/2}}{(l_c + i vt)^{1/2}}\,  
\int_{0}^{t} d\tilde{s}_2 \, 
\sqrt{\frac{\tilde{s}_2}{t - \tilde{s}_2 }} e^{i eV \tilde{s}_2/2}\,
\int_{0}^{\infty}\!\! d\tilde{s}_1 e^{i eV  \tilde{s}_1/2}
\frac{1}{\tilde{s}_1 \!+\! \tilde{s}_2} 
\sqrt{\frac{t}{\tilde{s}_1}}
\notag 
\\
&\qquad =\frac{2|w_u|^2}{ \pi v} \,\frac{l_c^{1/2}}{(l_c + i vt)^{1/2}}\, 
\left\{2\pi eV t -\frac{(\pi eV t)^{3/2}}{\sqrt{2}}\,
\big[J_0(eVt/4) - i J_1(eVt/4)\big]\, e^{i eVt/4-i\pi/4} \right\},
\end{align}
where $J_n(z)$ is the Bessel function.
Taking the large-$t$ asymptotics, we obtain 
\begin{align}
\left\langle e^{i \phi_u(L,t)/\sqrt{2}}\, e^{-i \phi_u(L,0)/\sqrt{2}}\right\rangle_2^{(D_2)}&= \frac{2|w_u|^2}{ v \, eV} \,\frac{l_c^{1/2}}{(l_c + i vt)^{1/2}}\, 
\left(i + 2 e^{i eV t/2}  \right)
=i \frac{4 I_u}{e^2V} \,\frac{l_c^{1/2}}{(l_c + i vt)^{1/2}}
+ \frac{8 I_u}{e^2V} \,\frac{l_c^{1/2}\, e^{i eV t/2}}{(l_c + i vt)^{1/2}}.
\label{D2-result}
\end{align}
The first term here is equal to the full result for subdomain ${D_1}$,  Eq.~(\ref{D1-result}). The second term represents the oscillatory contribution of the point $s_1=-L/v,\, s_2=t-L/v$ (blue dot in Fig.~\eqref{fig:diff_range_integrals});
such a contribution was absent in subdomain ${D_1}$ at this order in $(e V t)^{-1}$. 

Combining Eq.~\eqref{D2-result} with Eqs.~(\ref{D1-result}) and \eqref{App:eq81}, we finally arrive at 
\begin{equation}
\left\langle e^{i \phi_u(L,t)/\sqrt{2}}\, e^{-i \phi_u(L,0)/\sqrt{2}}\right\rangle_2
=\underbrace{-\frac{4 I_u t}{e} \frac{l_c^{1/2}}{(l_c + i vt)^{1/2}}}_\text{Region I}+\underbrace{i \frac{8 I_u}{e^2V} \,\frac{l_c^{1/2}}{(l_c + i vt)^{1/2}}}_\text{Region III}\, 
+ \, \underbrace{\frac{8 I_u}{e^2V} \,\frac{l_c^{1/2}}{(l_c + i vt)^{1/2}}\exp\left(\frac{i eV t}{2}\right)}_\text{Region II}.
\label{three-regions}
\end{equation}
This result reproduces the second-order term in the expansion of Eq.~\eqref{eq:ud_full_correlationsA} of the main text in powers of $w_u$, when $t > 0$. In the above calculation, we have also demonstrated that the Region-I contribution to the correlation function stems exactly from the line $-L/v<s_1=s_2<t-L/v$ within the main (square) domain, while the Region-II and Region-III terms are dominated by the areas $s_1\sim -L/v,\, s_2\sim t-L/v$ (near the red focal point of Fig.~\ref{fig:diff_range_integrals}) and $s_1,s_2\sim -L/v$ (near blue focal point of Fig.~\ref{fig:diff_range_integrals}), respectively, in the domains side-attached to the main square (cyan square of Fig.~\ref{fig:diff_range_integrals}).

For negative times, $t < 0$, the result for Region II of Eq.~\eqref{three-regions} remains unchanged. The Regions-I and Region-III terms, on the other hand, are multiplied by $\text{sgn}(t)$. Actually, when going back to Eq.~\eqref{eq:ft}, the influence of the negativity of $t$ can be included by shifting time variables $s_1$ and $s_2$ by $-t$.
After doing so, the only difference between positive and negative $t$ grounds in the tangling factor, Eq.~\eqref{quadrant}.
Indeed, when $t$ becomes negative, the phase factor of Eq.~\eqref{quadrant} changes sign, leading to the change of the sign of Region-I and Region-III terms of Eq.~\eqref{three-regions}.

\subsection{Interpretation of different-time connected contraction (Region II)}

Having derived the general formula for the second-order correlation function, we interpret the obtained results in terms of physical processes behind the terms corresponding to Region II and III, elucidating the role of anyons from the nonequilibrium diluted beam in each of them. Based on this consideration, we will then analyze the fate of these perturbative terms after the resummation of higher orders in the diluter tunneling processes, as we already did with the Region-I time-domain-braiding processes.  

Let us focus on Region II, where $s_1$ and $s_2$ are next to $-L/v$ and $t-L/v$, respectively (since $s_1\neq s_2$, we call it ``different-time'' contraction).
In this case, the two injected anyons arrive at the central QPC (located at $x=L$) when tunneling at the collider occurs. Thus, these injected anyons may participate in tunneling through the central QPC. 
To interpret the contribution of Region II, it is convenient to rewrite the integral Eq.~\eqref{second_order} as follows:
\begin{equation}
\begin{aligned}
& \big\langle e^{-i\phi_u (t^-,L)/\sqrt{2}} e^{i\phi_u (0^+, L)/\sqrt{2}} \big\rangle_2\\
&= \sum_{\eta_1\eta_2}\!\!  \iint \frac{ds_1 ds_2}{2 \pi l_c} \frac{- \eta_1\eta_2  |w_u|^2 l_c^{3/2} e^{-i eV ( s_1 - s_2)/2}}{\left\{[l_c + i (vt - vs_2 - L) \chi_{-,\eta_2} (t- s_2)] [l_c + i ( -v s_1 - L) \chi_{+,\eta_1} (- s_1)][l_c + i v(s_1 - s_2 )\chi_{\eta_1,\eta_2} (s_1 - s_2)]\right\}^{1/2}}\\
&\qquad \qquad \qquad \times  \frac{[l_c + i (vt - vs_1 - L) \chi_{-,\eta_1} (t - s_1)]^{1/2} [l_c + i ( - vs_2 - L) \chi_{+,\eta_2} ( - s_2)]^{1/2} }{(l_c + i vt)^{1/2} [l_c + i v(s_1 - s_2) \chi_{\eta_1,\eta_2} (s_1 - s_2)]^{1/2}} .
\end{aligned}
\label{eq:half_1st_expansion}
\end{equation}
Equation~\eqref{eq:half_1st_expansion} emphasizes that in Region II, we focus on the singularities at $s_1 \to -L/v$ and $s_2 \to t - L/v$, which are origins of the branch cuts. 
The last line of Eq.~\eqref{eq:half_1st_expansion} is a new ``tangling part'', designed specifically to yield a phase factor for the Region-II contraction.

Above, we have obtained the exact result for Region I by setting ``by hand'' $s_1=s_2$ in all places in the integrand of Eq.~\eqref{second_order}, where this did not lead to divergence of the integral. This prescription was later justified by a rigorous calculation that identified an emergent delta-function $\delta(s_1-s_2)$ in the full integral over the domain of Region I.
Here, we proceed in a similar way for Region II.
Indeed, by taking $s_1\to -L/v$ and $s_2 \to t-L/v$, the new tangling factor simplifies into
\begin{equation}
\begin{aligned}
& \frac{[l_c + i vt \chi_{-\eta_1} (t +L/v)]^{1/2} [l_c + i ( - vt) \chi_{+,\eta_2} ( - t + L/v)]^{1/2} }{(l_c + i vt)^{1/2} [l_c + i (-v t) \chi_{\eta_1,\eta_2} (-t)]^{1/2}}  = \frac{(l_c + i vt\, \eta_1)^{1/2} [l_c - i vt\, \eta_2]^{1/2} }{(l_c + i v t)^{1/2} [l_c - i vt \chi_{\eta_1,\eta_2} (-t)]^{1/2}} ,
\end{aligned}
\end{equation}
which equals $\exp[-i \pi (1 + \eta_2)/4]$ for $t>0$ and $\exp[-i \pi(-1 + \eta_1)/4]$ for $t<0$.
With these phase factors, Eq.~\eqref{eq:half_1st_expansion} becomes
\begin{equation}
\begin{aligned}
  \big\langle e^{-i\phi_u (t^-,L)/\sqrt{2}} e^{i\phi_u (0^+, L)/\sqrt{2}} \big\rangle_2^\text{(II)}&= -\frac{|w_u|^2 l_c^{3/2}}{2 \pi l_c\sqrt{l_c + ivt}}\sum_{\eta_1\eta_2} \iint ds_1 ds_2\frac{ \eta_1\eta_2 \,  e^{-i eV( \tilde{s}_1 - \tilde{s}_2)/2}}{\sqrt{[l_c + i v(t - \tilde{s}_2 ) \eta_2] [l_c - i e v\tilde{s}_1 \eta_1]}}  \\
& = \frac{8 I_u}{e^2 V} \frac{l_c^{1/2} e^{i eV t/2 } }{(l_c + i vt)^{1/2}} ,
\end{aligned}
\label{eq:dif_connec|w_A|^2}
\end{equation}
where we shifted the integration variables:  $\tilde{s}_{1,2}=s_{1,2}+L/v$. 
This result reproduces the last term in Eq.~(\ref{three-regions}), which is the last term of Eq.~\eqref{eq:ud_full_correlationsB}.

Note that in the sum over $\eta_1$ and $\eta_2$, here only ${\eta_1=+}$ and ${\eta_2=-}$ contribute: otherwise the contour integral vanishes.
Indeed, since $V>0$, the exponential factor $\exp[-i V ( s_1 - s_2)/2]$ requires the contour integral over $s_1$ and $s_2$ to go through the lower and upper half-plane, respectively, to capture the branch-cut contributions. Equation~\eqref{eq:dif_connec|w_A|^2} yields the last term of Eq.~\eqref{eq:ud_full_correlationsA} for the correlation function of  channel-$u$ operators.
The Region-II contribution to the correlation function of the tunneling operators in channel $d$ can be obtained in an analogous way, yielding
\begin{equation}
\begin{aligned}
    & \big\langle e^{i\phi_d (t^-,L)/\sqrt{2}} 
e^{-i\phi_d (0^+, L)/\sqrt{2}} \big\rangle_2^\text{(II)}
= \frac{8 I_d}{e^2 V} \frac{l_c^{1/2}}{(l_c + i vt)^{1/2}} \exp\left(\frac{i eV t}{2} \right).
\end{aligned}
    \label{eq:dif_connec|w_B|^2}
\end{equation}

It is worthwhile emphasizing that the integral in Eq.~\eqref{eq:dif_connec|w_A|^2} has precisely the same structure as the expression that one would obtain for a fermionic system, where two middle channels can only communicate via the direct tunneling of nonequilibrium fermions at the central QPC.
The difference would only be in replacing the anyonic square-root propagators with the fermionic Green's functions.
Thus, this ``connected'' contribution of Region II to the correlation function of the tunneling operators at the collider can be interpreted as describing direct tunneling of anyons from diluted beams.
As another important fact, both Eqs.~\eqref{eq:dif_connec|w_A|^2} and \eqref{eq:dif_connec|w_B|^2} contain a phase factor, $\exp (\pm i e V t/2)$, which potentially produces an infrared cutoff $(e V)^{-1}$ in the integrals over $t$ involving these correlation functions.
This cutoff greatly contrasts Region II from Region I, where the cutoff in the $t$-integral only appears after resummation over higher-order terms in the expansion in the powers of the diluter transmission [cf. Eq.~\eqref{eq:a_disconnected}]. Specifically, the cutoff of the $t$ integral is provided by the inverse of the nonequilibrium current and, hence, much longer times $t$ are relevant in Region I compared to Region II. We note, however, that in the product of two Region-II terms (coming from channels $u$ and $d$), the oscillatory factor cancels out for symmetric bias $V_u=V_d$, see discussion in the main text. 

Before finishing the analysis of the contribution of Region II to the correlation functions, we briefly discuss the possibility of interfering of the Region-II processes with those of Region I at higher orders in thediluter transmission upon resummation over disconnected nonequilibrium anyonic pairs. Without loss of generality, we consider the next-to-leading-order tunneling processes at the upper diluter (i.e., $\propto |w_u|^4$). In this case, the anyon field operators associated with the diluter have time arguments $s_1$, $s_2$, $s_3$, and $s_4$.
Let us choose the singularities at $s_1\to -L/v$ and $s_2 \to t-L/v$ according to the prescription for Region II, and $s_3 \to s_4$ (far away from $t-L/v$ and $-L/v$) corresponding to Region I.
This interplay of connected and disconnected processes leads to
the vanishing of the ``braiding phase'':
\begin{equation}
\mathcal{R}_{\eta_3,\eta_4}(s_3,s_4,t)
 \frac{[l_c \!+\! i v(s_1 \!\!-\!\! s_3) \chi_{\eta_1,\eta_3} (s_1 \!\!-\!\! s_3)][l_c \!+\! i v (s_2 \!\!-\!\! s_4) \chi_{\eta_2,\eta_4} (s_2 \!\!-\!\! s_4)]}{[l_c \!+\! i v (s_1 \!\!-\!\! s_4) \chi_{\eta_1,\eta_4} (s_1 \!\!-\!\! s_4)][l_c \!+\! i v(s_2 \!\!-\!\! s_3) \chi_{\eta_2,\eta_3} (s_2 \!\!-\!\! s_3)]}
\to \exp\! \left[\frac{i \pi (\eta_3 -\eta_4) }{2} \right]
\exp\!\left[ -\frac{i \pi ( \eta_3 - \eta_4 )}{2 }  \right] = 1.
\label{eq:phase_terms1}
\end{equation}
Since Eq.~\eqref{eq:phase_terms1} does not depend on $\eta_{1,2}$, this next-to-leading order term exactly vanishes after performing the summations over $\eta_3$ or $\eta_4$.
Consequently, one does not need to worry about resummation of higher-order terms that mix the Region-II and Region-I contributions, rendering Eqs.~\eqref{eq:dif_connec|w_A|^2} and \eqref{eq:dif_connec|w_B|^2} the full Region-II correlation functions.

\subsection{Interpretation of equal-time connected contraction (Region III)}

Finally, we move to Region III, where $s_1,s_2\to -L/v$ for $t>0$ or $s_1,s_2\to t-L/v$ for $t<0$: 
both injected anyons participate in the tunneling at the central QPC.
As a difference from Region II, the diluter anyonic operators in Region III have close time arguments, $s_1 \approx s_2$. 
As discussed above, the Region-III contribution to the integral for the correlation function 
is dominated by the vicinity of the branching point in the tangling factor in Eq.~(\ref{second_order}).

\subsubsection{Second-order correlation}

We begin with the case $t > 0$ and focus on branch cuts with $s_1, s_2 \to -L/v$.
In this region, variables $\tilde{s}_1 = s_1 + L /v$ and $s = s_1 - s_2$ are both much smaller in amplitude than $|t|$ and $L/v$.
With these variables, the second-order correlation for Region III can be cast in the form
\begin{equation}
\begin{aligned}
 \big\langle e^{-i\phi_u (t^-,L)/\sqrt{2}} e^{i\phi_u (0^+, L)/\sqrt{2}} \big\rangle_2^{\text{(III)}}
\!\!\approx\! -\frac{|w_u|^2 l_c^{1/2}}{2\pi(l_c + i vt)^{1/2}}\sum_{\eta_1\eta_2}  \int_{-s}^{s}\! d \tilde{s}_1 \int_{-\infty}^\infty \!\!ds\, \frac{ \eta_1\eta_2\, e^{i\frac{\pi}{4} (\eta_1 - \eta_2)} \, e^{-i eV (s +\tilde{s}_1) /2}}{ [l_c + i ( -v \tilde{s}_1) \eta_1]^{1/2}}  \frac{(l_c + i vs  \eta_2 )^{1/2}}{l_c + i v  s \chi_{\eta_1,\eta_2} (s)},
\end{aligned}
\label{eq:region_iii_deltas}
\end{equation}
where assumed $|\tilde{s}_1| < |s|$. 
To proceed, we simplify Eq.~\eqref{eq:region_iii_deltas} with the identity
\begin{equation}
\begin{aligned}
&\frac{(l_c + i vs  \eta_2 )^{1/2}}{l_c + i v  s \chi_{\eta_1\eta_2} (s)} = \frac{e^{-i\frac{\pi}{4} \eta_2}}{\sqrt{|l_c + i v  s|}}= \frac{e^{i\frac{\pi}{4} (-\eta_1 -\eta_2)}}{\sqrt{l_c - i v  s \eta_1}}, \quad s > 0,\\
&\frac{(l_c + i vs  \eta_2 )^{1/2}}{l_c + i v  s \chi_{\eta_1\eta_2} (s)} = \frac{e^{-i\frac{\pi}{4} (2\eta_1+\eta_2)}}{\sqrt{|l_c + i v  s|}}= \frac{e^{i\frac{\pi}{4} (-\eta_1 -\eta_2)}}{\sqrt{l_c - i v  s \eta_1}}, \quad s < 0,
\end{aligned}
\label{eq:remove_theta}
\end{equation}
leading to
\begin{equation}
\begin{aligned}
\big\langle e^{-i\phi_u (t^-,L)/\sqrt{2}} e^{i\phi_u (0^+, L)/\sqrt{2}} \big\rangle_2^{\text{(III)}} 
= & i\, \frac{|w_u|^2 l_c^{3/2}}{(l_c + i vt)^{1/2}} \sum_{\eta_1}  \int_{-\infty}^{\infty} d\tilde{s}_1 \int_{-\infty}^\infty \frac{ ds }{2 \pi l_c} \frac{ \eta_1  e^{-i eV (s +\tilde{s}_1) /2}}{ (l_c - i v s \eta_1)^{1/2}(l_c - i v  \tilde{s}_1 \eta_1)^{1/2} } \\
= & i\, \frac{4 |w_u|^2}{veV} \frac{ l_c^{1/2}}{(l_c + i vt)^{1/2}}.
\end{aligned}
\label{eq:region_iii_deltas_3}
\end{equation}
In Eq.~\eqref{eq:region_iii_deltas_3}, we have extended the integral over $\tilde{s}_1$ from $(-s, s)$ to infinite limits, since the integrand is symmetric with respect to exchanging $s$ and $\tilde{s}_1$. 
Equation ~\eqref{eq:region_iii_deltas_3} reproduces Eq.~\eqref{three-regions} of Appendix~\ref{app:correlation_derivations} and Eq.~\eqref{eq:ud_full_correlationsA} of the main text for $t>0$:
\begin{equation}
    \big\langle e^{-i\phi_u (t^-,L)/\sqrt{2}} e^{i\phi_u (0^+, L)/\sqrt{2}} \big\rangle_2^\text{(III)}\Big|_{t>0} = i \frac{8 I_u}{e^2V} \frac{ l_c^{1/2}}{(l_c + i vt)^{1/2}}.
    \label{eq:region_iii_positive_t}
\end{equation}

It is also instructive to analyze in the same manner the
contribution of the other branching point to the correlation function at $t>0$, setting $s_1, s_2 \to t - L/v$. It turns out that the corresponding integral [an analog of Eq.~\eqref{eq:region_iii_deltas_3}] vanishes after summing over $\eta_1$, since the integrand is independent of $\eta_1$.

For negative times, $t < 0$, phase factors of the last line of Eqs.~\eqref{eq:region_iii_deltas} 
change from $\exp[i\pi (\eta_1 - \eta_2)/4]$ to $\exp[-i\pi (\eta_1 - \eta_2)/4]$.
When $s_1 , s_2 \to -L/v$, the integral becomes modified:
\begin{equation}
\begin{aligned}
\text{Eq.~\eqref{eq:region_iii_deltas}}
\underbrace{\longrightarrow}_{t < 0} &  -\frac{|w_u|^2 l_c^{3/2}}{(l_c + i vt)^{1/2}}\sum_{\eta_1\eta_2}  \eta_1\eta_2 \int_{-s}^{s} d \tilde{s}_1 \int_{-\infty}^\infty \frac{ ds }{2 \pi l_c} \frac{ e^{-i\frac{\pi}{2} \eta_1}  e^{-i eV (s+\tilde{s}_1) /2}}{ (l_c - i v \tilde{s}_1 \eta_1)^{1/2}(l_c - i v  s \eta_1)^{1/2} } =  0
\end{aligned}
\end{equation}
(it vanishes after the summation over $\eta_2$).
Now the Region-III contribution is determined by $s_1 , s_2 \to t-L/v$:
\begin{equation}
\begin{aligned}
\big\langle e^{-i\phi_u (t^-,L)/\sqrt{2}} e^{i\phi_u (0^+, L)/\sqrt{2}} \big\rangle_2^{\text{(III)}} \
\underbrace{\longrightarrow}_{t < 0}\ & -\frac{|w_u|^2 l_c^{1/2}}{2\pi(l_c + i v t)^{1/2}} \sum_{\eta_1\eta_2 } \eta_1\eta_2 \int_{t-s}^{t+s} d\tilde{s}_2 \int_{-\infty}^\infty ds\, \frac{  e^{-i\frac{\pi}{2}\eta_1} e^{-i eV (s -\tilde{s}_2)/2} e^{-ieV t/2}}{(l_c + i v (t - \tilde{s}_2)\eta_2)^{1/2}(l_c + i v  s \eta_2)^{1/2}}  \\
= & \frac{i|w_u|^2 l_c^{1/2}}{2\pi(l_c + i v t)^{1/2}} \sum_{\eta_2 } \eta_2 \int_{-\infty}^\infty d\tilde{s}_2 \int_{-\infty}^\infty ds\, \frac{   e^{-i eV (s -\tilde{s}_2)/2} e^{-ieV t/2}}{(l_c + i v (t - \tilde{s}_2)\eta_2)^{1/2}(l_c + i v  s \eta_2)^{1/2}}\\
= & -i \frac{4 |w_u|^2}{veV} \frac{ l_c^{1/2}}{(l_c + i vt)^{1/2}} = -i \frac{8 I_u}{e^2V} \frac{ l_c^{1/2}}{(l_c + i vt)^{1/2}}.
\end{aligned}
\label{eq:region_iii_6}
\end{equation}

Combining Eq.~\eqref{eq:region_iii_deltas_3} for $t > 0$ and Eq.~\eqref{eq:region_iii_6} for $t < 0$, we arrive at the final result for Region III,
\begin{equation}
    \big\langle e^{-i\phi_u (t^-,L)/\sqrt{2}} e^{i\phi_u (0^+, L)/\sqrt{2}} \big\rangle_2^\text{(III)} = i\, \text{sgn} (t)\frac{8 I_u}{e^2V} \frac{ l_c^{1/2}}{(l_c + i vt)^{1/2}},
    \label{eq:region_iii_final}
\end{equation}
which equals Eq.~\eqref{eq:ud_full_correlationsA} of the main text, if keeping only leading-order contribution of $I_u/V$.
The above consideration underscores that the Region-III contribution to the correlation function of tunneling operators is indeed associated with direct tunneling of anyons from the diluted beam, similar to Region II and in contrast to Region I.

\subsubsection{Resummation over higher-order tunneling events at the diluters}

Similar to Region I and Region II, we need to analyze the role of higher-order processes with multiple injected nonequilibrium anyonic pairs in Region III.
We once again begin with the fourth-order term in the correlation function, where there are four nonequilibrium anyons, with time arguments $s_1$ to $s_4$.
Without loss of generality, we assume that $s_1, s_2 \to -L/v$ or $t-L/v$ are the time arguments of two nonequilibrium anyons that tunnel at the central QPC.
The other two nonequilibrium anyons, with time arguments $s_3 $ and $s_4$, bypass the central QPC.
The interplay between these two pairs of anyonic operators and tunneling operators at the central QPC produces the correlation factor
\begin{equation}
\begin{aligned}
&\mathcal{R}_{\eta_3,\eta_4}(s_3,s_4,t)\,
\frac{[l_c \!+\! i v(s_1 \!-\! s_3) \chi_{\eta_1,\eta_3} (s_1 \!-\! s_3][l_c \!+\! i v(s_2 \!-\! s_4) \chi_{\eta_2,\eta_4} (s_2 \!-\! s_4)]}{[l_c \!+\! i v(s_1 \!-\! s_4) \chi_{\eta_1,\eta_4} (s_1 \!-\! s_4)][l_c \!+\! i v(s_2 \!-\! s_3) \chi_{\eta_2,\eta_3} (s_2 \!-\! s_3)]}\\
 &\quad \to \frac{[ l_c + i v(t-s_3'  )\eta_3]^{1/2} [ l_c + i (-vs_3'  ) \eta_4 ]^{1/2} }{[ l_c + i v(t-s_3' )\eta_4 ]^{1/2} [ l_c + i (-vs_3')  \eta_3 ]^{1/2} } \, \frac{[l_c + i ( - vs_3') \chi_{\eta_1,\eta_3} ( - s_3')][l_c + i (-v s_3') \chi_{\eta_2,\eta_4} (- s_3')]}{[l_c + i ( - vs_3') \chi_{\eta_1,\eta_4} ( - s_3')][l_c + i (- vs_3') \chi_{\eta_2,\eta_3} ( - s_3')]}\\
&\quad \to \exp \left[\frac{i \pi}{2} (\eta_3 -\eta_4) \right].
\end{aligned}
\label{eq:phase_terms2}
\end{equation}
\end{widetext}
Crucially, in contrast to Eq.~(\ref{eq:phase_terms1}) for the Region-II processes,  Eq.~\eqref{eq:phase_terms2} now depends on Keldysh indexes ($\eta_3$ and $\eta_4$) of two disconnected nonequilibrium anyonic operators (with time arguments $s_3$ and $s_4$), meaning that the summation over $\eta_3$ and $\eta_4$ leads to a finite result.
As a consequence, higher-order tunneling processes in Region III yield non-vanishing correlations.

To perform the resummation of higher-order perturbative terms, we consider an expansion in diluters' transmissions at the $2n^\text{th}$ order. This expansion involves two connected operators (i.e., both operators with time argument equal to $-L/v$ or $t-L/v$, for positive and negative $t$, respectively), and $n-1$ pairs of disconnected ones (i.e., anyonic pairs with time arguments that are distant from $-L/v$ and $t-L/v$).
When the latter $n-1$ anyonic pairs are independent of each other, the integral over time arguments of these pairs equals $\left( -4 I_u |t|/e \right)^{n-1}$, following similar arguments for resummation in Region I.

In addition, as discussed in the case of Region I, when performing resummation, one needs to compute the number of options to pair up anyonic operators.
For Region III, the choice of $n-1$ connected pairs can be done in
\begin{equation}
   \frac{2^{n-1}}{(n-1)!} C_{2n-2}^2 C_{2n-4}^2 \cdot \cdot\cdot C_2^2 = \frac{(2n-2)!}{(n-1)!}
\end{equation}
ways, where $ C_{2n-2}^2$ chooses two operators out of $2n-2$ ones, and the factor of 2 in front refers to the option to choose the creation operator from a pair of operators. The denominator $(n-1)!$ is the number of permutations.
As a result, for Region III, the resummation leads to
\begin{equation}
\begin{aligned}
   &\sum_{n=1}^\infty \left[ -\frac{I_{u}}{ e/2} \left( 1 - e^{i\pi} \right) |t| \right]^{n-1} \frac{(2n-2)!}{(n-1)!} (2n-1) 2n \frac{1}{(2n)!}\\
   &=\sum_{n=1}^\infty  \frac{\left[ -\frac{I_{u}}{e/2} \left( 1 - e^{i\pi} \right) |t| \right]^{n-1}}{(n-1)!} =\exp \left( -4\frac{I_{u}}{e} |t| \right),
\end{aligned}
\label{eq:resummation_factor}
\end{equation}
where the factor $(2n-1) 2n$ refers to the number of options to choose two connected operators (i.e., those with time arguments equal to $-L/v$ or $t - L/v$), and $2n!$ on the denominator is the prefactor when performing the Keldysh expansion. With the leading-order expression, Eq.~\eqref{eq:region_iii_final} and the suppression factor from resummation, Eq.~\eqref{eq:resummation_factor}, we have arrived at the last term of Eq.~\eqref{eq:ud_full_correlationsA}.

\section{Corrections to correlation functions}
\label{app:corrections}

Correlation functions Eqs.~\eqref{eq:ud_full_correlationsA} and \eqref{eq:ud_full_correlationsB} both include corrections from resummation over higher-order tunneling events at the diluters. However, these expressions are not exact, since at each order of perturbative expansion, we neglected details of ``braiding'' between nonequilibrium anyonic pairs, as well as the possibility of simultaneous tunneling of nonequilibrium anyons through diluters. 
In this section, we estimate the corrections to the correlation functions, Eqs.~\eqref{eq:ud_full_correlationsA} and \eqref{eq:ud_full_correlationsB}, stemming from the neglected processes.

\begin{figure}[ht] \begin{center} 
\vspace*{0.5cm}
\includegraphics[width=6cm]{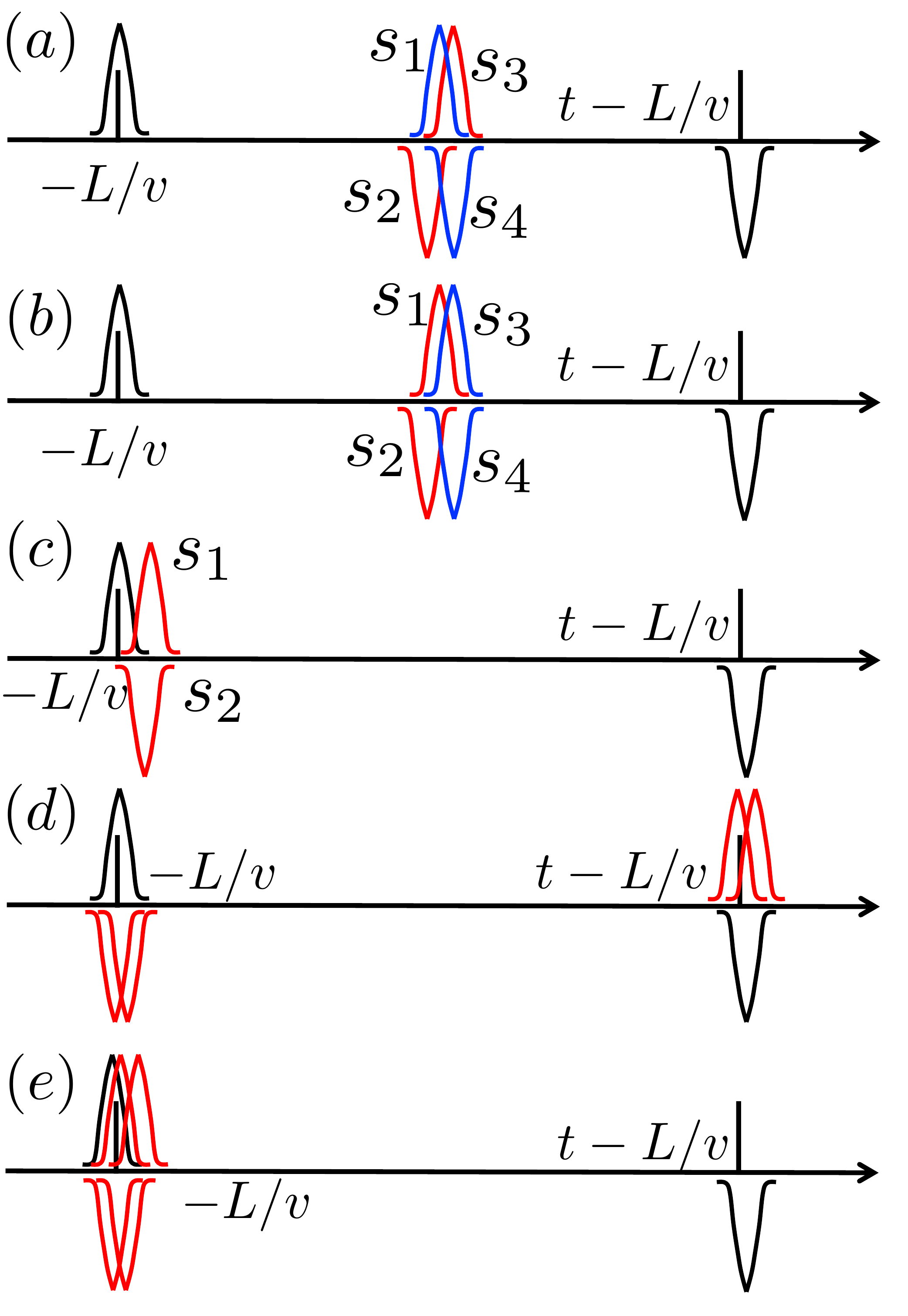}
\end{center}
\caption{Illustration of processes giving rise to corrections to correlation functions. In all panels, the horizontal axis refers to the time at which nonequilibrium anyons are generated at the diluter. Times $t-L/v$ and $-L/v$ correspond to the tunneling events at the central QPC involving anyons injected by the diluters (assuming their propagation with velocity $v$).
Time arguments of the tunneling operators at the central QPC  are marked by black pulses. Anyons injected from sources are denoted by red or blue pluses.
Panels (a) and (b): Correction for Region I, induced by partial overlap between nonequilibrium anyons. Here, anyonic pulses marked by the same color (red or blue) are contracted with each other.
Both contractions shown in (a) and (b) lead to finite contributions, thus leading to ambiguity in contracting nonequilibrium pairs.
Panel (c): The situation considered by Refs.~\cite{IyerX2023, ThammBerndPRL24}, where two nonequilibrium anyons (red pulses) overlap with the tunneling operators of the central QPC (black pulses). The situation is different from that in panels (a) and (b), where only nonequilibrium anyonic pairs overlap with each other. Note that the situation described in panel (c) corresponds to the leading-order term for Region III and is thus accounted for in our calculations.
Panel (d): Correction in Region II. Here, two nonequilibrium anyons tunnel simultaneously at the central QPC. This corresponds to a \textit{collision} of these anyons at the collider. Panel (e): Correction in Region III, where two nonequilibrium anyons tunnel through the central QPC simultaneously (again, a collision of anyons).
}
\label{fig:corrections}
\end{figure}

The contribution of 
Region I to the correlation function receives correction from the overlap between nonequilibrium anyonic pairs, when resummation of perturbative series is performed.
Briefly, the result for Region I, Eq.~\eqref{eq:a_disconnected}, assumes the injected anyons to be spatially distant from each other. This is, however, not necessarily true, especially for weaker dilution.
Indeed, as shown in Figs.~\ref{fig:corrections}(a) and \ref{fig:corrections}(b), when two nonequilibrium pulses partially overlap, there is an ambiguity in defining disconnected pairs. 
The overlap between injected anyonic pairs leads to corrections to the Region-I correlation function.
This correction starts appears at the fourth-order ($\propto w_u^4$ or $w_d^4$, with two pairs of nonequilibrium anyonic operators), as an overlap between anyonic pairs requires the presence of at least four nonequilibrium anyonic operators.
Consequently, correction to the Region-I correlation function will be of order $O(I_u^2/V^2)$ and $O(I_d^2/V^2)$ in the prefactor.

It is worth noticing that the correction induced by the overlap of nonequilibrium anyons is different from Refs.~\cite{IyerX2023, ThammBerndPRL24}, where corrections originate from the overlap of nonequilibrium anyons with anyons that tunnel at the central QPC [cf. Fig.~\ref{fig:corrections}(c)]. Actually, the corrections addressed in Refs.~\cite{IyerX2023, ThammBerndPRL24} are similar to the leading-order contribution of Region III.
A more rigorous connection between the results of  Refs.~\cite{IyerX2023, ThammBerndPRL24} and our Region-III contribution remains to be established.
In addition, corrections in Region I, induced by the overlap between nonequilibrium anyons, appear in all orders of perturbation theory. Consequently, it should also appear in the argument of the exponential function upon resummation.

Now, we move to Region II, for which the main correction stems from the possibility that two nonequilibrium anyons, from either $su$ or $sd$, tunnel through the central collider simultaneously [cf. Fig.~\ref{fig:corrections}(d)].
This amounts to a direct anyonic collision at the collider.
The corresponding modification of the Region-II correlator will be of order $O(I_u^2/V^2)$ and $O(I_d^2/V^2)$.

Finally, we discuss corrections to the Region-III correlator.
Similar to that of Region I, it involves resummation of higher-order terms and thus also receives corrections from the overlap between nonequilibrium anyonic pairs.
As the leading contribution of Region III is proportional to $I_u/V$ or $I_d/V$, anyonic-pair overlap leads to corrections of the order of $O(I_u^3/V^3)$ and $O(I_d^3/V^3)$, even smaller than corrections in Regions I and II mentioned above.
In addition to this small correction, Region III also receives corrections from events where two nonequilibrium anyonic pairs simultaneously tunnel at the central collider [cf. Fig.~\ref{fig:corrections}(e)], leading to corrections of the order of $O(I_u^2/V^2)$ and $O(I_d^2/V^2)$.

To conclude, following the above discussions, all three Regions receive leading corrections of the order of $O(I_u^2/V^2)$ and $O(I_d^2/V^2)$. These corrections generate the terms of the order of $O(I_u^2/V^2)$ and $O(I_d^2/V^2)$ in the prefactors of the exponential terms in the correlation functions and of order $O(I_u^2|t|/V)$ and $O(I_d^2|t|/V)$ in the arguments of the exponentials. Therefore, these corrections to correlation functions Eqs.~\eqref{eq:ud_full_correlationsA} and \eqref{eq:ud_full_correlationsB} can be neglected in the strongly diluted limit, when the tunneling current and tunneling-current noise are calculated.

\end{document}